# CROWDSOURCING: A FRAMEWORK FOR USABILITY EVALUATION

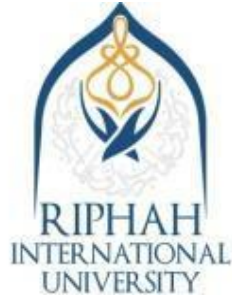

By

Muhammad Nasir
Registration #: F14C04P01004
SAP #: 2697

Supervised by
Prof. Dr. Naveed Ikram

A thesis submitted in partial fulfillment of the requirements for the degree of
Doctor of Philosophy
In
Computing
at
Riphah International University,
Islamabad, Pakistan

May 2022

## RIPHAH INTERNATIONAL UNIVERSITY, ISLAMABAD

### APPROVAL SHEET

### SUBMISSION OF HIGHER RESEARCH DEGREE THESIS

*The following statement is to be signed by the candidates 'supervisor (s), Dean/ HOD, and must be received by the COE prior to the dispatch of the Thesis to the approved examiners.*

**Candidate's Name & Reg#:** Muhammad Nasir, Registration #: F14C04P01004, SAP # 2697

**Program Title:** PhD  Computing

**Faculty/Department:** Faculty of Computing

**Thesis Title:** Crowdsourcing: A Framework for Usability Evaluation

*I hereby certify that the above candidate's work, including the Thesis, has been completed to my satisfaction and that the Thesis is in a format and of an editorial standard recognized by the faculty/department as appropriate for examination. The Thesis has been checked through Turnitin for plagiarism (test report attached).*

Signature (s):

Principal Supervisor: 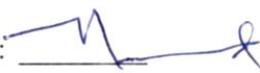

Date: ___________________

The undersigned certifies that:

1.      The candidate presented at a pre-completion seminar an overview and synthesis of the thesis's major findings and that the research is of a standard and extent appropriate for submission as a thesis.

2.      I have checked the candidate's Thesis, and the faculty/department recognizes its scope, format, and editorial standards as appropriate.

3. The plagiarism check has been performed. Report is attached

Signature (s):

Dean/Head of Faculty/Department: 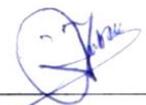

Date: ___________________

# DECLARATION OF AUTHENTICATION

*I certify that the research work presented in this Thesis is my own to the best of my knowledge. All sources used and any help received in the preparation of this dissertation have been acknowledged. I hereby declare that I have not submitted this material, either in whole or in part, for any other degree at this or any other institution.*

*Signature………………* 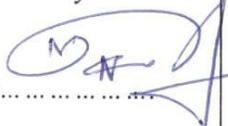

## <u>ACCEPTANCE CERTIFICATE</u>

# Crowdsourcing: A Framework for Usability Evaluation

By
## Muhammad Nasir

A thesis submitted in the partial fulfillment of the requirements for the degree of
**Doctor of Philosophy in Computing**
We accept this thesis as conforming to the required standard.

**Examination Committee:**

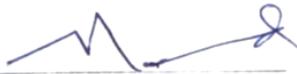

**Supervisor**
**Dr. Naveed Ikram**
Professor
Faculty of Computing
Riphah International University
Islamabad

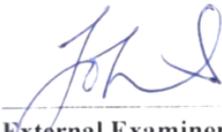

**External Examiner**
**Dr. Muhammad Zohaib Iqbal**
Professor
National University of Computer and
Emerging Sciences (FAST)
Islamabad

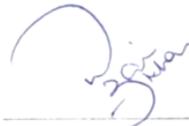

**External Examiner**
**Dr. Muhammad Uzair Khan**
Associate Professor
National University of Computer and Emerging
Sciences (FAST)
Islamabad

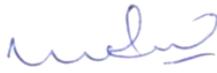

**Dr. Musharraf Ahmed**
**Head, Faculty of Computing**
Riphah International University
Islamabad

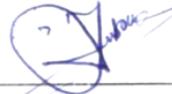

**Prof. Dr. Muhammad Zubair**
**Dean, Faculty of Computing**
Riphah International University
Islamabad

**FACULTY OF COMPUTING**
**RIPHAH INTERNATIONAL UNIVERSITY,**
**ISLAMABAD, PAKISTAN**
**2022**

## CERTIFICATE OF APPROVAL

This is to certify that the research work presented in this thesis, entitled "**Crowdsourcing: A Framework for Usability Evaluation**" was conducted by **Mr. Muhammad Nasir** under the supervision of "**Dr. Naveed Ikram**". No part of this thesis has been submitted anywhere else for any other degree. This thesis is submitted to the "Faculty of Computing" in partial fulfillment of the requirements for the degree of Doctor of Philosophy in Computing,

**Faculty of Computing, Riphah International University, Islamabad, Pakistan.**

Student Name: **Muhammad Nasir**                    Signature: 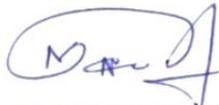

**Examination Committee:**

    a)    **External Examiner 1:**
        **Dr. Muhammad Zohaib Iqbal**              Signature: 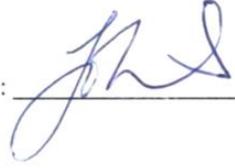
        Professor,
        National University of Computer &
        Emerging Sciences (FAST), Islamabad

    b)    **External Examiner 2:**
        **Dr. Muhammad Uzair Khan**               Signature: 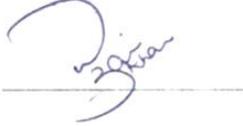
        Associate Professor,
        National University of Computer &
        Emerging Sciences (FAST), Islamabad

    c)    **Internal Examiner:**
        **Dr. Musharraf Ahmed**                   Signature: 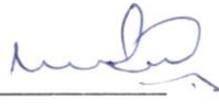
        Assistant Professor,
        Faculty of Computing
        Riphah International University, Islamabad

Supervisor Name(s):

        **Prof. Dr. Naveed Ikram**                Signature: 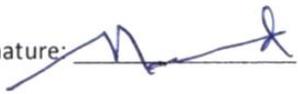

Head, Faculty of Computing:
        **Dr. Musharraf Ahmed**                   Signature: 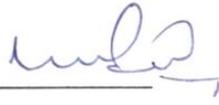

Dean, Faculty of Engineering & Applied Sciences:
        **Prof. Dr. Muhammad Zubair**             Signature: 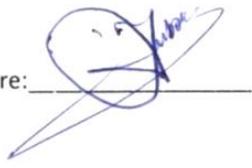

# ACKNOWLEDGEMENT

All glory and gratitude be to Allah, the most merciful and generous, who gave us the ability to reason and the courage to venture out into the realms of wisdom. Peace be upon His prophet Muhammad ﷺ and his faithful followers, who are a source of knowledge for those who seek it.

I am thankful to my supervisor, Dr. Naveed Ikram, for his consistent efforts, encouragement, and help to complete this Thesis.

I want to pay my heartfelt gratitude to my parents, whose prayers and support always inspired me. I am also grateful to my wife, Zakia Jalil, for her unwavering support during this period. Besides, I would like to express my gratitude to my elder brothers, Amjad Ali and Azhar Mehmood, and my younger sister, who has always been very supportive.

I would also like to express my sincere gratitude to my ex-colleague Dr. Muhammad Usman for introducing me to my supervisor Dr. Naveed Ikram and encouraging and motivating me in my Ph.D. studies.

Finally, I would like to thank my colleague, friend, and class-fellow Muhammad Waseem for insightful research discussions.

# DEDICATION

I dedicate this research endeavor to my family, friends, and teachers…

*"It is He (الله) who made the stars so that they can guide you in land and sea when it is dark. We (الله) have made the signs clear for those who have knowledge" – (Al-Quran 6:97)*

# PUBLICATIONS

The publications made as part of this thesis work are listed below:

1. Nasir, M., Ikram, N., & Jalil, Z. (2022). Usability inspection: Novice crowd inspectors versus expert. *Journal of Systems and Software*, 111122. **– Impact Factor 2.829**

# STUDIES IN SUBMISSION PROCESS

1. Nasir, M., Ikram, N., Jalil, Z. Framework for crowd usability inspection: A case study.

2. Nasir, M., Ikram, N. Usability inspection methods: A systematic mapping study.

# TABLE OF CONTENTS

# LIST OF TABLES

# LIST OF FIGURES/ ILLUSTRATIONS

# ABSTRACT

**Objective:** The objective of this research study is to explore and exploit crowdsourcing for software usability evaluation.

**Background:** Usability evaluation studies are necessary to design software products that are easy to use and deliver high user satisfaction. However, traditional usability evaluation methods are time-consuming and expensive. Crowdsourcing has emerged as a cost-effective and quick means of remote software usability evaluation platform. Nevertheless, getting quality feedback from the crowdsourcing platform is still an outstanding challenge. Research studies examining the wisdom of the crowd for usability evaluation are few and far between. We have not found any study that would have devised a framework for crowd usability evaluation.

**Method:** To satisfy the stated objective, we first conducted a systematic mapping study to examine the current research evidence for usability evaluation. Later, a multi-experiment study was conducted to explore the wisdom of crowd using expert heuristic evaluation as a benchmark to compare novice crowd usability inspectors with experts. The findings of the study were used to design a framework for crowd usability inspection. Subsequently, a third study was conducted to validate the proposed framework in practical settings through a case study, and findings were presented.

**Results:** The systematic mapping study showed that expert heuristic evaluation is a widely employed usability evaluation method with websites as the most investigated application type. Furthermore, the results of the experimental study showed that, on average, novice crowd usability inspection guided by a single expert's heuristic usability inspection (novice crowd usability inspection henceforth) in comparison with expert heuristic evaluation can identify the same usability problems with respect to content, quality, severity, with time efficiency. The results of the case study indicated that the framework for crowd usability inspection enabled software usability practitioners to effectively conduct novice crowd usability inspections and interpret their findings into a successful re-design of their website. The self-validating iterations of novice crowd usability inspections in the case study indicated that in less than three iterations, with 3-5 novice crowd usability inspections, can neutralize key usability issues in a software product.

**Conclusion:** This thesis established that crowdsourcing is effective for usability evaluation and gives promising results comparable to expert heuristic evaluation. Moreover, a framework for crowd usability inspection was proposed, and its use in practical settings was demonstrated in a medium-sized budget software company. This thesis concludes that novice crowd usability inspection employing the framework for crowd usability inspection is an effective and efficient substitute for budget-constrained software development organizations.

*Keywords*: crowdsourcing, crowd usability evaluation, expert heuristic evaluation, framework.

# Chapter 1

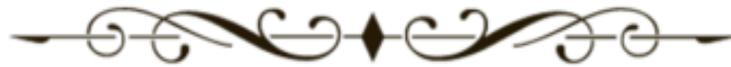

# 1. Introduction

Usability evaluations are necessary for developing software products with high user satisfaction; however, acceptance of usability evaluation practices is quite inadequate in resource-constraint software development enterprises (Bak et al., 2008; Ardito et al., 2011).

Several studies have been conducted to highlight the obstacles in adopting software usability practices. These obstacles mainly include resistance towards adopting usability practices by software developers, difficulty in understanding the concepts of usability practices, and resource limitations (Bak et al., 2008; Ardito et al., 2011; Häkli, 2005).

The management often overlooks usability practices to cut costs for software products (Cajander, Gulliksen, & Boivie, 2006). A survey was conducted in a small geographic area in Denmark where representatives from 17 software development organizations were interviewed to find the obstacles in adopting software usability practices in software development organizations. The results revealed that the main obstacle is resource constraint (Bak et al., 2008). A replication survey was conducted in Italy three years later (Ardito et al., 2011). The major obstacle identified in adopting usability practices was resource constraint.

Nevertheless, usability evaluation is costly (Nielsen, 1994a), but they have a significant return on investment and revenue increases (Juristo, Moreno, & Sanchez-Segura, 2007; Nielsen, 2003; Donahue, 2001; Chrusch, 2000). After collecting data from 863 projects, it was found that if we spend 10% of the project's budget on usability practices, then we can get 135% on the average increase in desired metrics. The results for an increase in desired metrics are different depending upon the metric. For example, web projects' sales increased 100%, traffic or visitor count increased 150%, user performance or productivity increased 161%, and use of specific features increased 202% (Nielsen, 2003).

There are several usability evaluation methods. Some of them collect data from the user, i.e., usability testing, while others are based on expert evaluations, i.e., usability inspections. Traditional usability testing methods are considered more effective, but they are expensive (Nielsen, 1994a; Rosenbaum, 1989; Nielsen, 1989). Usability inspection methods were introduced as low-cost methods compared to usability testing (Nielsen, 1994a). Usability inspection methods work fine with 3-5 experts. However, hiring multiple usability experts for budget-constrained software development organizations is also not affordable. Therefore, there was a need to develop and adopt such usability practices that are effective, efficient, and affordable for low-budget software development organizations.

A couple of studies have been conducted to review the knowledge on usability inspection methods (Hollingsed & Novick, 2007; Insfran & Fernandez, 2008). However, there was a need to review the recent knowledge on usability inspection methods. Therefore, as part of this thesis work, we first conducted a systematic study to gather knowledge on usability inspection methods in one place.

Some of the authors have explored crowdsourcing for usability evaluations to overcome the costs in traditional usability evaluations (Bruun and Stage, 2015; Guaiani and Muccini, 2015; Liu et al., 2012). However, this limited evidence is not adequate. Several challenges are involved in crowdsourcing, including poor quality of work (Peer, Vosgerau, & Acquisti, 2014; Goodman, Cryder, & Cheema, 2013), the validity of demographic and professional details (Liu et al., 2012), bogus work (Kittur, Chi, & Suh, 2008). Therefore, there was a need to examine the potential of the wisdom of the crowd for usability evaluation. In this Thesis, a multi-experiment study was conducted to compare the novice crowd inspector's usability evaluation with expert heuristic usability evaluation. Later, we proposed a framework for crowd usability evaluation to represent our findings on crowd usability inspection. We then validated the proposed framework by means of a case study to examine its effectiveness in practical settings. In other words, this Thesis aims to propose an effective and economical alternative method for budget-constrained software development organizations for usability evaluation.

This thesis makes the following contributions, employing a combination of primary (experiment, case-study) and secondary studies (systematic mapping study):

1. Collected and reported evidence on usability inspection methods.
2. Examined the wisdom of the crowd for usability evaluation and identified and discussed different factors that influence crowd usability evaluation.
3. Proposed a framework for novice crowd usability inspection.
4. Validated the proposed framework for crowd usability inspection in practical settings by means of a case-study.

The rest of the thesis is arranged as follows: In chapter 2, I discussed the background of the thesis, including a brief introduction to different concepts of software engineering investigated in thesis work. In chapter 3, I described the research gaps, contributions, and the research questions that have been investigated in this thesis. In chapter 3, I also briefly describe the research methods employed in the studies conducted as part of this thesis work. In chapter 4, I present a systematic mapping study discussing trends for usability inspection methods. In chapter 5, I reported a multi-experiment study comparing novice crowd usability inspection with expert heuristic usability inspection. In chapter 6, I presented a framework for crowd usability inspection and validates it with a real case from the software industry. In chapter 7, I presented results and discussion. The validity threats are discussed in chapter 8. Finally, in chapter 9, I presented the conclusion and future work.

# Chapter 2

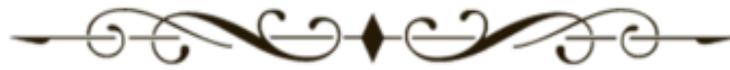

# 2. Background

This Thesis involves different software engineering concepts, including usability evaluation methods, the wisdom of the crowd, crowdsourcing platforms, usability evaluation using crowdsourcing. These concepts are discussed in section to provide an overview of the concepts in which this Thesis made research contributions.

## 2.1 Usability evaluation methods

In the last three decades, researchers and practitioners have reported different usability evaluations methods (UEMs). A couple of secondary studies (survey, systematic reviews, and mapping studies) have also been conducted to present an overview of the literature on usability evaluation methods (Hollingsed & Novick, 2007; Insfran & Fernandez, 2008; Rivero, Barreto, & Conte, 2013; Bastien, 2010). There are two main categories of the UEMs, i.e., user testing and usability inspections (Rocha & Baranauska, 2003).

### 2.1.1 User Testing

Usability testing involves testing the system with real users performing pre-defined tasks based on the user-centered interaction model. The usability evaluators collect the testing data from end-users through observation or post-test interviews and questionnaires. The evaluators can later analyze the test data to identify the usability problems and improve the system's interactive design. However, usability testing methods are not cost-effective and require special arrangements, i.e., laboratory, end-users with different profiles (Matera, Rizzo, & Carughi, 2006). Moreover, usability testing applies in later stages of the system development life cycle where the system is complete or partially complete, making the system's changes more expensive.

There are several usability testing methods with procedural variations, e.g., Traditional laboratory testing, Think-a-loud protocol (Boren & Ramey, 2000), Remote usability testing (synchronous and asynchronous), etc. (Bastien, 2010). In contrast to laboratory usability testing, synchronous remote usability testing is economical and gives consistent results, e.g., cuts travel expenses (Bastien, 2010). However, finding a sufficient number of actual users with different authentic profiles is still challenging, and user hiring cost is also like laboratory user testing. Nevertheless, remote usability testing gives freedom from laboratory requirements and saves time.

### 2.1.2  Usability Inspection

Since the early 1990s, when usability evaluation became a major research concern, a core argument was the effectiveness of usability testing compared to the cost-effectiveness of usability inspection methods (Jeffries et al., 1991; Desurvire, Kondziela & Atwood, 1992; Jeffries & Desurvire, 1992; Karat, Campbell & Fiegel, 1992). Usability inspection methods were introduced as a cost-effective substitute for software development organizations (Nielsen, 1994a; Hollingsed & Novick, 2007). Usability inspection methods require fewer resources and are applicable in the early stages of the system development process, lowering the cost of change in software development (Rocha & Baranauska, 2003).

Usability inspection methods involve multiple usability experts experienced in usability practices and based their evaluations on usability guidelines and heuristics (Rivero, Barreto, & Conte, 2013). Researchers introduced several usability inspection methods in the last three decades, including expert heuristic usability inspection, cognitive walkthrough, pluralistic usability walkthrough, formal usability inspections, and perspective-based inspections (Hollingsed & Novick, 2007; Rivero, Barreto, & Conte, 2013).

Although usability inspection methods were introduced as an economical substitute for usability testing, hiring 3 to 5 usability experts is still not affordable for budget-constrained software organizations. In this regard, some researchers have explored the potential of novice inspectors.

**Expert Heuristic Usability Inspection**

The expert heuristic usability inspection method was developed by Jakob Nielsen (Nielsen, 1992). In this method, usability experts use heuristics to evaluate a user interface design for usability problems. Usability heuristics are the rules of thumb (a set of general principles) in interactive design to ensure and measure usability, e.g., visibility of system status, the match between the system and the real-world, user control and freedom, etc. Usability experts examine the compliance of user interfaces with heuristics and report its violations, if found, and associate it with relevant heuristics. Nielsen's ten usability heuristics are the most widely used in interaction design and evaluations in general (Nielsen, 1994b). However, researchers proposed several other domain-specific heuristics later as well, see subsection 4.3.2 for more details.

**Expert Review**

Expert reviews as the name suggest involving usability experts. Expert review is not only based on usability heuristics but also engages other recognized usability guidelines, concepts relevant to human-computer interaction design and psychology, as well as the reviewer's expertise and previous experience in the subject.

**Cognitive Walkthrough**

A cognitive walkthrough is a non-user-based method for usability evaluation introduced by Polson et al., (1992). A group of inspectors based on their skills and experience evaluate the usability of a system using pre-defined tasks from the perspective of a new user. It is a task-based method, unlike usability heuristics. A group of inspectors, walkthrough each task and document each stage of the workflow and answers the pre-defined questions. The purpose of the cognitive walkthrough is to identify those areas of the user interface that are difficult to learn and use for new users. Other examples of Cognitive walkthroughs can be found in the studies (Clayton, Biddle, & Tempero, 2000) and Filgueiras et al., (2009). Examples of the questions that are answered for each task (patient check-in, online payment) are as below:

1. Would users notice that the correct action is available to achieve the goal?
2. After performing the action, would the user notice the progress is made towards achieving the goal?
3. Would users be able to link correct action with desired results?

## 2.2    Wisdom of the crowd

Crowdsourcing is the idea of taking benefit from the wisdom of the crowd and performing intelligent tasks using it. Surowiecki talked about this in his famous book The Wisdom of Crowd for the first time in 2005. He maintained that the intelligence and wisdom of a large group of people collectively are more promising than hiring a few experts for decision-making or problem-solving. He termed the large group of people as a crowd and argued that it is not the case that the crowd will appear wise in all situations. Instead, a wise crowd carries some personality traits and characteristics; they will have diversity in their opinion because every individual has their thought process based upon their knowledge, so diversity is a must. Additionally, the wise crowd must have the ability to motivate people so that individual opinions are converted into a collective decision.

Crowdsourcing has been used extensively in the field of computing for the last decade. In recent years, many empirical studies have been conducted on this topic (Ambreen and Ikram, 2016). It has also been investigated for usability evaluation (Bruun and Stage, 2015; Guaini and Muccini, 2015; Liu et al., 2012). The integrated definition of crowdsourcing was first provided by Estell' es-Arolos et al. (2012). It defines it as a collaborative activity conducted online so that an individual or organization announces a task via an open call to a group of people with varied knowledge, skills, and experience on a crowdsourcing platform. The posted jobs can be of different complexity levels and offer individual financial benefits, demanding different knowledge levels. In return, that individual or organization who posted the job will get the job done.

Uber, a famous international ride-hailing service is an example of crowdsourcing in transportation. Uber connects available drivers with people who need a ride. The riders get the ride, and the drivers get paid for it. Another well-known example of voluntary-based crowdsourcing is Wikipedia. Wikipedia is a free encyclopedia authored and maintained by a group of volunteers with an open collaboration model of crowdsourcing. There are different types of crowdsourcing, e.g., Crowdfunding, raising small donations from a larger number of people to finance a new business initiative.

## 2.3    Crowdsourcing platforms

There are different online platforms available for crowdsourcing, including Amazons Mechanical Turk (mTurk) (mTurk, 2022), uTest (uTest, 2022), and Upwork (Upwork, 2022), formerly known as oDesk, etc. Usability evaluations on crowdsourcing platforms can be performed in a questionnaire containing quantitative and qualitative questions that a user must fill after performing tasks on a provided software system. Although crowdsourcing has significant opportunities, there are still some challenges.

Challenges faced by mTurk are no different than other crowdsourcing platforms like the task you post on mTurk must be designed carefully to make it harder to cheat. However, answering open-ended questions or containing user ratings is often harder to classify, whether from a valid user or a malicious one. Another challenge faced by mTurk is the validity of demographic details provided by the mTurk worker. Moreover, it's hard to control the ecological settings for an experiment on mTurk (Kittur, Chi, & Suh, 2008). Lastly, mTurk works for limited countries due to billing and tax issues.

uTest is a crowdsourcing platform specifically designed for software testing. Controlling the ecological settings (e.g., specific software type and version & hardware, software deployment environment) and demographic information (e.g., education, experience, and profile of the user, etc.) is easier on uTest than mTurk as workers on mTurk are anonymous or job requester must trust the self-reported demographic details by a worker (Liu et al., 2012). Although the workflow model of uTest is more promising than mTurk, the pricing scheme on uTest is a bit expensive. A typical HIT spanning over an hour on mTurk costs USD 1-2, while the minimum payment on uTest is USD 25-30 for a task that requires approximately one hour of work (Kittur et al., 2008; uTest, 2022). To receive payments on uTest, a tester needs to set up either a Payoneer or a PayPal account.

Upwork is one of the largest freelancing online platforms. Nevertheless, finding a suitable freelancer on Upwork at an economical rate is not easy. It takes quite a bit of time to interview different freelancers and negotiate with them on pricing. Preferably, you should have a credit card to make payments to freelancers on Upwork. Other payments methods include Payoneer, PayPal, and local bank transfers. A typical usability expert with some certification for usability evaluation and above 80% success rate for job completion on Upwork charges more than USD 40 per hour. Unless a job posted by a client offers a very handsome payment package, freelancers are not attracted to it. Besides, freelancers with strong profile do not opt for smaller tasks; instead, they choose large projects.

## 2.4.    Usability studies using crowdsourcing.

A few usability evaluation studies are investigating the wisdom of the crowd. A study to compare traditional usability testing with remote usability testing was conducted using Amazon's Mechanical Turk by Liu et al. (2012). A school's website was taken as an object of the experiment. The school's current students participated in the conventional usability testing part of the experiment, giving them an edge of familiarity with the website. Whereas the crowd usability testing participants were not familiar with the said website, negatively affecting their performance. However, crowd usability testing was found economical, fast, and easy.

We believe that by controlling some factors, the methodology of the study could have been improved. For instance, both sides of the participants should have the same experience level with the website's usage. Additionally, the selected crowdsourcing platform, the CrowdFlower, does not allow discarding the poor-quality results submitted by the crowd user, resulting in no incentive for the quality of work.

Another study found that teams can be formed dynamically on a crowdsourcing platform and named flash teams (Retelny et al., 2014). The flash team consists of experts in a particular domain, and they collaborate to complete a project. One of the tasks in the experiment conducted by Retelny et al. (2014) was the heuristic usability evaluation. They have concluded that flash teams can be more time-efficient than the normal teams on crowdsourcing platforms. However, it is quite expensive to hire an expert on crowdsourcing platforms. Secondly, the study was focused on the wisdom of experts using crowdsourcing.

Bruun and Stage (2015) conducted a study and introduced a technique called Barefoot. The technique suggests that local software practitioners may be trained to conduct usability testing in resource-constrained organizations. Later, they compared Barefoot with crowdsourced usability testing while hiring university students by sharing minimalist usability training material online. The Barefoot approach was found more suitable for small budget organizations that cannot afford full-time usability experts. Besides, crowd usability testing is not suitable without equipping real users with HCI competencies. Moreover, crowd usability testing requires end-user reports analysis.

We, however, recommend a few different settings for this study. We assume that if the sample population for crowdsourcing had been hired through a crowdsourcing platform, results might have differed as crowdsourcing platforms also have experienced crowd-workers (Estell' es-Arolas et al., 2012). Moreover, since they argued that crowd usability testing works well with the crowd's competencies in HCI and usability testing, hiring a skilled crowd could still be more economical than hiring a usability expert. It can be accomplished by technically designing use-cases for the crowd workers, significantly increasing their performance. Moreover, this study did not address crowd usability inspection in comparison with expert heuristic usability inspection.

# Chapter 3

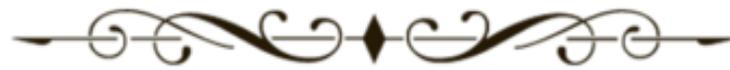

# 3. Research Design

## 3.1 Research Gap

Many studies have identified that one of the major obstacles to adapting usability practices is a budget constraint in small and medium-size software development organizations (Bak, Nguyen, Risgaard, & Stage, 2008; Ardito et al., 2011; Häkli, 2005). Budget-constrained organizations are often reluctant to adapt usability practices to reduce software development costs. Traditional usability evaluation methods are expensive and time-consuming, i.e., usability testing and expert evaluation (Nielsen, 1994a; Rosenbaum, 1989; Nielsen, 1989). Therefore, there is a need for a usability evaluation framework that is economical and effective.

The crowd usability evaluation methods are exposed to a higher risk of failure due to the unavailability of a guiding framework for crowd usability evaluation. A methodological framework for research using crowdsourcing has been designed by Keating and Furberg (2013). Although this framework is quite insightful concerning crowdsourcing, it is too abstract to use for crowd usability evaluation.

However, keeping in view the importance of usability evaluation practices (Nielsen, 2003, Donahue, 2001, Chrush, 2000) and the obstacles faced by medium and small-sized organizations (Ardito et al., 2011; Bak et al., 2008; Häkli, 2005, Cajander et al., 2006), there is a need for a framework for usability evaluation that is economical, less time-consuming, and easy to understand and implement.

## 3.2 Research Objectives & Contributions

The following research objectives were achieved in my Ph.D. thesis:

**Objective 1:** To explore the current research front in the area of crowd usability inspection.

**Contribution:** A Systematic Mapping Study was performed to investigate the current research evidence on software usability inspection methods.

**Objective 2:** To explore the wisdom of the crowd for usability inspection.

**Contribution:** A multi-experiment study was conducted to examine the effectiveness and efficiency of crowd usability inspection.

**Objective 3:** To design a framework for crowd usability inspection.

**Contribution:** A framework for crowd usability inspection was designed in light of the findings of a multi-experiment study.

**Objective 4:** To evaluate the framework for crowd usability inspection in practical settings.

**Contribution:** An industrial case study was performed to validate the crowd usability inspection framework.

## 3.3    Research Questions

The research questions followed by relevant research studies are as below:

**RQ 1:** What evidence has been reported in the literature regarding usability inspection methods?

**RQ 2:** Is the wisdom of crowd effective for usability evaluation?

- **RQ 2a:** Which of the following technique is better in terms of time spent on usability evaluation, cost incurred, number of usability problems found, and severity level of usability problems.

  1. Expert Usability Evaluation
  2. Crowd Usability Evaluation based on usability heuristics and use cases

- **RQ 2b:** What factors are important in using crowdsourcing for effective and efficient usability evaluation, i.e., a profile of crowd-worker, usability heuristics, use-cases, crowdsourcing platform, payment method (i.e., fixed price, per hour)?

**RQ 3:** Is the crowd usability inspection framework effective in practical settings?

Table 3.1 presents the studies with relevant research questions that have been addressed in this thesis work.

Table 3.1 – Studies and research questions

| Research Questions | Study#1 | Study#2 | Study#3 |
|---|---|---|---|
| RQ#1 | ✓ | | |
| RQ#2a | | ✓ | |
| RQ#2b | | ✓ | ✓ |
| RQ#3 | | | ✓ |

## 3.4    Research Strategy

To address the research objectives and questions defined in subsections 3.2 and 3.3, a multipronged research strategy was adopted as shown in figure 3.1. The main theme of this research endeavor was to enable small and medium-sized budget-constraint software development organizations to adopt software usability practices economically. Usability inspection methods were proposed as discount usability evaluation (Nielsen, 1989). Therefore, as part of this research work, usability inspection methods were investigated using a systematic mapping study. The objective of this systematic mapping study was to examine the current research trends in the use of inspection methods and explore any new methods and techniques for discount usability evaluation. This mapping study further led us to crowdsourced and novice usability evaluation for discount usability evaluation. Later in study 2, a comprehensive literature review on crowdsourced usability evaluation was conducted as part of multiple experimental studies. This experimental study focused on exploiting novice usability inspection in combination with crowdsourcing as an alternative means for usability evaluation for budget constraint software organizations. Later, in study 3, a framework for crowd usability inspection was proposed based on the findings of the experimental study, i.e., study 2. This framework was then validated in practical settings using a case study. This case study was designed in a way that with 2-3 iterations of novice crowd usability inspections employing 3 to 5 crowd inspectors could eliminate key usability issues in a software product.

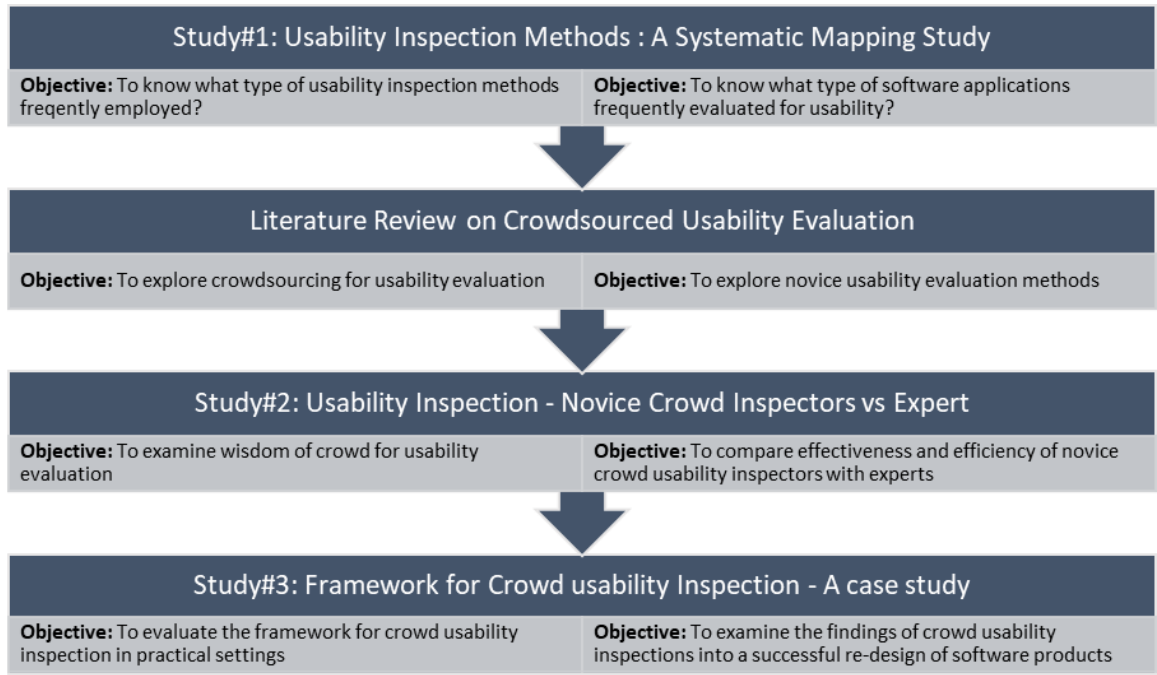

Figure 3.1 – Research Strategy

## 3.5    Research Method

The Thesis has used multiple research methods to answer the research questions. Therefore, we considered a mixed-method research strategy (Shull, Singer, Sjøberg, 2007). Research methods were selected based on their suitability for each research question. Table 3.2 shows the research methods employed in relevant studies. These methods are discussed in detail in relevant studies. An overview of the research methods is provided in the following subsections.

Table 3.2 – Research methods and studies

| Research Methods | Study#1 | Study#2 | Study#3 |
|---|:---:|:---:|:---:|
| Systematic Mapping Study | ✓ | | |
| Experiment | | ✓ | |
| Case Study | | | ✓ |

### 3.5.1 Systematic Mapping Study

A systematic mapping study identifies primary studies on a particular research topic and provides broad research trends on that research topic (Kitchenham & Charters, 2007). The results of the mapping study can be further used to start other primary studies. A mapping study begins with a protocol design to conduct the study, i.e., defining the research questions, search string, inclusion/exclusion criteria, and data extraction process. Finally, the study is executed, and results are reported. However, all of these steps are performed at a broad level. Unlike systematic literature review, mapping studies are less rigorous and do not synthesize the primary studies in detail (Kitchenham & Charters, 2007).

Study 1 (Chapter 9) reports a systematic mapping study to identify the current research front in usability inspection methods. The study's goal was to get an overview of the usability inspection methods reported in the research.

### 3.5.2 Experimental Study

The objective of the experimental research is to investigate a phenomenon in laboratory settings by manipulating one independent variable and controlling others to study its effects on the dependent variable (Wohlin et al., 2012). In experimentation, subjects are given different treatments randomly. However, when random treatments are not possible, then we may employ quasi-experiments. In quasi-experiments, treatments are not assigned randomly; instead, treatments emerge from the characteristics of the subjects or objects (Wohlin et al., 2012). The effect of manipulation is measured using statistical analysis. In other words, statistical inference is used to determine the statistical significance of one technique being better than the other. The process of the experimentation includes activities, i.e., definition, planning, operation, analysis & interpretation, and presentation of results.

Study 2 (Chapter 5) reports an experimental study. We examined the crowd's wisdom with respect to usability evaluation while using expert usability evaluation as a benchmark. We investigated how novice crowd inspectors perform in comparison with expert heuristic evaluation.

### 3.5.3    Case Study

A case study investigates a phenomenon in its actual context in a particular period when the boundary between the phenomenon and its context is not clearly defined (Runeson, Host, Rainer, & Regnell, 2012). Multiple sources of data are used to study the phenomenon. A case study design may be single, multiple, embedded, or holistic based on the circumstances. Case studies are used mainly for exploratory purposes; however, they can be performed for explanatory and descriptive purposes. Case studies are good for studying software engineering methods and tools in industrial settings.

Case studies are conducted when:

- You cannot control the variables.
- You cannot separate a phenomenon from its context.
- Context is more important.
- You want to know whether a hypothesis would be true in practical settings.

In comparison with experiments, case studies are easier to plan, and conduct and are more realistic. However, the case study results are hard to generalize over a large population (Wohlin et al., 2012). The findings of the multi-experiment study (Chapter 5) were used to design a framework for crowd usability evaluation. Study 3 validates this framework using a single case study with the following case.

- Esoft case: Esoft is a medium-sized budget (64,000 USD per annum) software vendor with 20 employees in Pakistan. We examined the use of the crowd usability inspection framework on a project developed by a small-sized software development organization named Esoft to evaluate its effectiveness. The project was a website developed for the client to sell its products online. For confidentiality, we are not using the actual name of the organization.

Figure 3.2 shows the overview of the research methods, questions, and research contributions in this thesis work.

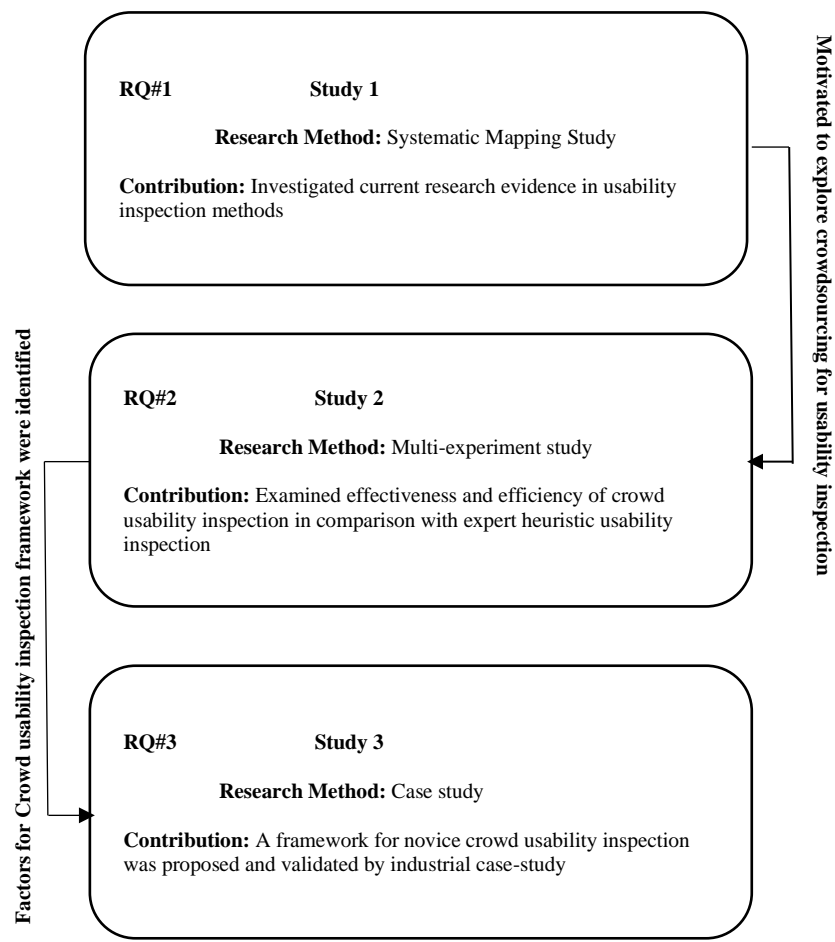

Figure 3.2 – Thesis overview, studies, research methods, and questions

# Chapter 4

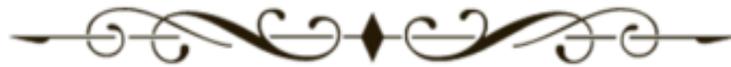

# 4. Study 1 - Software Usability Inspection Methods - A Mapping Study

## 4.1 INTRODUCTION

The usability of any software system is one of the critical software quality attributes. Software usability refers to the ability of a software system that enables it to be used with ease, efficiency, and satisfaction in a particular context of use (Abran, Khelifi, Suryn, & Seffah, 2003). Several methods and tools have been developed to evaluate the usability of a software system. There are four major types of methods for usability evaluation, i.e., a) Automatic (measuring usability using software tools), b) User Testing (assessing usability through testing software systems with real users), c) Formal Methods (calculating usability using some precise formulas and models), c) Informal Methods (Assessing usability based on some rules of thumb, skills, and experience of usability experts) (Nielsen, 1994a).

So far, automatic and formal methods cover only a few usability aspects (i.e., system response time, task completion time, etc.) and are difficult to employ (Nielsen, 1994a). User testing is a widely adopted method to evaluate software systems. However, real users are most of the time difficult or costly to hire in appropriate numbers to test all the usability aspects of a software system (Nielsen, 1994a; Nielsen, 1989; Rosenbaum, 1989). Besides project budgets and schedules often impose restrictions on large-scale user testing (Nielsen, 1994a). In the meantime, usability inspection methods (informal methods) have emerged as cost-effective means of usability evaluation (Cockton, Lavery, & Woolrychn, 2003; Nielsen, 1994a). Numerous studies have shown that usability inspection methods can identify usability problems that have not been observed by user testing and vice versa (Desurvire, Kondziela, & Atwood, 1992; Karat, Campbell, & Fiegel, 1992; Nielsen, 1989; Nielsen, 1994a).

Usability inspection refers to cost-effective methods employing usability evaluators to identify usability problems in a software system (Jeffries & Desurvire, 1992). Usability inspection methods work well when employed using multiple usability experts (Jeffries & Desurvire, 1992). Other drawbacks of usability testing include being employed later in the software development life cycle while inspection methods are employed earlier in the software development life cycle (Jeffries & Desurvire, 1992).

### 4.1.1 Research Motivation

A few systematic mapping studies, systematic reviews, and surveys were conducted in the past to investigate the trends in the evaluation of usability evaluation methods (Fernandez, Abrahão, & Insfran, 2012; Fernandez, Insfran & Abrahão, 2011; Hollingsed & Novick, 2007; Insfran & Fernandez, 2008; Rivero & Conte, 2012). However, these studies are either not recent or merely focused on web usability evaluation methods while ignoring usability evaluation methods in other essential application domains, i.e., mobile applications. In the last decade, usability evaluation of mobile phone applications has become an emerging research area (Ahmadi & Kong, 2012; Heo et al., 2009; Lee et al., 2006; Nayebi, Desharnais, & Abran, 2012; Zhang & Adipat, 2005). Therefore, to determine trends in evaluating software usability inspection methods, a comprehensive systematic mapping study must cover all vital software application domains, i.e., Mobile phone applications, Desktop, and Embedded applications.

A survey was conducted in 2007 by Hollingsed & Novick in 2007 on usability inspection methods to determine what usability inspection methods were still being researched and what methods lead to a dead-end. The study covered a period of the last 15 years duration. The study did not discuss the type of software applications frequently evaluated for usability inspection methods. A systematic literature review was reported in 2008 (Insfran & Fernandez, 2008). The authors only covered web usability evaluation methods and their association with the web development process. An extended version of this systematic review was later presented by Fernandez et al. in 2011 (Fernandez et al., 2011). In this version, the search string was improved to include more research papers, and more bibliographic sources were added to the search process.

Another systematic review was performed in 2012 to find out the effectiveness of web usability evaluation methods (Fernandez et al., 2012). A systematic mapping study was also conducted by Rivero and Conte (2012), based on the work of another systematic mapping study (Fernandez et al., 2011). They figured out what new usability inspection methods have been developed to evaluate web artifacts and how they have been employed in the web development process. However, all these systematic reviews and mapping studies were focused on web usability evaluation methods.

Our research focuses on the usability inspection methods investigated by researchers over the last decade to evaluate different software applications. Our mapping study is different than Fernandez et al.'s work in a way that we are focused on usability inspection methods evaluating different types of software applications (Fernandez et al., 2011). In contrast, Fernandez et al. focused on different usability evaluation methods, e.g. (including usability inspection methods and usability testing, etc.) evaluating only web artifacts.

Our work has brought insight into usability inspection studies evaluating other types of software applications, i.e., Mobile applications, desktop applications, embedded applications, including web applications, and presented research evidence on it. This study would be helpful for researchers and practitioners to look into usability inspection methods concerning different types of software applications.

This paper is organized as follows. Section 4.2 is about the research method. Section 4.3 presents the results. Section 4.4 is about discussion, and section 4.5 contains the conclusion and future work. Section 4.6 includes references, and Appendix A contains an external web link for references to included studies.

## 4.2    RESEARCH METHOD

Systematic mapping studies are performed to provide a broad view of a research topic and determine whether research evidence exists on the topic or not, with an indication of its quantity (Kitchenham & Charters, 2007). We designed our mapping study by following the guidelines provided by Kitchenham & Charters (2007). A systematic mapping study involves several steps and activities. We defined a protocol to conduct a systematic mapping study in the planning stage.

This protocol includes research questions, identification and selection of primary studies, and inclusion/exclusion criteria. We performed this systematic mapping study in three steps: Planning, Executing, and Reporting. Planning and Execution steps are covered in this section, while results have been reported in section 4.3.

### 4.2.1    Research Question

This systematic study aims to analyze the evaluation of usability inspection methods from the perspective of the following research question.

"*What researchers have investigated regarding software usability inspection methods from 2004 to 2013*"?

### 4.2.2   Search Strategy

The research publications included in this mapping study were only limited to journal and conference papers. We extracted data from different data sources. The primary sources of research publications were as follows:

1. Compendex (EI Village)
2. IEEExplorer
3. ACM Digital Library
4. ScienceDirect

We kept Compendex EI as our primary data source to discard the duplicate studies. Kitchenham & Charters (2007) mentioned in their guidelines that the search string for a mapping study should be less narrowed than systematic reviews so that it might extract a more extensive data set. Therefore, we kept our search string very broad to extract a more significant number of studies. We tested a couple of different search strings to extract the relevant research articles. It was observed that the following search string fetched the most relevant records. We applied this search string to metadata (e.g., Title, Abstract, and keywords).

> *(Usability AND Inspection)*

A period of 10 years was reviewed from 2004 to 2013. The exact search string extracted relevant records from all the databases mentioned above. We did not search conferences or journals for relevant publications manually.

### 4.2.3   Inclusion and Exclusion Criteria

Research papers were included based on the Title, Abstract, and Keywords. Each research paper's abstract section was read, and based on the research question discussed above, articles were sorted out. However, sometimes it was not evident in the abstract or title section whether a particular research study investigates a usability inspection method to evaluate a software system. Likewise, often the type of software application evaluated was not mentioned. We downloaded the complete paper and went through the conclusion and introduction sections.

Studies meeting the following inclusion criteria were included.

1. The publications investigating software usability inspection methods

2. The publications, including and between 2004 and 2013, were selected (see Appendix A for complete references of included studies).

3. The complete research papers

Studies meeting the following criteria were not included.

1. The publications discussed other usability evaluation methods, i.e., User Testing, Interviews, Focus groups, Remote Usability Testing, Metric based methods, Think-aloud protocol, etc.

2. The publications that reported mapping studies and systematic literature reviews

3. The publications that discussed basic concepts about usability, guidelines, principles, or recommendations about usability

4. The publications that are not written in the English language

### 4.2.4 Data Extraction Process

We extracted the data based on our research question defined in sub-section 4.2.1. We applied the same data extraction policy to all the research publications included. The following information was extracted from included research publications:

1. The name of the usability inspection method investigated by the researcher

2. The information that the technique explored was tool supported or manual

3. The information whether the study was empirical or not

4. The information about the type of the application evaluated, e.g., Web Application, Desktop Application, Mobile Application, etc

To extract the type of inspection method investigated, we classified a usability evaluation method as an inspection method if usability evaluation is performed by an expert (Matera, Rizzo & Carughi, 2006). Examples of usability inspection methods are Heuristic Evaluation, Cognitive Walkthrough, etc. We did not include those studies that involve a user in the usability evaluation process, i.e., think-aloud protocol. We also did not consider those studies that measured usability based on usability metrics, i.e., Eye Tracking, etc.

To classify the information about the research study, the usability evaluation method it investigated was empirical or not. We considered a study practical if it validated the software usability inspection method by employing an experiment, case study, or survey. Some researchers only thought of user testing as an empirical approach, but this is not the case in this systematic mapping study.

To sort out the information about whether the inspection method was tool supported or manual? We considered an inspection method tool supported if it presented a software tool to support usability inspection partially or entirely by helping an expert in the evaluation process. A software tool either records the usability problems documented by an expert or provides guidelines, checklists, and heuristics to evaluate the software system.

| Table 4.1 – Results of Systematic Mapping Study | | |
|---|---|---|
| Information Extracted | Parameters | Results |
| Frequently investigated usability inspection method | Heuristic Evaluation | 52% |
| | Cognitive Walkthrough | 10% |
| | Multiple Method Studies (excluding Heuristic Evaluation and Cognitive Walkthrough) | 3% |
| | Other Inspection Methods | 16% |
| Type of usability evaluation methods investigated | Manual | 91% |
| | Tool supported | 9% |
| Type of studies | Empirical | 94% |
| | Non-Empirical | 6% |
| Type of software applications frequently evaluated with usability inspection methods | Web Application | 52% |
| | Desktop Application | 25% |
| | Mobile Application | 8% |
| | Embedded Application | 2% |
| | Type of Application Unknown | 13% |

To extract the information about the type of application evaluated, we did not consider the difference between a website, web application, or web portal. We just considered all these categories under a single label of a web application. Besides if the type of application evaluated was not mentioned or ambiguous, we considered it unknown.

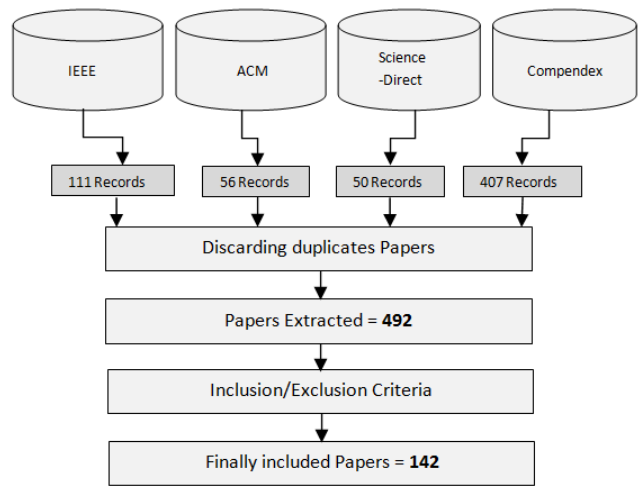

Figure 4.1 – Data Extraction Strategy

#### 4.2.5    Executing Mapping Study

The search operation was performed on the databases, i.e., Compendex, IEEExplorer, ACM digital library, and Science Direct, in July 2014. The search operation returned 624 potentially relevant results. The specific number of papers found in each database is depicted in figure 4.1. Duplicate studies were discarded by reviewing the Title and Abstract part of each research article.

If we found that a conference publication was later converted into a journal publication, we kept the journal publication and discarded the conference version. Besides, if the same study was improved and later published again, we kept the latest version and discarded the older version. After discarding the duplicate papers, the total number of papers was reduced to 492. We then applied the inclusion/exclusion criteria and reduced the total number of publications to 142 only.

### 4.3    RESULTS

The results of this systematic study are summarized in table 4.1. The list of included publications and their references are provided on a web link in Appendix A. The results indicate that the most frequently investigated usability inspection method is Heuristic evaluation, with 52% of studies reporting it. The second most investigated inspection method is Cognitive Walkthrough. The results show that 10% of studies investigated Cognitive walkthroughs for usability evaluation. Other prominent usability inspection methods reported are Web Design Perspective (WDP) with 3% studies and Metaphor of human Thinking with 2% studies, Semiotic inspection method, 2% studies, Discount usability inspection method, 2% studies. The investigation of remaining usability inspection methods was reported as 16% collectively. Main usability inspection methods and multi-method studies are indicated in figure 4.2. It is also worth noting that 3% of publications reported investigation of multiple inspection methods, excluding those multiple method studies that investigated Heuristic Evaluation and Cognitive walkthrough. A total of 35 different inspection methods were reported in included studies.

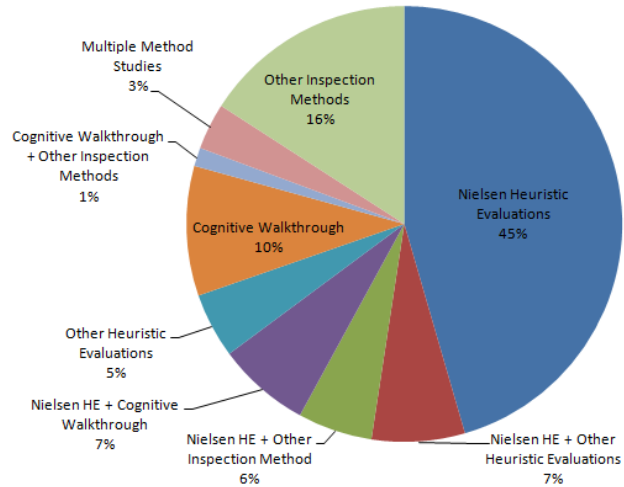

Figure 4.2 – Type of Usability Inspection Methods Investigated

The results also indicated that 94% of studies were empirical. Nevertheless, 6% of studies did not empirically evaluate the investigated inspection methods. The results also revealed that web applications are the most frequently evaluated application type, i.e., more than 52% of studies reported evaluation of web applications, as shown in figure 4.3. The other applications evaluated were desktop applications with 25% studies, mobile applications with 8% studies, and embedded applications with 2% studies reporting them. Moreover, in 13% of studies, the type of application evaluated was not mentioned. Out of 142 included research publications, 116 were conference papers, while only 26 journal publications contributed to the results, see appendix B.

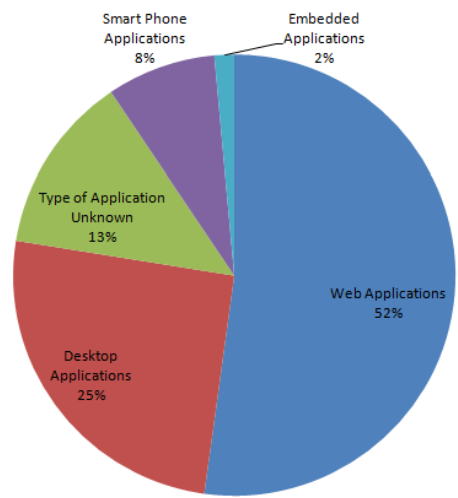

Figure 4.3 – Type of Software Applications Evaluated

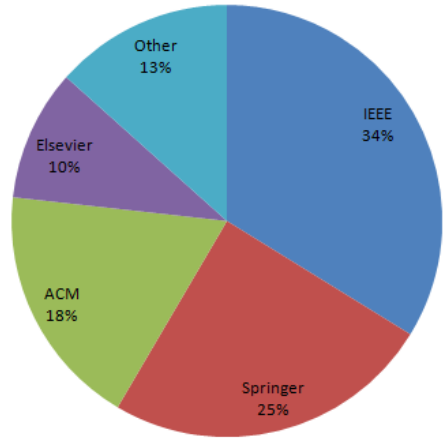

Figure 4.4 – Prominent Publishers

It was clear that the usability inspection method was a research area with growing interest till 2009, after plotting the included publication's data year-wise,. After that, there has been a gradual decrease significantly in the last two years, i.e., 2012 and 2013, see figure 4.5.

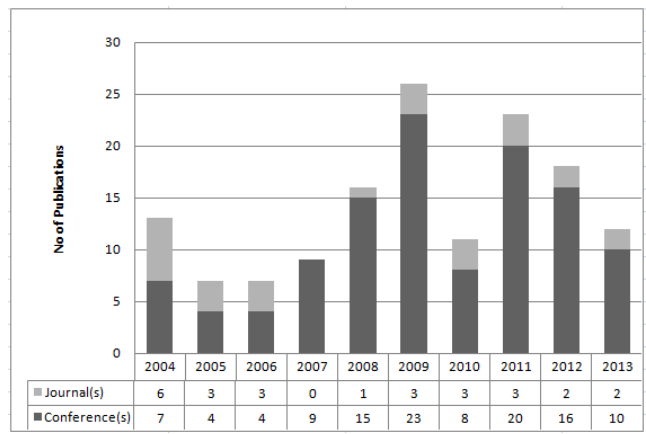

Figure 4.5 – Usability Inspection Publications from 2004 – 2013

### 4.3.1   Usability Heuristics' Violations

It is critical to know what usability problems are frequently identified. Once we have main usability problem areas identified with their magnitude, this information can be beneficial for usability researchers and practitioners to avoid these most frequently occurring usability problems. A similar study has been conducted by Pinelle et al. on usability heuristic violations in PC video games (Pinelle, Wong, & Stach, 2008). Pinelle et al. took reviews from popular gaming websites on 108 different video games from 2001 to 2008. After analyzing the studies, Pinelle et al. identified twelve standard classes of usability problem areas for PC video games and proposed their own set of heuristics for PC video games. However, our scope was not limited to PC video games, and we wanted to find the most frequently occurring usability problems for all types of software applications.

Therefore, after addressing our systematic mapping study's basic questions in subsection 3.4, we decided to synthesize further 142 included research publications to distribute usability heuristics violations. We read each research paper thoroughly and selected only those publications that reported the number of usability heuristics' violations with their nature, i.e., what sort of usability heuristic has been violated and how many times.

Table 4.2 – Mapping of Studies Violating Usability Heuristics

| Study ID | Website/Web App/Web Portal | Desktop Application | Smart Phone/PDA Application | Embedded System | Application Domain |
|---|---|---|---|---|---|
| S52 | Yes | X | X | X | e-Learning System |
| S79 | Yes | X | X | X | e-Learning System |
| S138 | Yes | X | X | X | e-Learning System |
| S116 | Yes | X | X | X | e-Learning System, & e-Commerce Website |
| S23 | Yes | X | X | X | Educational Social Network |
| S90 | Yes | X | X | X | Social Networking Website |
| S31 | Yes | X | X | X | e-Govt Website |
| S118 | Yes | X | X | X | e-Govt Website |
| S133 | Yes | X | X | X | Online Medical Information System |
| S135 | Yes | X | X | X | Online Shopping Websites |
| S15 | Yes | X | X | X | Online Travel Agent |
| S130 | Yes | X | X | X | Journal, Conference, & Event Management System |
| S106 | Yes | X | X | X | Virtual World Games |
| S55 | Yes | X | X | X | Anonymity Network |
| S24 | X | Yes | X | X | Desktop Video Game |
| S60 | X | Yes | X | X | Video Game |
| S127 | X | Yes | X | X | Health Management |
| S12 | X | Yes | X | X | API (Application Program |
| S95 | X | Yes | X | X | Intrution Dectection |
| S108 | X | Yes | X | X | Grid Computing |
| S121 | X | X | Yes | X | Safety Inspection Drone Technology Application |
| S36 | X | X | X | Yes | Robotic System |
| Total | 14 | 6 | 1 | 1 | |

We also recorded the severity of usability heuristics violations, but only a few research papers mentioned the severity of usability heuristics. We found only 22 studies that reported several usability heuristics violations with their nature. Several usability heuristics were reported with their nature and intensity of violation.

We found only 22 research publications because the studies reported no statistics about usability problems, sometimes either not mentioning the number of usability problems, the nature of usability problems, or the severity level of the usability problems. Initially, we decided to include research publications that would report all these three parameters about the usability problems, i.e., No. of usability problems found, nature of usability problems, and Severity level of usability problems. After mapping the data from 142 publications, we found that only 39% of studies mentioned the number of usability problems found, and 20% reported the nature of usability problems. Only 3% of studies reported the usability problems' severity level. Therefore, to increase the number of research publications in this analysis, we made our inclusion criteria a little more lenient and included those usability publications that reported several usability problems found with their nature only.

In this way, our initial set of usability publications increased to 15%, i.e., 22 publications, instead of 3%, i.e., four publications. Some publications used other scales to report the usability problems instead of the number, e.g., a study S123 used percentage compliance with usability heuristics instead of reporting usability problems found. Four studies (i.e., S124, S48, S68, S11) used a category scale to report the usability problems found instead of the number. We did not include studies reporting usability problems with category scale or percentage compliance as that does not reveal an exact number of usability problems found. In this way, our shortlisted publications were just 22 because of the lack of statistical data that included 142 usability publications. After analyzing the four publications that reported the severity level of the usability problems found, it was revealed that most of the issues were of minor severity level. Then we have a few problems of major severity level. Finally, the least number of usability problems were of critical severity level.

We picked up usability heuristic violations from each study and mapped them into a spreadsheet. Later, we synthesize them to discover the main broad categories of heuristic violations. We observed that the broad primary categories of the heuristics proposed are general heuristics by Nielsen (1992), web design perspectives-based heuristics (Conte, Massolar, Mendes, & Travassos, 2009), Metaphors of Human Thinking (Hornbæk & Frøkjær, 2003; Hornbæk & Frøkjær, 2004a; Hornbæk & Frokjaer, 2004b; Hornbæk & Frøkjær, 2005; Frøkjær & Hornbæk, 2008), Video Game Heuristics by Pinelle et al. (Pinelle, Wong, & Stach, 2008), and few other heuristics. Wherever the authors mentioned usability heuristic violations with the same name as in the broad category, we mapped it in front of that main category (see figure 4.7). However, if the author named the heuristic violation with a synonym and has the exact definition as the main category, we mapped it in the sub-category column in front of the main category. For example, Visibility of System Status is the main category of heuristic violation, and sub-categories are Gaming Status, Display of Information, Informative Feedback, Feedback, and Confirmation (see figure 4.7). Some sub-categories have significantly reported more usability problems than other categories like Feedback in the Visibility of System Status. Such a high frequency of usability heuristic violations is that some studies conduct multiple experiments with the same or different heuristic evaluation methods. In this way, reported heuristic violations significantly increase sub-categories more than other sub-categories since heuristic violations from a particular study may contribute to different broad categories.

Table 4.3 – Mapping of Usability Heuristics

| Usability Inspection Method | Main Heuristics | Heuristic Violations in Main Category | Studies IDs | Sub Heuristics | Heuristic Violations in Sub Heuristics | Studies IDs | Total |
|---|---|---|---|---|---|---|---|
| Nielsen's Heuristics | H1 - Visibility of System Status | 141 | S135, S121, S23, S95, S108, S118, S55, S36 | Gaming Status | 37 | S24, S60 | 374 |
| | | | | Display of Information | 31 | S36, S95 | |
| | | | | Informative Feedback | 30 | S31 | |
| | | | | Feedback | 132 | S138, S79 | |
| | | | | Confirmation | 3 | S15 | |
| | H2 - Match between the System and the Real World | 106 | S135, S121, S95, S108, S52, S133 | Use Natural Cues | 7 | S36 | 252 |
| | | | | Speak User's Language | 117 | S138 | |
| | | | | Compatibility | 22 | S127, S31 | |
| | H3 - User Control and Freedom | 100 | S135, S23, S95, S108, S118, S55 | Control | 45 | S127, S24, S52, S60 | 259 |
| | | | | Flexibility & Control | 21 | S31 | |
| | | | | Customization | 14 | S23, S24, S60 | |
| | | | | Accessibility | 8 | S108, S90 | |
| | | | | Independence | 19 | S116 | |
| | | | | Clearly Marked Exits | 41 | S138 | |
| | | | | Search | 11 | S15 | |
| | H4 - Consistency and Standards | 128 | S135, S121, S23, S95, S108, S118 | Consistency | 387 | S133, S12, S24, S31, S52, S60, S95, S138 | 515 |
| | H5 - Help Users Recognise, Diagnose, Recover from Errors | 107 | S135, S121, S23, S95, S108, S118, S55 | Error Management | 3 | S52 | 162 |
| | | | | Good Error Messages | 41 | S138 | |
| | | | | Error Handling & Exceptions | 11 | S12 | |
| | H6 - Error Prevention | 82 | S135, S127, S121, S23, S31, S95, S55 | Prevent Errors | 64 | S138 | 146 |
| | H7 - Recognition rather than Recall | 94 | S135, S23, S95, S108 | Defaults | 1 | S52 | 165 |
| | | | | Minimize the users' memory load | 63 | S138 | |
| | | | | Minimum Memory Load | 7 | S133 | |
| | H8 - Flexibility and Efficiency of Use | 88 | S135, S121, S23, S31, S108, S118 | Flexibility | 56 | S36, S52, S95 | 144 |
| | | | | Shortcuts | | S138 | |
| | H9 - Aesthetic and minimalist design | 56 | S135, S138, S23, S95, S108, S55 | Visual Clarity | 33 | S31 | 130 |
| | | | | Simple and natural dialogue | 205 | S138 | |
| | | | | Design & Aesthetics | 41 | S106 | |
| | H10 - Help and documentation | 181 | S135, S138, S23, S52, S95, S108 | Training & Help | 48 | S24, S60 | 291 |
| | | | | Documentation | 35 | S12 | |
| | | | | User Guidance & Support | 25 | S31 | |
| | | | | Learnability | 2 | S52 | |
| Web Design Perspectives | Conceptual | 17 | S130 | Cognition | 10 | S52 | 38 |
| | | | | Predictability | 4 | S60 | |
| | | | | Conceptual Correctness | 7 | S12 | |
| | Presentation | 62 | S130 | Graphical Design | 78 | S79 | 283 |
| | | | | Explicitness | 9 | S31 | |
| | | | | Language | 20 | S133 | |
| | | | | Content | 114 | S31, S79, S90 | |
| | Navigation | 182 | S130, S23, S31, S52, S79, S95, S106, S90 | | | | 182 |
| | Structural | 2 | S130 | Layout & Formatting | 24 | S90 | 26 |
| Metaphors of Human Thinking | Habit Formation | 124 | S138 | | | | 124 |
| | Thinking | 169 | S138 | | | | 169 |
| | Awareness | 141 | S138 | | | | 141 |
| | Utterances | 66 | S138 | | | | 66 |
| | Knowing | 73 | S138 | | | | 73 |
| Video Game Heuristics | Response Time | 28 | S24 | | | | 28 |
| | Artificial Intelligence | 23 | S24 | | | | 23 |
| | View Mismatch | 27 | S24, S60 | | | | 27 |
| | Command Sequences | 20 | S24 | | | | 20 |
| | Virtual Representation | 16 | S24 | | | | 16 |
| | Input Mappings | 23 | S24, S60 | | | | 23 |
| | Skip Content | 5 | S24 | | | | 5 |
| Other Heuristics | Security & Privacy | 34 | S31, S90, S135 | | | | 34 |
| | Functionality | 40 | S31 | | | | 40 |
| | User Interaction & Sociability | 69 | S135, S52, S90 | | | | 69 |

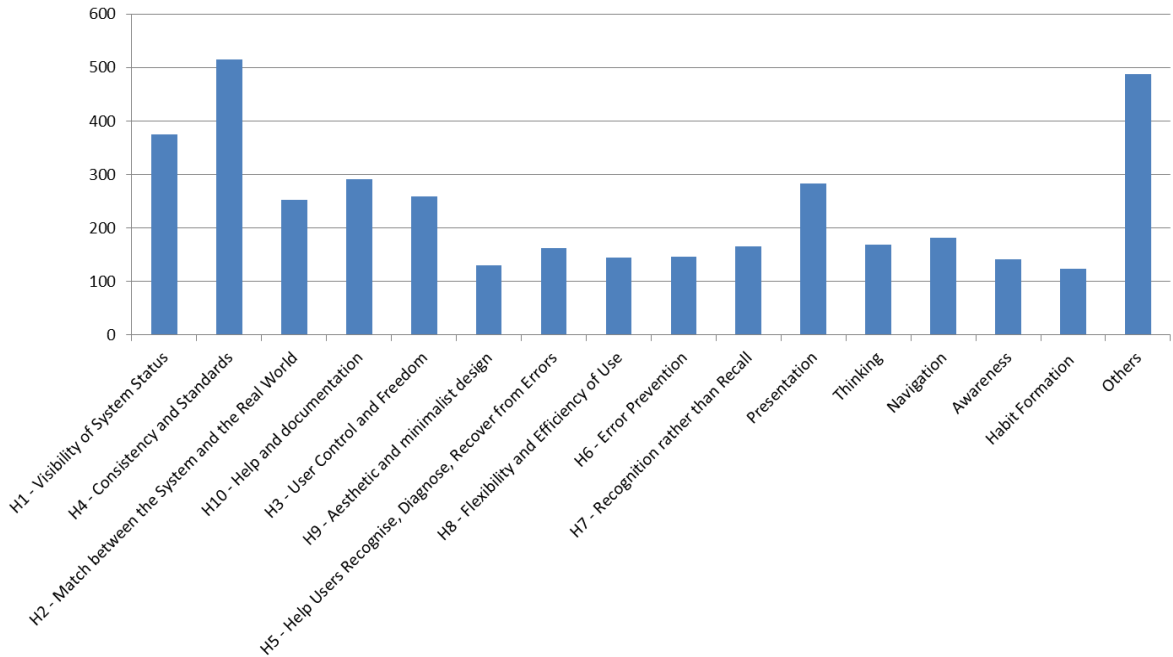

Figure 4.6 – Distribution of Usability Heuristics' Violations

For example, study S79 has reported 89 usability violations of sub-category Feedback. The other heuristic violations in study S79 were relevant to Content, Graphical Design, Technical problems, and Navigation. The heuristic violations in a particular study may belong to one usability inspection method. In the usability evaluation study S79, we mapped the heuristic violations relevant to feedback in Nielson's heuristics. In contrast, other heuristic violations relevant to Navigation, Graphical Design, and Content were mapped to another usability inspection method, i.e., Web Design Perspectives.

The extracted results for usability heuristics violations show that Consistency & Standards and Visibility of System Status are the top-most violated usability heuristics with 515 and 374 violations, respectively see Figure 4.8. The usability evaluation results in studies [S118] [S31] also showed a similar trend that Consistency & Standard and Visibility of the System Status are the top violated usability heuristics.

We will discuss both heuristic violations in later sections in detail. Other most violated usability heuristics were Help & Documentation and Presentation. The results also revealed that 18% of studies reporting Heuristics violations were e-Learning Management systems, 9% were social networking websites, and 9% were health-related web applications (see figure 4.6).

#### 4.3.1.1 Consistency & Standard

Consistency means expressing the same thing in the same way throughout the system by consistent interfaces, uniform command syntaxes, theme, colors, etc. (Nielsen, 1994b). Desurvire et al. describe consistency in video games as that a game should respond to the user in a consistent, struggling, and exciting manner to the player's activities (Desurvire, Caplan, & Toth, 2004). The usability problems reported by different studies are as below:

1. The page layout is not consistent [S135]
2. Optional fields in data entry forms are not correctly identified [S135]
3. Formatting conventions are different in various sections of the website [S135]
4. The home button is not shown at the expected place, i.e., top left corner [S135]
5. Actions do not reveal the same results, e.g., Pressing the "Follow" icon sometimes lead to a profile and sometimes to the contents of the user [S135]
6. Low hit detection and inconsistent response to user input in video games [S24]

#### 4.3.1.2 Visibility of System of Status

Visibility of system status implies that the system should keep the user informed about its current state through proper feedback (Nielsen, 1994b). The problems reported regarding visibility of system status are of one of several types [S135]:

1. Users cannot foresee how much the system will take to complete a task.
2. What is the alternative course of action available in a particular context?
3. The system does not show the system's current state, i.e., the user cannot identify where he is standing in the interaction process with the system.

In another study [S23], evaluators faced problems regarding the visibility of system status, like the system does not show the file uploading progress. In studies [S60, S24] regarding video games, evaluators identified usability problems regarding the visibility of the system. A system does not give enough information about the game's characters, game context, or enemies. Visual communication, like maps, provides information about everything except the user's current location [S60, S24].

#### 4.3.2 Prominent Usability Inspection Methods

The primary usability inspection methods frequently investigated by researchers in our systematic mapping study are discussed below.

#### 4.3.2.1 Heuristic Evaluation

The heuristic usability evaluation method was developed by Jakob Nielsen (1992). In this method, usability experts use heuristics to evaluate a user interface design for usability problems. Heuristics are the typical characteristics of a usable software interface. Usability experts examine the compliance of user interfaces with heuristics and report its violations, if found, and associate it with relevant heuristics. Other examples of the Heuristic Evaluation method are reported in different studies (Allen, Currie, Bakken, Patel, & Cimino, 2006; Nielsen, & Loranger, 2006; Oztekin, Nikov, & Zaim, 2009). Other types of Heuristic usability inspection methods found in the included studies of this systematic study are discussed below.

#### 4.3.2.2 HE+ & HE++

HE-Plus & HE++ was proposed by Chattratichart & Lindgaard (2008). HE-Plus employs a similar set of heuristics as Nielson's Heuristic Evaluation [HE] method (Nielsen, 1992). The factor that makes HE+ different from HE is the usability problems' profile. This usability problem profile was based on the User Action Framework, a structured repository of usability concepts, issues, and concerns (Andre, Hartson, Belz, & McCreary, 2001). Chattratichart & Lindgaard (2008) experimented with evaluating the results with two HE variants HE+ and HE++. Both new variants performed better than HE in reliability and effectiveness (Chattratichart & Lindgaard, 2008).

#### 4.3.2.3 Domain-Specific Usability Inspection

Roobaea et al. proposed a methodological framework that can be employed to develop a domain-specific usability evaluation method (AlRoobaea, Al-Badi, & Mayhew, 2013). This study also suggested a usability evaluation method for social media websites to validate this framework. The proposed usability evaluation method identified heuristics for seven usability areas.

#### 4.3.2.4 Masip's Heuristics

In another study, existing heuristics published from 1982 to 2011 were summarized in 16 categories (Masip, Granollers, & Oliva, 2011). Later, Masip's et al. compared these 16 categories with 14 sets, including Nielsen's heuristics, plus four more categories added by Masip et al.'s for interactive systems, i.e., public Kiosks and virtual assistants. The experiment results showed that 14 types of heuristics identified significantly more problems than 16 sets of heuristics.

### 4.3.2.5 Critic Proofing

Critic Proofing is a modified version of the heuristic evaluation method that prioritizes the list of heuristic violations based on severity level within a game genre, e.g., frequency of the problem and impact (Livingston, Mandryk, & Stanley, 2010). This approach employs critical reviews in the usability evaluation process to eliminate all potential criticisms before release. Critic Proofing Heuristic evaluation was based on usability game heuristics and game genres (Pinelle et al., 2009).

### 4.3.2.6 Cognitive Walkthrough Ergonomic Inspection Method

The cognitive walkthrough Ergonomic Inspection method supports the ergonomic evaluation of collaborating software systems (Mahatody, Kolski, & Sagar, 2009). It exploits different versions of the Cognitive Walkthrough method and consists of three major steps: preparation of the evaluation, Inspection, and analysis of the results.

### 4.3.2.7 Semiotic Inspection Method

The semiotic Inspection Method proposes heuristics for evaluating the semiotic content of websites (Bolchini, Chatterji, & Speroni, 2009). Semiotic content includes labels, icons, or signs to present a concept or an idea. The semiotic Inspection method also considers ontology aspects of the semiotic content. Ontology refers to a particular domain of knowledge. The semiotic inspection method brought a new concern to the usability evaluation of information-intensive websites.

### 4.3.2.8 Formal Specifications of Usability Heuristics

Another study discussed formal specifications of usability heuristics to improve the usability evaluation (Jimenez et al., 2012). This study considers the development of patterns for the specification of usability heuristics. A pattern consists of the name of the pattern, the problem it solves, motivation, examples, and consequences of its use. This study finds the applicability of patterns in defining heuristics based on empirical evidence gathered through query techniques. The study results show that patterns can expedite the heuristic evaluations no matter if experienced evaluators use it or novice evaluator.

### 4.3.2.9 Web Design Perspectives (WDP)

This method was proposed by Conte et al. (2009). The Web Design Perspective method defined new heuristics for web applications, called web design perspectives. WDP method focuses on design perspectives of a web application, i.e., Conceptual, Presentation, Navigation, and Structural elements, to evaluate a web application. Other examples of perspective-based methods are WebTango Methodology (Ivory & Hearst, 2002) and Abstract-Tasks Inspection (Costabile, & Matera, 2001).

**4.3.2.10 Grid Computing Heuristics**

Roncagliolo et al. proposed heuristics to evaluate Grid computing applications (Roncagliolo, 2011). The heuristics proposed in this study are mapped with Nielsen's heuristics. However, Roncagliolo et al.'s heuristics are not significantly different than Nielsen's (1992) heuristics.

**4.3.2.11 Cognitive Walkthrough**

The cognitive walkthrough was another method proposed for usability evaluation based on user goals (Polson, Lewis, Rieman, & Wharton, 1992). This method is an expert's evaluation. In this method, experts try to achieve a user goal by going through a combination of tasks. Other examples of Cognitive walkthroughs are also reported in different studies (Clayton, Biddle, & Tempero, 2000; Filgueiras, Martins, Tambascia, & Duarte, 2009).

**4.3.2.12 Metaphors of Human Thinking (MOT)**

Metaphors of Human Thinking (MOT) involves psychology in usability evaluation. This method focuses on the evaluator's mental activities that simulate the user and try to achieve a goal by completing tasks and using them for usability evaluation. This technique has emerged from classical Heuristic and Cognitive walkthrough usability evaluation methods. Several studies have investigated metaphors of human thinking in usability evaluation (Hornbæk & Frøkjær, 2003; Hornbæk & Frøkjær, 2004a; Hornbæk & Frokjaer, 2004b; Hornbæk & Frøkjær, 2005; Frøkjær & Hornbæk, 2008).

## 4.4    DISCUSSION

This section presents trends in the investigation of usability inspection methods. We also discussed the limitations of this systematic study that may threaten its validity.

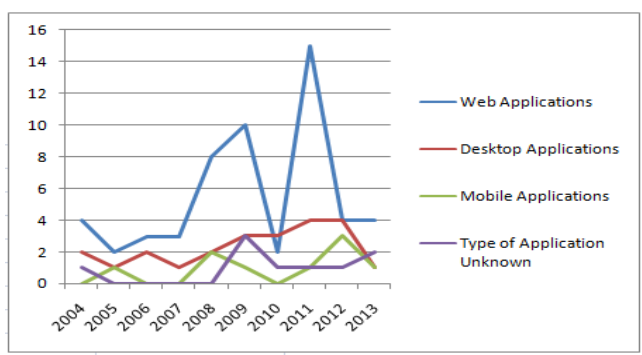

Figure 4.7 – Heuristic Usability Evaluations: Type of Application and Year-wise

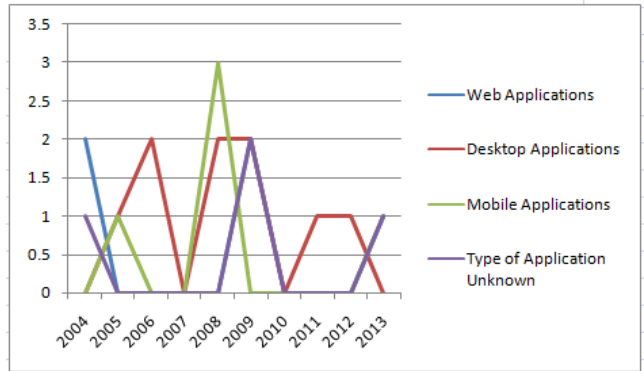

Figure 4.8 – Cognitive Usability Evaluations: Type of Application and Year-wise

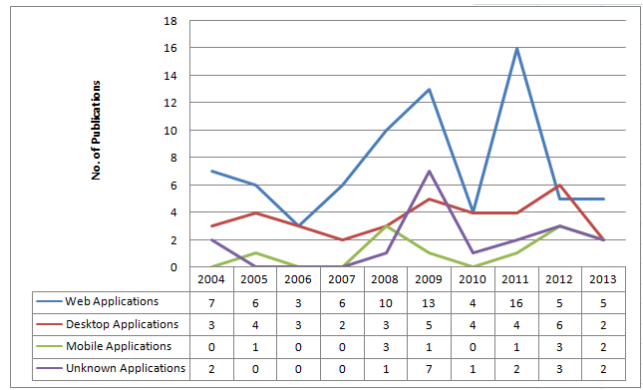

Figure 4.9 – Software Usability Inspections: Application Type and Year-wise

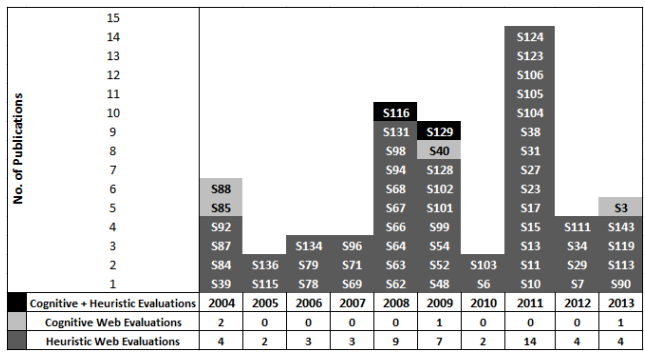

Figure 4.10 – Web Usability Evaluations: Heuristic vs. Cognitive Method and Year-wise

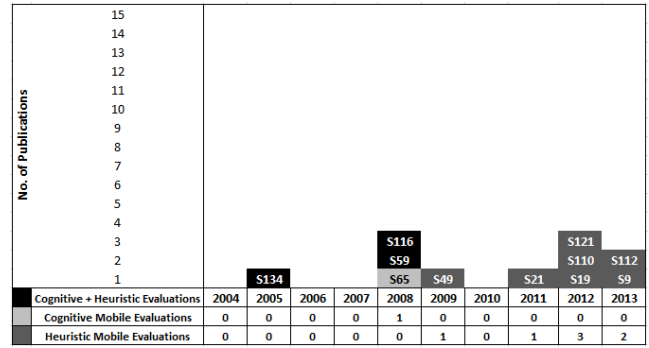

Figure 4.11 – Mobile App Usability Evaluations: Heuristic vs. Cognitive Method & Year-wise

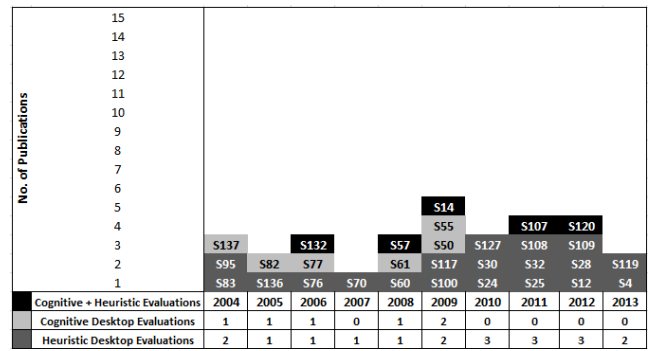

Figure 4.12 – Desktop App Usability Evaluations: Heuristic vs. Cognitive Method & Year-wise

### 4.4.1  Trend Analysis

This systematic study aimed to find out the trends in the investigation of usability inspection methods regarding different software application types in the last ten years, i.e., 2004 to 2013. The trends observed in this systematic study are as below:

1. The most frequently violated usability heuristics are Consistency & Standards and Visibility of System Status (see Figure 4.8).

2. Inspection-based usability evaluations significantly increased in 2009 and 2011; see figure 4.5.

3. Usability inspections significantly decreased in 2010 and 2013; see figure 4.5.

4. Heuristic usability evaluation of mobile applications has increased in the last couple of years except for a decrease in 2013, see Figures 4.9 and 4.13.

5. Heuristic usability evaluations have overall increasing trends except in 2010, 2012, and 2013 (see figure 4.9).

6. Cognitive usability inspections have overall downward trends. Besides, cognitive usability inspections considerably increased in 2008 and 2009, see figure 4.10.

7. In contrast to the web and desktop applications, mobile application usability evaluations have an increasing trend in 2011 and 2012; see Figures 4.11 and 4.13.

8. Web applications have been more frequently evaluated than desktop and mobile applications, see Figures 4.11 and 4.12.

9. Desktop usability evaluations increased in 2009, 2011, and 2012; see figure 4.14.

If we carefully review the above trends, we can easily observe that inspection-based usability evaluations had a sharp downfall in 2010. A slightly similar downfall trend can be observed in 2012 and 2013. After careful investigation of inconsistent trends in the year 2010, 2012, and 2013, it was found that some journals and conferences that contributed their publications in the selected list of studies for this Systematic Mapping study did not take place in the years 2010, 2012, and 2013, e.g.

**Journals:**

1. International Journal of Medical Informatics did not publish any journal paper in 2010 & 2012.

2. International Journal of Human-Computer Studies did not publish any journal paper in 2012.

3. International Journal of Human-Computer Interaction did not publish any journal paper in 2012.

**Conferences:**

1. ITI – Information Technology Interfaces did not take place in the year 2010.

2. The Human-Centered Software Engineering conference did not take place in the year 2013.

3. The Human-Centered Design conference did not take place in the year 2013.

Therefore, we can infer that inconsistent trend in 2010, 2012, and 2013 were due to situational factors.

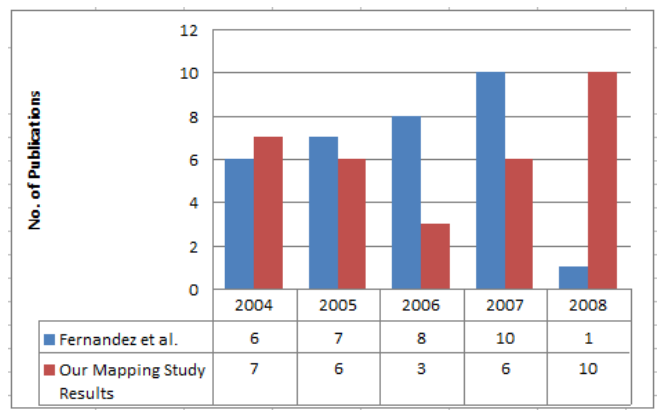

Figure 4.13 – Web Usability Evaluations: Fernandez et al. vs. Our Mapping Study

The number of publications for web applications in our data set is slightly less than Fernandez et al.'s systematic review results from 2004 to 2007 (see figure 4.13) (Fernandez et al., 2011). This slight decrease is that we only included publications for Usability Inspection Methods while Fernandez et al. included publications for User Testing. Besides, Fernandez et al. added papers through manual search in various relevant conferences and journals (Fernandez et al., 2011). Therefore, increasing the number of publications in their study for web applications is justified. Fernandez et al. did not cover the whole year of 2008 as they performed their search in March 2008 (Fernandez et al., 2011). Therefore, the number of publications reported by Fernandez et al. in 2008 is significantly less than our results (see figure 4.15).

Besides that, Fernandez et al. reported that 69% of studies employed manual usability evaluation methods while only 31% reported using tool-supported usability evaluation methods. However, our study shows that 91% of studies employed manual usability evaluation methods while only 9% employed tool-supported methods. One reason for the increase in the number of tools supported by Fernandez et al.'s systematic review results is that they included studies using User Testing, Metric based Usability Evaluation methods, and Formal Usability Evaluation Methods. It is relatively easy to employ tool support in User Testing than in Usability Inspection methods that evaluate the usability of a software system based on usability experts' knowledge, skills, and experience. Usability Testing methods often employ tool support through performance measurement, user activity analysis, etc. (Ivory & Hearst, 2001). To validate it further, we went through a total of 26 usability inspection-based studies in Fernandez et al.'s data set identified by Rivero & Conte, (2012). We observed that only 19% of studies were tool-supported, while 79% employed manual usability inspection. Therefore, we can infer that usability inspection-based studies used less tool support than other usability evaluation-based studies.

Likewise, Fernandez et al.'s study reported that 53% of studies empirically validated usability evaluation methods while 47% did not employ empirical validation. Our results show that 94% of studies empirically validated the usability evaluation method, while only 6% were not validated empirically. The trend of empirical validation is gradually increasing in usability evaluations, see figure 4.16. The sudden decline in the graph in 2010, 2012, and 2013 is due to situational factors discussed above. Fernandez et al.'s Systematic Mapping Study included publications from 2009 (Fernandez et al., 2011). Therefore, we could not compare our results for recent years with Fernandez et al.'s work (Fernandez et al., 2011).

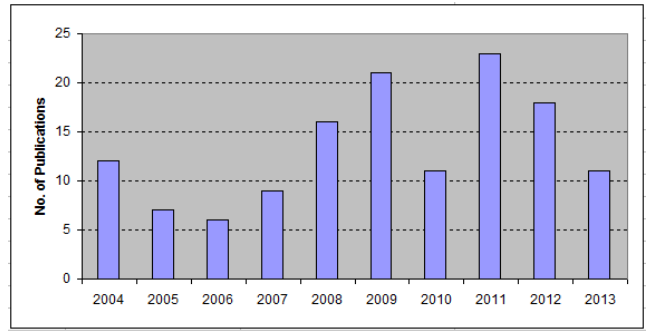

Figure 4.14 – Empirical Usability Evaluations

Additionally, the finding that the Heuristic and Cognitive methods are the most frequently investigated usability inspection methods is common in our results and systematic mapping study conducted by Rivero & Conte (2012). However, Rivero et al.'s Mapping Study on Usability Inspection methods was based on 26 publications extracted from Fernandez et al.'s Systematic Review data set. Moreover, we did not find any other Systematic Review or Mapping Study to compare our results.

## 4.4.2 Secondary Studies on Usability Evaluation between 2014 – 2022

Several systematic reviews and mapping studies have been performed on usability evaluation between 2014 and 2022. However, two distinct categories are apparent in this regard, a) General systematic reviews & mapping studies b) Domain-specific systematic reviews and mapping studies. We will discuss these secondary studies in both categories.

### a) General secondary studies on usability evaluation methods

Paz et al. conducted a systematic mapping review to investigate what types of usability methods are frequently employed besides the type of software systems (Paz & Pow-Sang, 2015). This short mapping study covered research articles between 2012 and 2015. A total of 228 research articles were selected out of 1169. The study results showed that questionnaires, user testing, heuristic evaluation, interviews, and think-aloud protocol are the most frequently employed usability evaluation methods. Moreover, health informatics, education, software development, e-commerce, and gaming are the most frequently evaluated application types.

A systematic literature review was conducted to investigate the state of the art in defect reporting in usability evaluation and software engineering between 2000 and 2016 (Yusop, Grundy, & Vasa, 2016). The results extracted from 57 selected studies identified several limitations, including mixed data, inconsistency in terms and values of defect data, and insufficient attributes to classify usability defects. The study also identified that usability defect reporting had received less attention. Existing research on usability defect reporting has primarily focused on the shortcomings of open-source defect repositories in supporting usability defects. The online forums have become an alternative interaction hotspot for users to discuss usability defects, especially in open-source development communities. However, the linear sequence of communication makes it difficult to extract the contextual information for developers to fix the problems (Yusop, Grundy, & Vasa, 2016).

A thorough review of usability studies published between 1990 and 2016 was performed, leading to the identification of 150 studies (Sagar & Saha, 2017). The results show that usability testing, heuristic evaluation, and questionnaires are the most used methods for evaluating usability. This research shows that 71 percent of studies address usability issues during the design phase of the software development life cycle. Furthermore, usability evaluation is commonly employed in the Web domain.

**b) Domain-specific secondary studies on usability evaluation methods**

The secondary studies conducted after 2017 were more focused on specific domains, e.g., health systems, mobile applications, learning management systems, etc. an overview of these systematic literature reviews and mappings studies is presented below:

**1) Health systems**

Zapata et al., conducted a systematic review to analyze the empirical usability evaluation techniques mentioned in 22 selected studies for mHealth applications (Zapata, Fernández-Alemán, Idri, & Toval, 2015). The results of the study show that 73% of the selected studies employed interviews or questionnaires for usability evaluation of mHealth applications. The study reported the necessity for automated evaluation tools.

Wronikowska et al. (2021) reviewed the literature between 1986 to 2019 on usability evaluation methods, metrics, and associated measurement procedures that have been documented to evaluate the usability of the systems developed for inpatient hospitals. A total of 1336 studies were found in the initial search results. There were eleven different usability evaluation methods discovered in the 51 papers considered. Seven usability metrics were reported as part of these usability evaluation approaches. ISO9241-11 and Nielsen's components were the most used measures. Almost 40% of the studies included in the study incorporated "usefulness" as a metric. The study concluded that there is little consistency in the usability evaluation of the electronic health record systems. The variety in usability methodologies, metrics, and reporting is highlighted in this study.

## 2) E-Commerce Websites

A systematic literature review was conducted to identify the metrics to assess the usability of e-commerce websites between 2014 and 2019. The study concluded with a summary of several usability evaluation metrics for e-commerce websites (Díaz, Arenas, Moquillaza, & Paz, 2019).

## 3) ATM Interfaces

A thorough review of approaches for evaluating the usability of ATM interfaces and tools is presented in a study (Sahua & Moquillaza, 2020). A total of 132 studies were discovered, with 12 finally included studies. The study results show that no usability studies exist on the usability evaluation of ATMs. This paper aimed to provide ideas for usability assessments of ATM interfaces and information on the present status of the Systematic Review and potential studies that can aid in the performance of usability evaluations for ATMs. The authors intended to provide a usability evaluation process for ATM apps based on this information in the future.

Falconi et al., presented a systematic literature review on usability and security criteria for ATM interfaces (Falconi, Zapata, Moquillaza, & Paz, 2020). The study contributed a set of 160 metrics separated into 13 categories for evaluating the security and usability of an Internet banking system. The study proposes to assess the relevance of these recommendations in ATM systems in a future study.

## 4) Mobile Applications

Even though numerous usability evaluation methods are available, most of them are focused on traditional computer use and are not fully compatible with mobile phone use (Lin, Véliz, & Paz, 2019). As a result, a systematic literature assessment was performed to identify usability evaluation guidelines for mobile educational games that target primary school students as users. This secondary study proposed a set of usability guidelines for evaluating mobile educational games for students in primary school.

Al-Razgan et al. (2021) conducted a thorough literature review of the usability of mobile applications for visually impaired people. The authors reviewed 60 studies out of a total of 932 that were published within the last six years. This study presented varied trends, themes, and evaluation findings of numerous mobile applications generated between 2015 to 2021.

## 5) Learning Management Systems

LMS (Learning Management System) platforms have become increasingly prevalent in the last two decades, with widespread acceptance by educational institutions throughout the world, demonstrating the necessity of designing systems that are easy to use. In light of the present pandemic, where long-distance education is no longer a choice but a necessity, having a learning management system that meets usability criteria becomes even more critical. Júnior et al. presented a systematic mapping study of the available literature, highlighting the most popular methodologies used to evaluate usability and UX in LMS throughout the last decade (2010–2020) (Júnior, Hernández-Ramírez, & Estima, 2021). The authors looked at around 80 selected studies. Their findings reveal that most use methodologies entirely based on Jakob Nielsen's heuristics or combine them with other methods. The authors identify the need for automated usability evaluation methods for LMS.

### 4.4.3  Limitations of Systematic Mapping Study

The significant limitations of this systematic mapping study are publication bias, selection bias, inaccurate data extraction, and classification. Publication bias refers to what is likely published among what is available Kitchenham & Charters (2007). In other words, we are more likely to include publications that report positive results on our hypothesis instead of negative results. We searched all the major databases and publishers to address this threat, i.e., Compendex EI, IEEExplorer, ACM Digital Library, and ScienceDirect. We only included conference and journal papers. However, we did not include grey literature, i.e., un-published industry work, MS or Ph.D. thesis reports, etc. Selection bias refers to the problems in selection criteria used to include papers. We addressed this problem by defining inclusion and exclusion criteria.

# Chapter 5

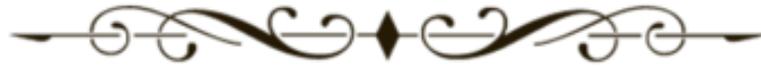

# 5. STUDY 2 - USABILITY INSPECTION: NOVICE CROWD INSPECTORS VS EXPERT

## 5.1 INTRODUCTION

Software usability is the degree to which a software system can be used with satisfaction, efficiency, and effectiveness in a particular environment (Abran et al., 2003). There are different types of usability evaluation methods (UEMs), but the prominent two are usability testing and usability inspection. Conventional usability testing is time-consuming and expensive. The real users are not easily available in sufficient number or costly to hire. Moreover, it is difficult to set up an environment to test all aspects of software usability (Nielsen, 1994a; Rosenbaum, 1989; Nielsen, 1989). Usability inspection is an alternative method employing usability experts to evaluate a software system's usability based on their skills, experience, and rules of thumb, also called usability heuristics (Nielsen, 1994a). Usability inspection methods require multiple usability evaluations, i.e., 3-5 expert evaluations (Jeffries and Deservers, 1992), and work well with usability experts only. Although usability inspection methods were introduced as a low-cost means of usability evaluation, there is still a need to find more cost-effective UEMs. Therefore, crowdsourcing has been investigated for usability evaluation to overcome the cost and time delays in traditional UEMs (Sarı et al., 2019).

Crowdsourcing is based on the idea of using the wisdom of the crowd. Surowiecki, in 2005 first introduced the concept of the wisdom of the crowd in his famous book, *The Wisdom of Crowds*. He argued that a large group of people (i.e., crowd) are collectively more intelligent than a couple of experts in problem-solving and decision making. Surowiecki further argued that crowds are not always wise; rather, wise crowds have some common characteristics. The wise crowds should have diverse opinions, and every individual in the group should have an independent view based on their knowledge. Finally, the wise crowd should be able to convert individual opinions into a collective decision.

Crowdsourcing has considerably matured in computing with empirical research (Ambreen and Ikram, 2016), and it has been investigated for usability evaluation (Bruun and Stage, 2015; Guaiani and Muccini, 2015; Liu et al., 2012). Estell' es-Arolas et al. (2012) provided an integrated definition for crowdsourcing. It is a collaborative online activity in which an individual or an organization announces a work task via an open call to a group of people with varying knowledge, skills, and experience on a crowdsourcing platform. The published job can be of variable complexity and completing the task would let the individual earn money, experience, and knowledge. At the same time, the outsourcer will get the job done in return (Estell' es-Arolas et al., 2012).

### 5.1.1      Research Motivation

Usability engineering has quite rich history spanned over the last couple of decades. However, budget-constrained organizations are often reluctant to adopt usability practices to reduce software development costs. Many studies have identified that one of the major obstacles in adapting usability practices is the budget constraint in small and medium-size software development organizations (Bak et al., 2008; Ardito et al., 2011; Häkli, 2005). In this regard, crowdsourcing has brought new opportunities for usability engineering by overcoming the shortcomings of traditional UEMs (Bruun and Stage, 2015; Guaiani and Muccini, 2015).

Nevertheless, researchers and practitioners face the challenge of crowd workers' poor-quality work (Peer, Vosgerau, & Acquisti, 2014; Goodman, Cryder, & Cheema, 2013). Another problem is the validity of demographic and professional details that crowd worker reports (Liu et al., 2012). Crowdsourced workers sometimes complete tasks multiple times to increase their reward (Kittur, Chi, & Suh, 2008). Therefore, it was worthwhile to explore the wisdom of the novice crowd from the perspective of usability evaluation. One way of exploring the novice crowd's wisdom was to compare usability inspection using crowdsourcing with expert heuristic usability inspection. In this regard, we did not find any research article reporting the comparison between expert heuristic usability inspection and novice crowd usability inspection.

Therefore, there was a need to conduct a research study investigating the novice crowd's wisdom by comparing the novice crowd usability inspection with the expert heuristic inspection. In other words, we wanted to discover the possibility of getting comparable results from novice crowd usability inspection as expert heuristic usability inspection. In this regard, we compared expert heuristic usability inspection with novice crowd usability inspection using crowdsourcing.

We choose expert heuristic usability inspection as a benchmark for comparison as usability inspection methods are economical than traditional usability testing (Nielsen, 1994a; Hollingsed and Novick, 2007). The expert heuristic usability inspection finds more usability problems than any other usability inspection method (Jeffries et al.,1991; Desurvire, Kondziela, & Atwood,1992; Hollingsed and Novick, 2007). Moreover, usability inspection methods can be employed at any development stage of an interactive system (Botella et al., 2013). Nevertheless, some authors support that usability inspection cannot fully substitute usability testing (Hollingsed and Novick, 2007). Besides, some studies have suggested that a combination of end-user and experts-based methods may be employed for more thorough and effective results for usability evaluation (Yen & Bakken, 2009; Hasan et al., 2012). While other authors affirm that there are no real differences between usability testing and inspection methods in terms of usability problems found; and usability testing also misses even critical problems like other usability evaluation methods (Molich and Dumas, 2008). And usability testing is not a gold-standard method to test all other usability evaluation methods (Molich and Dumas, 2008).

### 5.1.2  Related Work

**5.1.2.1    Comparative usability studies.** Comparative usability studies have been conducted for the past three decades.

Gray and Salzman (1998) conducted a comprehensive literature survey covering comparative usability studies before 1997. They found that most of the studies suffered from methodological flaws, e.g., (a) low statistical power and wildcard effect that implies that the real differences are not noted, while differences observed are not real, (b) Comparing the outcome of analytical UEMs with empirical UEMs based on the number of usability problems found, instead of the content of usability problems, etc. They also introduced the concept of false alarms, misses, hits, and correct rejections while comparing the UEM's. These flaws challenged the validity of comparative usability studies.

Molich et al. (2004) conducted a comparative usability study by comparing usability testing results conducted by nine independent teams of the same website, i.e., www.hotmail.com. Around 75% of the usability problems were uniquely identified, i.e., no two teams found the same problem. The study observed vast differences in identified usability problems, methodology, tasks, and usability reports submitted by each team. The study concluded that the usability tests' outcome relies highly upon the selected tasks, methodology, and the test controller. Moreover, it concluded that a traditional website might have numerous usability problems, and a typical usability test can only find a small portion of all usability problems. The study emphasized that rather than finding all the usability problems, iterative testing should focus on finding the most critical usability problems.

Molich and Dumas (2008) conducted a comparative usability evaluation study. The study involved 17 teams for the usability evaluation of a hotel's website. Nine teams used usability testing, and eight teams performed an expert evaluation. The study concluded that there are no differences between usability testing and expert reviews in terms of the problems identified, and usability testing is not a benchmark method to compare all other techniques. Usability testing also overlooks problems like any other UEMs, even critical problems. Besides, no evidence was found for the existence of false alarms in expert reviews.

**5.1.2.2    Comparative usability studies involving novices.** Comparative usability studies involving novices started to appear in the early 90s. Jackob Nielsen performed a comparative study in 1992 to investigate the evaluator effect on heuristic evaluation with varying levels of usability expertise of evaluators, i.e., novice, regular usability experts, and double usability experts. The results showed novice evaluators performed worst, while double experts performed better than regular experts. However, in this case, novice evaluators did not have any experience or knowledge of usability evaluation. Nevertheless, some of the severe usability problems were even found by half the novice evaluators.

Koutsabasis et al., in 2007, conducted a comparative study investigating the performance of novice usability evaluators, having wide background knowledge in interactive design as they graduated in disciplines like graphic design, arts, etc. The study compared nine usability evaluation teams employing different usability evaluation methods, i.e., Heuristic evaluation, Cognitive walkthrough, Think-aloud protocol, and Co-discovery learning. Each group consisted of 3 MSc students studying interaction design as a course. Besides, all the students were familiar with usability evaluation of the websites as part of their coursework previously. The novices first conducted individual evaluations, and later they gathered their findings to interpret the results. The results showed that novice usability evaluators could identify thorough and valid usability problems with recommendations, i.e., multiple novice usability evaluations must be performed, etc. This study, however, compared the results of the novice usability evaluations with novices.

Law and Hvannberg (2008) conducted an exploratory study to determine how novice evaluators merge and rate the severity of usability problems individually and in collaborative settings. In this regard, the participants were given already collected observational reports of the user testing to examine. The study results show that novice usability evaluation in collaborative settings leads to undesirable effects, i.e., inflation (erroneously increasing severity ratings) and deflation (leniently merging dissimilar usability problems). However, the study compared the performance of novice usability evaluators with novices.

Another study was conducted to explore the effect of using business goals in usability evaluation with novice evaluators (Hornbæk & Frøkjær, 2008). The study results show that novice evaluators employing business goals identified less relevant usability problems than the control group. Moreover, the study compared the results of the novice evaluators with novices.

Howarth et al. (2009) argued that usability evaluation methods do not provide enough guidance to support usability practitioners, especially novice usability evaluators, to perform reliable usability evaluations. They proposed usability problem (UP) instances as a desirable feature to help novices analyze raw usability data (Howarth et al., 2009). Their study shows that usability problem instances can support novice usability evaluators to perform more reliable usability evaluations. However, the study compared the results of novice usability evaluators with novices.

Følstad et al., in 2010, compared the work-domain experts with usability experts in terms of validity and thoroughness for usability evaluation. All the fifteen domain experts were Ph.D. students with less than a year of experience in the relevant domain; none of them studied HCI as a course or had any experience with usability evaluation previously. The twelve usability experts were hired from different IT companies working as interaction designers or usability consultants. Both usability experts and domain experts did not have any previous experience of using the test objects. However, domain experts had domain knowledge and extensive computer experience. The usability inspection method employed was a group-based expert walkthrough. Usability testing was performed as a benchmark to compare results. The study results showed that work-domain experts could produce equally valid but less thorough results for usability inspection compared with usability experts. The study concluded that work-domain experts might be used for usability inspections without compromising validity. Nevertheless, usability inspections with multiple work-domain experts would generate a comparable level of thoroughness as usability experts (Følstad et al., 2010). However, the focus of the study was to investigate usability inspection using work-domain experts, not novice usability inspection.

In their work, Botella et al. (2013) presented a framework involving the concept of design patterns to support novice evaluators performing heuristic evaluations. They proposed that experts' recurring usability problems and relevant solutions to fix them would work as a design pattern to assist novices in their evaluations. However, they did not validate this framework in practical settings.

Borys and Laskowski (2014) investigated whether a large group of novice evaluators can achieve comparable results to a couple of usability experts using heuristic evaluation. The study results rejected the hypothesis that novice usability evaluation results match with expert usability evaluation. However, novice usability evaluators, in this case, had little or no knowledge of usability. All the novice evaluators were computer science students and had attended a training session of an hour to understand the concept of usability evaluation.

De Lima Salgado et al., in 2018, surveyed 38 novice usability evaluators and found that "context of use" is the most challenging usability aspect for novices to understand. The authors suggested that novices may use scenarios, storyboards, and domain-specific principles to conduct heuristic evaluations. Moreover, the authors suggested using collaborative heuristic evaluation for novices, where usability experts analyze usability problem descriptions' quality to shortlist false alarms and duplicate usability problems. However, this study was based on a survey and did not contain any practical comparative evaluation of novice usability evaluations with other usability evaluation methods.

**5.1.2.3      Design critique-based comparative studies using crowdsourcing.** Some studies have compared experts with novices based on critique/feedback to design UI elements using crowdsourcing. Although design feedback/critique is not the same as usability inspection, the findings of these studies are valuable and relevant to discuss in the perspective of comparing the novice crowd with an expert using crowdsourcing.

Xu et al. (2015) conducted a classroom-based study to investigate the value of feedback received from crowd-workers (non-experts) on posters designed by students as part of an introductory visual design course. The feedback was collected through the structured and free-form method and was compared with experts to judge its quality. The crowd-feedback received was then employed to improve the design of the posters. Later, another iteration of crowd-feedback elicitation was performed to see the effects of earlier changes on posters. The study concludes that the range and depth of expert feedback do not match with crowd-workers. However, crowd-feedback serves as a supplement rather than a replacement for expert feedback.

Moreover, it was found that structured feedback was more interpretive, diverse, and critical than free-form feedback. Nevertheless, crowd workers in this study did not have any design expertise. Besides, this study does not address the challenges of crowdsourcing (discussed in subsection 1.1) and their solution except comparing the quality of the feedback of crowd-workers with experts.

Yuan et al. (2016) conducted a study to investigate whether a rubric of design principles can help novice designers to provide design critique comparable to experts? In this regard, multiple novices and expert designers were recruited from different crowdsourcing platforms. In this experiment, a total of fifteen undergraduate students from a design course were asked to design a dashboard for weather applications. The study results reveal that with the help of expert rubrics, novices can provide a design critique comparable to experts. Besides, this study does not address the other challenges of crowdsourcing (discussed in subsection 5.1.1) except comparing the quality of the feedback of novice crowd designers (equipped with rubrics of design principles) with experts.

**5.1.2.4      Usability comparative studies using crowdsourcing.** Usability evaluation studies examining the wisdom of the crowd are few and far between. These studies are discussed in this subsection with their limitations.

Liu et al. (2012) conducted a study on Amazon's Mechanical Turk to compare traditional usability testing with remote usability testing. The object of the experiment was a school's website. In the conventional usability testing part of the experiment, all the participants were experienced users of the website as they were current students at the same school. However, participants in the crowd usability testing were not familiar with the school's website. The results of the crowdsourced usability testing were not as good as traditional usability testing. Nevertheless, crowdsourced usability testing was less expensive, faster, and easy to perform.

In our opinion, the methodology of this study (Liu et al., 2012) could have been improved by controlling some factors. For example, the test group for traditional usability testing was already experienced users of the website that may have influenced the results in favor of conventional usability testing. Moreover, we assume that the crowdsourcing platform selected for this study, i.e., CrowdFlower, also affected results in favor of traditional usability testing as CrowdFlower does not allow to discard crowd workers' poor-quality work. Hence everyone in the crowd usability evaluation received equal payments, despite the quality of the work submitted.

Retelny et al. (2014) introduced the concept of flash teams that can be formulated dynamically using a crowdsourcing platform. The team members in a flash team are experts in a domain that can collaborate to complete a project using a runtime manager named Foundry. One of the tasks in the experiment was the heuristic usability evaluation of a prototype. This study showed that flash teams could complete their work in less time than normal teams on a crowdsourcing platform. However, hiring an expert on a crowdsourcing platform is equally expensive as hiring it conventionally. Besides, this study did not examine the crowd's wisdom for usability evaluation compared to experts.

Bruun and Stage (2015) proposed a usability testing technique called Barefoot. In this technique, Bruun and Stage (2015) suggested that local software practitioners may be given short training to conduct usability testing in resource-constrained organizations. Later, they compared Barefoot with crowdsourced usability testing while hiring university students by sharing minimalist usability training material online. Bruun and Stage (2015) found that the Barefoot approach suits small budget organizations that cannot afford full-time usability experts. Besides that, crowd usability testing is not suitable without HCI competencies. Crowd usability testing requires end-user reports analysis.

In our opinion, the study (Bruun and Stage, 2015) could be improved by controlling the study settings differently. For instance, although asynchronous usability testing (Bruun et al., 2009) is like crowdsourcing in sharing usability evaluation tasks online and completing them remotely (either online or offline) and then submitting them back to the outsourcer online. However, as per the crowdsourcing definition in (Estell' es-Arolas et al., 2012), refers that a work task, i.e., usability evaluation, in this case, may be announced via an open call to a group of people available online having varying skills, knowledge, and experience. We assume that if the sample population for crowdsourcing had been hired through a crowdsourcing platform, results might have differed as crowdsourcing platforms have experienced crowd-workers. Moreover, considering their argument that crowd usability testing does not work well without a crowd's competencies in HCI and usability testing, as valid, we assume that hiring a novice crowd with HCI competencies would not be as expensive as hiring a usability expert. Moreover, this study did not address crowd usability inspection in comparison with expert heuristic usability inspection.

We suggest that designing the tasks (i.e., use-cases) smartly for crowd workers can significantly increase their performance. For example, while developing the use-cases for crowd usability inspection, instead of merely focusing on the system's functionality, use-cases may be designed while keeping in mind the usability guidelines and heuristics to uncover potential usability problems.

**5.1.2.5    Overview of related work and uniqueness of this study.** The studies discussed in sub-section 5.1.2 contain insightful work about comparative usability evaluation, crowdsourcing, or novice usability evaluation. However, these studies have some differences compared to this study—the study settings of the current literature differ in one or more of the following aspects.

1.  The study does not involve novices (Gray & Salzman,1998; Molich et al., 2004; Molich & Dumas, 2008).

2.  The study does not involve crowdsourcing (Nielsen, 1992; Koutsabasis et al., 2007; Law and Hvannberg, 2008; Howarth et al., 2009; Følstad et al., in 2010; Botella et al., 2013; Borys & Laskowski, 2014; De et al., in 2018).

3.  Novice usability evaluators have no experience and knowledge of usability evaluation (Nielsen, 1992; Borys and Laskowski, 2014)

4.  Novice usability evaluation has been compared with novices (Koutsabasis et al., 2007; Law and Hvannberg, 2008; Hornbæk & Frøkjær, 2008; Howarth et al., 2009).

5.  The study presents a theoretical framework for novice usability evaluation without validation (Botella et al., 2013).

6.  The study presents suggestions based on a survey of novice usability evaluators (De et al., in 2018).

7.  The study does not address the challenges of crowdsourcing or proposed any solution for it (Liu et al., 2012; Bruun & Stage, 2015; Xu et al., 2015; Yuan et al., 2016).

8.  The studies are based on the design critique of UI elements but do not specifically compare usability inspection methods (Xu et al., 2015; Yuan et al., 2016).

9.  The studies support novices with features different from our proposed study, i.e., UP

instances, domain knowledge, a rubric of design principles, short usability training, etc. (Howarth et al., 2009; Folstad et al., 2010; Botella et al., 2013; Bruun & Stage, 2015).

10. The study settings could have been improved in some cases, e.g., the control group and the treated group could have the same learning experience with test objects to avoid any biases, etc. (Liu et al., 2012).

We will discuss the overview of each subsection of related work with their limitations here further. The studies discussed in subsection 5.1.2.1 brought valuable insights regarding methodological flaws and misconceptions while comparing usability evaluation methods. However, these studies were not focused on novices.

Subsection 5.1.2.2 discusses comparative usability studies involving novices. However, these studies were conducted in traditional settings and did not involve crowdsourcing. Besides, the study conducted by Jackob Nielsen in 1992 involved novices that did not have any experience or knowledge of usability evaluation. It has been established in several studies that novices that do not have any practical experience or knowledge of usability evaluation would not be able to perform significantly in usability evaluation (Borys and Laskowski, 2014). Another limitation of most of the studies discussed in subsection 5.1.2.2 is that they compared the performance of novice usability evaluators with novices (Koutsabasis et al., 2007; Law and Hvannberg, 2008; Hornbæk & Frøkjær, 2008; Howarth et al., 2009). Other studies included in subsection 5.1.2.2 either proposed a theoretical framework for novice crowd-workers without validation (Botella et al., 2013) or presented suggestions based on a survey (De et al., in 2018).

Subsection 5.1.2.3 discusses studies that compared the design critique of novice crowd-workers/designers in comparison with experts involving crowdsourcing. However, these studies did not compare usability evaluation methods (Xu et al., 2015, Yuan et al., 2016). Besides, these studies did not address the challenges of crowdsourcing except the quality of the feedback of crowd-workers.

Subsection 5.1.2.4 is focused on comparative usability studies involving crowdsourcing. The study conducted by Liu et al. (2012) concluded that remote usability testing employing crowdsourcing did not perform well compared to traditional usability testing. However, the participants in conventional usability testing were already experienced users of the test object, while this was not the case with crowd usability testers. In another study by Retelny et al. (2014) suggested that a group of experts can work more productively using crowdsourcing than traditional teams. However, this study did not involve novice crowd usability evaluators, neither they addressed the challenges of crowdsourcing. Bruun and Stage (2015) concluded that crowd usability testing does not perform well without skills and experience in HCI. However, in our opinion, hiring novice crowd inspectors with HCI skills and experience are not as expensive as usability experts. Moreover, these studies did not address the challenges of crowdsourcing except the quality of the work of crowd-workers.

Besides, we have not found any other study comparing novice usability inspectors with experts involving crowdsourcing. Most of the usability studies (discussed in subsection 5.1.1) involving crowdsourcing do not address the challenges of crowdsourcing nor provide any reasonable solutions for these challenges to the best of our knowledge except comparing results of crowd-workers with control groups.

To overcome the limitations of the existing literature discussed earlier, this study investigates novice usability inspection in comparison with expert heuristic usability inspection using crowdsourcing while addressing the challenges of crowdsourcing and proposing a reasonable solution for these challenges.

The rest of the paper is organized as follows: section 5.2 discusses Research Method, section 5.3 is Analysis and Discussion, section 5.4 is the Limitations and Validity Threats. Section 5.5 is the Conclusion and Future Work.

## 5.2    RESEARCH METHOD

Our chosen research method is experimentation in this research study since we want to study the cause-and-effect relationship between expert heuristic usability inspection and novice crowd usability inspection with use-cases. The complete experiment design is illustrated in Figure 5.1.

### 5.2.1   Goal and Definition

The goal statement of this experiment is as below:

"*Analyzing Crowdsourcing for the purpose of Usability Inspection with respect to Wisdom of crowd from the point of view of Researcher using two websites and a web dashboard.*"

The artifacts of the experiment definition are as below:

**The object of study.** The object of the study is crowdsourcing. In this regard, we explored different crowdsourcing platforms like Amazons Mechanical Turk (MTurk), uTest, and Upwork. We found that MTurk does not work outside the US due to the payments and tax issues (MTurk, 2022). Moreover, MTurk allows crowd-workers to submit their work anonymously (mTurk, 2022). Therefore, it is impossible to generalize the findings to a target population (Stritch et al., 2017). Besides, MTurk allows outsourcing work in microtasks which is not suitable for large assignments such as usability inspections.

Besides exploring uTest quite a bit, we were not able to find how to post our project on uTest (uTest, 2022). Other people have also reported similar issues with uTest (glassdoor, 2021). Finally, we selected Upwork, as it was easy to learn and simple to use (Upwork, 2022).

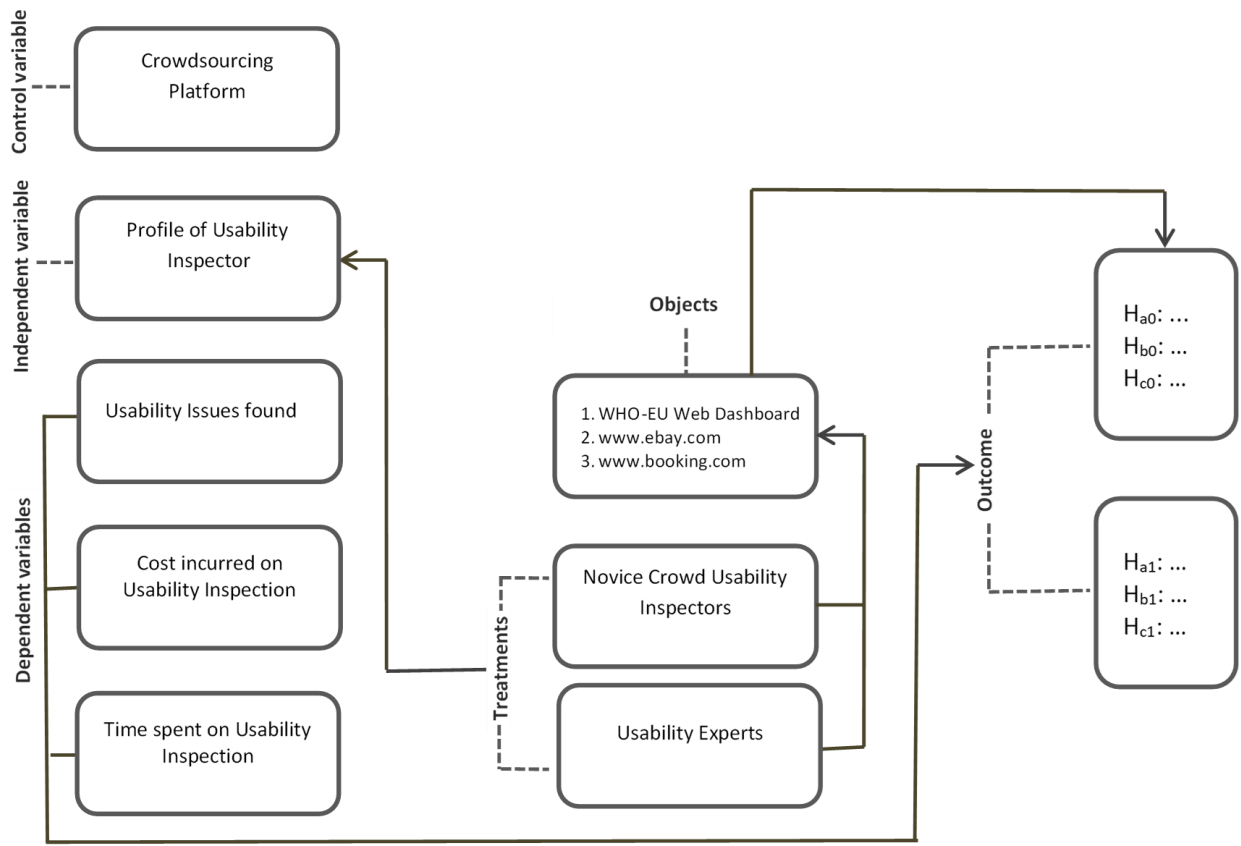

Figure 5.1 – Experiment Design

**Purpose.** The purpose of this experiment is to compare the expert heuristic usability inspection with novice crowd usability inspection.

**Quality Focus.** The quality focus of this experiment is to evaluate the wisdom of the crowd. We want to see whether novice crowd usability inspection can give us the results that are equally good as expert heuristic usability inspection or not? Other quality foci are time and cost incurred on both usability inspection methods.

**Perspective.** We are conducting this experiment from the perspective of the researcher.

**Context.** This experiment is conducted using a crowdsourcing platform, i.e., Upwork. Usability experts, as well as novice crowd inspectors, were hired using Upwork. Details about the test objects and subjects are discussed in subsection 5.2.2.3 instrumentation.

## 5.2.2 Planning

The experiment's planning stage consists of hypothesis formulation, variable selection, subject selection, design selection, instrumentation, and validation.

**5.2.2.1 Hypothesis Formulation.** Following hypotheses have been formulated for this experiment:

$H_{a0}$ – Novice crowd usability inspection on average finds different usability issues (w.r.t. content & quantity) compared to expert heuristic usability inspection.

*$H_{a1}$* – Novice crowd usability inspection on average finds the same usability issues (w.r.t. content & quantity) as expert heuristic usability inspection.

*$H_{b0}$* – Novice crowd usability inspection incurs the same cost as expert heuristic usability inspection.

*$H_{b1}$* – Novice crowd usability inspection incurs less cost than expert heuristic usability inspection.

*$H_{c0}$* – Novice crowd usability inspection, on average, takes the same time as expert heuristic usability inspection.

*$H_{c1}$* – Novice crowd usability inspection, on average, takes less time than expert heuristic usability inspection.

**5.2.2.2 Variables and Subjects' Selection.** There is only one independent variable in this experiment, i.e., the profile of the usability inspector. There are two treatments for it, i.e., usability experts and novice crowd usability inspectors. The dependent variable that we are measuring are usability issues found, cost incurred, and time spent on usability evaluations, as illustrated in Figure 5.1. We blocked the effect of two control variables in this experiment, i.e., the crowdsourcing platform and the evaluator effect. We used the same crowdsourcing platform, i.e., Upwork, for the whole experiment to block its effect. Moreover, to block the evaluator effect, we hired multiple usability evaluators, i.e., at least five usability inspectors, for each experiment trial for expert heuristic evaluation and novice crowd usability inspection (Hertzum & Jacobsen, 2001). Details about the usability experts and novice crowd inspectors will be discussed in subsequent subsection 5.2.2.3 Instrumentation.

**5.2.2.3 Instrumentation.** There are two parts to the experiment. Therefore, we designed separate instruments for each section. However, to compare the UEMs, we need a standard coding scheme to identify and classify usability problems.

**Coding Scheme:**

The coding scheme used in this comparative study, provided in Appendix G, is derived from Molich and Dumas's (2008) work. The focus of the coding scheme is to transform raw usability descriptions into usability problems. In this regard, the coding scheme has classified raw usability descriptions into different artifacts, i.e., atomic comment, problem comment, false positive, minor problem, serious problem, critical problem, issue, a key issue, etc. The coding scheme's application with a complete data analysis method is described in subsection 5.3.1, Analyzing usability inspection reports.

**Test Objects:**

In the first trial of this experiment, we evaluated the usability of a web dashboard of the World Health Organization designed for Europe (WHO EU Health e-Atlas). We chose a web dashboard as it contains a variety of complex visual elements. A web dashboard plays a significant role in data mining (Kostkova et al., 2014). A web dashboard aims to present large data in a concise way to increase its understanding. A well-designed dashboard helps in decision making and allows users to see the data from different perspectives and co-relate different factors in it (Kostkova et al., 2014). The significance of web dashboards in the public health sector is even more vital as they can help us monitor disease outbreaks (Lechner and Fruhling, 2014). Therefore, we chose the WHO-EU web dashboard (WHO-EU, 2016) designed to provide information about core health statistics, including demographics, health status, risk factors, health resources and their utilization, and expenditure in 53 countries of Europe.

We diversified the same experiment with two more trials to add more confidence in results, using two general-use websites from different domains, i.e., www.ebay.com and www.booking.com. We chose these two websites as these are typical websites of public use. eBay.com is a website for online shopping and auction with business operations in 30 countries. The website does not charge anything to buyers, but sellers are charged for listing items after a couple of free listings and selling items. Booking.com provides travel and accommodation services in 43 different languages. The website has more than 28 million listings for accommodations to stay. The weblinks for all three test objects are provided in Appendix C.

**Expert Usability Inspection's Instrument:**

A criterion was defined for the selection of usability experts. A typical usability expert must have:

- Master's or Ph.D. degree in HCI or relevant field
- At least five years of practical experience in usability evaluations
- Preferably be a certified usability expert, i.e., HFI CUA (Human Factors International Certified Usability Expert)

Weblinks for the test objects were shared with usability experts for expert heuristic evaluation, see Appendix D.

**a)** ***Usability Heuristics.*** A list of usability heuristics was shared with usability experts; see Appendix D. The heuristics for the WHO-EU web dashboard were focused on usability inspection of the web dashboard and information visualization. These heuristics were previously validated in an MS thesis study (Ghaffar & Nasir, 2016). Ghaffar & Nasir (2016) examined the existing literature comprehensively to collect all the heuristics that have been reported for designing web-dashboard and information visualization. The authors divided these heuristics into two sets, i.e., common heuristics (set 1) and common plus other heuristics (set 2). The common heuristics were grouped using Nielson's (1994b) and Schneiderman's (1996). The heuristics that did not match Neilson's and Schneider's heuristics were considered in other heuristics. The study's objective was to find out the most suitable heuristics for designing web dashboards and information visualization. An experiment was conducted to compare the effectiveness of both sets of heuristics. In this experiment, four students, each containing one undergraduate student of 8th semester from Bachelor of Software Engineering, were randomly selected. Group A was treated with heuristic set 1, while group B was given heuristic set 2 to design web dashboards to represent information for POLIO vaccination in Pakistan. The usability testing of the developed web dashboards revealed that the web dashboard designed with heuristic set 2 (common+other) were better in usability.

The rest of the two test objects were evaluated using ten usability heuristics for user interface design by Jakob Nielsen, see Annex A (Nielsen, 1994b).

**b)**    ***Payment Criterion***. Usability experts were offered a fixed price contract based on their previous experience, skills, and qualification for performing heuristic usability inspections on Upwork. In this regard, we thoroughly checked their profiles on Upwork, including information like jobs completed previously on Upwork, payments, and feedback received to verify their skills and per hour rate.

**c)**    ***Time Management:*** Each expert was requested to record the time it took to perform the evaluation.

**d)**    ***Roles:*** Experts were asked to evaluate the test objects with the following roles:

   a)  Buyer for eBay.com

   b)  Guest/Tourist looking to book accommodation for Booking.com.

   c)  Any visitor seeking health-related information and trends for WHO-EU web dashboard.

**e)**    ***Format for Evaluation:*** Experts were asked to document usability problems with at least the following information.

- Title for the problem
- Brief description of the problem
- Heuristic(s) violated.
- Type: Minor problem, Serious problem, Critical problem
- Snapshot of the area of UI where the problem was found.

**Novice Crowd Usability Inspection's Instrument:**

The following criterion was used for the selection of novice crowd usability inspectors. As per this criterion, a typical novice crowd usability inspector preferably has:

- A bachelor's degree/diploma in computing or relevant field/attended a course relevant to HCI.
- 1-2 years of experience in software usability evaluation/software quality assurance testing/software testing

Botella et al. (2014) proposed a classification scheme for usability evaluators based on experience and university degrees. They suggested that a novice usability evaluator has a university degree, attending one course related to HCI with at least a few hours of usability evaluation experience.

However, it has been established in several usability studies that novice evaluators with no or few hours of usability evaluation training/experience cannot perform comparably to usability experts (Nielsen, 1992, Borys & Laskowski, 2014). On the other hand, the studies that employed novices with extensive computer experience and background knowledge in usability evaluation produced more encouraging results than usability experts (Koutsabasis et al., 2007, Howarth et al., 2009, Folstad et al., 2010).

Therefore, motivated by the existing literature, we set the criterion for novices to preferably have attended a course in HCI and at least 1-2 years of practical experience in usability evaluation/software quality assurance/software testing. However, Botella et al. (2014) classified a person without a university degree attending several courses on HCI or with less than 2500 hours of practical experience in usability evaluation as a beginner. On the other hand, studies that employed novices with less than a year's practical experience in usability evaluation with a course in HCI also referred to them as novices (Howarth et al., 2009). Therefore, the boundary between the novice and beginner is overlapping. Whether we call it a more experienced novice or beginner, we considered the university degree/diploma/certification in usability evaluation/HCI, besides the practical usability experience of the evaluators, as an independent variable in the design of our study. Therefore, we referred to it as the profile of the usability inspector, see Figure 5.1. Moreover, we used the term novice for the selection criterion set earlier in this study for novice crowd usability inspectors.

The artifacts that were shared with novice crowd usability inspectors to perform usability inspection are as below:

*a)*     ***Usability Guidelines***. The guidelines were provided to identify usability problems in the test objects. The guidelines contained a list of usability heuristics, the same as shared in expert usability evaluation, besides instructions about how to attempt questionnaires.

*b)*     ***Questionnaires for Novice Crowd Usability Inspection***. The existing evidence on novice usability evaluation supports the argument that without assisting tools (expert rubrics, design patterns, storyboards, domain-specific principles), novice usability inspectors may not be able to perform comparably to usability experts (De et al., 2018; Borys & Laskowski, 2014, Botella et al., 2013, Hornbæk & Frøkjær, 2008). Therefore, the authors designed the questionnaires for each test object to support novice inspectors. Each questionnaire contained 20 Use-Cases. The questionnaires were based on the usability inspection report of a single expert for each test object. However, not all the use-cases were based on the expert heuristic evaluation. Instead, half of the total use-cases were general use-cases based on the different functionalities of the system. Moreover, expert heuristic usability inspection involved five usability experts, while the questionnaire for novice inspectors was based on a single expert's usability inspection report. Therefore, the questionnaire for novice usability inspection may not entirely depend on the expert's findings and, consequently, the results.

Other studies have also suggested supporting novices with usability experts or artifacts extracted from expert's usability inspections reports (Botella et al., 2013; De et al., 2018). For example, Botella et al. (2013) proposed a framework based on the recurring usability problems identified by usability experts to assist novices. Another study by De et al. (2018) suggested collaborative usability inspection where the novices may be supported with usability experts to analyze the quality of usability problems identified by novices, identifying false alarms and duplicate usability problems. They also suggested supporting novices with scenarios, storyboards, and domain-specific principles.

The WHO-EU web dashboard questionnaire contained a couple of prerequisite tasks for novice crowd inspectors to familiarize themselves. We did not include any prerequisite tasks for eBay.com and Booking.com as these test objects were easy to use. After performing each use case, novice crowd inspectors were supposed to provide feedback regarding any usability problems found.

The questionnaire for the first trial of novice crowd usability inspection was developed using Google Forms. However, for the rest of the two novice crowd usability inspection trials, questionnaires were designed in MS Word, as Google Forms does not save a questionnaire before submission. As a result, the user must complete it in one sitting. The weblinks for questionnaires are provided in Appendix E.

The steps to design questionnaires are listed below:

1. Extract usability problems from an expert heuristic evaluation using the coding scheme provided in Appendix G. Detailed process to transform usability descriptions into usability problems is discussed in section 5.3.1, Analyzing Usability Inspection Reports.

2. Take a screenshot of the UI area where the usability problem was reported by an expert and mark that area on the screenshot.

3. Develop a general use case to guide the novice crowd inspector to execute a task relevant to the usability problem.

4. After performing the task, the crowd inspector shall be requested to document any usability problem found in the space provided below the use case.

The following example of a use-case from eBay.com explains how a usability problem comment can be employed to develop a use-case for a questionnaire.

**Example:**

*Usability problem comment:*

*A user is not able to select multiple items within one refiner category (E.g., Select AT&T and T-Mobile under Network). It is a UX glitch that hinders the refiner experience.*

**Use Case:** *Select multiple choices from side filters. E.g., For a mobile phone, you can choose Apple, LG, etc. Were you able to select multiple categories in filters easily?*

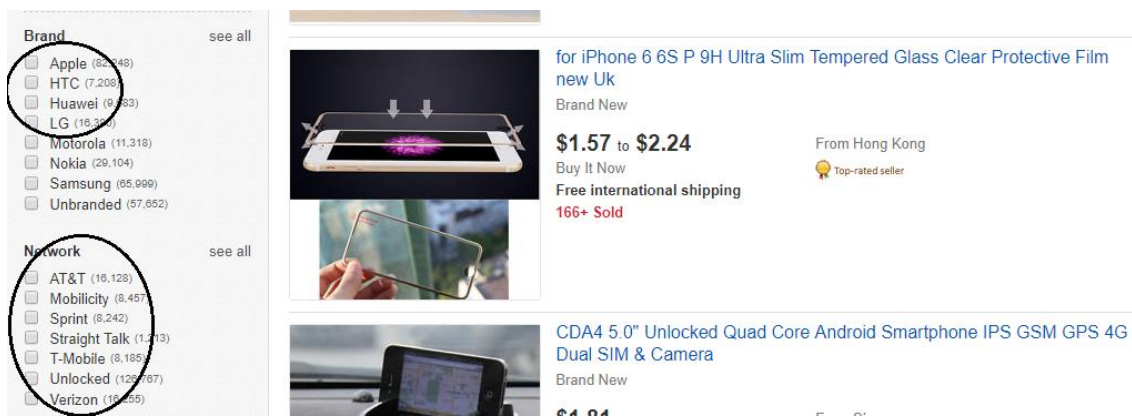

After performing this task, if you found any usability problem(s), please document them below:

It has been already validated in research that without assisting tools (expert rubrics, design patterns, storyboards, domain-specific principles), novice usability inspectors cannot perform to a level comparable to usability experts (De Lima Salgado et al., in 2018; Borys and Laskowski, 2014, Botella et al., 2013, Hornbæk & Frøkjær, 2008). Therefore, we did not need to validate the questionnaires employing an independent group of novice crowd inspectors that were not treated with the questionnaire. Hence, we considered the questionnaire part of the novice crowd usability inspectors' profile instead of a separate intervening variable.

*c)* **Payment Criterion**. We set the payment criterion for novice crowd usability inspection based on performance as below:

▪ A novice crowd inspector would earn 1 USD for the identification of 3 valid usability issues.

- A novice crowd inspector can earn a maximum of 10 USD to identify 30 valid usability issues.

*d)* ***Roles and Time Management***. Crowd inspectors were asked to evaluate test objects with the same roles as experts. Besides, crowd inspectors were requested to record time for questionnaire-based usability evaluation.

## 5.2.3 Execution of Usability Evaluations

We conducted heuristic usability inspection first since we were planning to use the results of one of the five expert's heuristic evaluations to design questionnaires for novice crowd usability inspection. All the evaluators, including experts and crowd usability inspectors, were asked to submit their English language evaluations. The authors ensured that the test objects did not change through the duration of expert and crowd evaluations. Consequently, evaluators may get the same test objects to evaluate. Time durations for usability evaluations were as below:

    a. WHO-EU web dashboard – January 2016

    b. [www.ebay.com](www.ebay.com) – February 2019

    c. [www.booking.com](www.booking.com) – February 2019

### 5.2.3.1 Heuristic Usability Inspection

In expert heuristic usability inspection, a total of 8 usability experts were self-selected based on a set criterion, discussed in subsection 5.2.2.3 instrumentation. Each usability expert was assigned an ID starting from E1 to E8. Two experts, i.e., E1 and E2, participated in all three experiment trials, while experts for trials 2 and 3 were the same, as shown in Table 5.2. A single expert performed each heuristic usability inspection.

***Qualification:*** Three experts (E1, E2, and E3) had certification in usability evaluation, one expert (E5) had a Ph.D. in cognitive science, one expert (E6) had a master's in Psychology and four experts (E3, E4, E7, and E8) had master's degree in HCI.

***Experience:*** Usability experts had an average practical usability evaluation experience of 7.2 years, with a minimum of 5 years and a maximum of 15 years (see Table 5.1).

***Payments:*** The fixed-price contracts ranged from USD 200 to USD 400, with an average of USD 280 per expert heuristic usability inspection (see Table 5.4).

***Diversity in hiring:*** Experts were invited worldwide, and the final selection included experts from Asia, Europe, North and South America.

***Familiarity with test objects:*** None of the experts was familiar with the WHO-EU web dashboard before the experiment. However, expert E1 hardly used eBay.com and used Booking.com only a couple of times. Expert E2 visited eBay.com a couple of years ago before evaluation but never bought anything. Besides, Expert E2 never used Booking.com; instead used a similar website Blocket.se that is commonly used in Sweden for booking accommodation. Expert E6, E7, and E8 visited these websites once in a blue moon before evaluation and were not frequent users of these websites.

***English language skills:*** All the experts were non-native English speakers except E5, E6, and E7. However, none of the experts had any communication problems in the English language during evaluation or communication with the authors.

Each expert performed a heuristic evaluation independently and submitted evaluation reports to the authors for further analysis.

### 5.2.3.2 Novice Crowd Usability Inspection

A total of 11 novice crowd usability inspectors were self-selected based on a set criterion discussed above in subsection 5.2.2.3 instrumentation. We assigned each novice crowd usability inspector an ID starting from C1 to C11. Five novice crowd inspectors (C1 to C5) participated in the first trial of the experiment. The other six novice crowd inspectors (C6 to C11) participated in the experiment's second and third trials, as shown in Table 5.2. A single novice crowd usability inspector performed each inspection.

***Qualification:*** A total of 10 out of 11 novice crowd inspectors had a bachelor's degree, i.e., three in Computer Sciences (C3, C9, and 10), two in IT (C2, C7), and one in each, Computer Applications (C4), Software Engineering (C11), Informatics (C5), Arts (C6), and in Business Administration (C8). Only one novice inspector had a Diploma in Computer Science (C1).

***Experience:*** All novice crowd usability inspectors, except C8, were familiar with HCI/usability/quality assurance testing. Besides, each novice crowd inspector had 1-2 years of experience in software testing except C3 (see Table 5.2).

***Payments:*** The payment for each novice crowd inspector ranged from USD 1 to USD 10, with an average amount of USD 6, see Table 5.4.

***Diversity in hiring:*** Novice crowd usability inspectors were hired on Upwork from different parts of the world, including Russia, India, UK, Turkey, Pakistan, Armenia, Ukraine, Kenya, Norway see Table 5.2.

***Familiarity with test objects:*** In the $1^{st}$ trial of the experiment, none of the novice crowd inspectors was familiar with the test object, i.e., the WHO-EU web dashboard. In the $2^{nd}$ and $3^{rd}$ trials of the experiment, except for C8, all the novice crowd inspectors (C6, C7, C9, C10, and C11) were familiar with at least similar test objects (see Table 5.2).

***English Language Skills:*** All the crowd inspectors were non-native English speakers, except C8 (see Table 5.2). The usability comments of crowd inspectors were understandable primarily.

However, if any confusion was observed in the language of usability comments, we clarified them from crowd inspectors using Upwork chat.

Table 5.1 – Profiles of Usability Experts

| Resource ID | Country | Education | Certification in Usability | Academic Experience (Years) | Practical Usability Experience (Years) | Currently Working as |
|---|---|---|---|---|---|---|
| **Experiment I – WHO-EU Web Dashboard** | | | | | | |
| **E1** | India | MSc in Computing | HFI-CUA | - | 6.5 | Senior UX Designer |
| **E2** | Sweden | MSc in Cognitive Science | HFI-CUA/CXA | 5 | 15 | UX/Usability Expert |
| **E3** | Brazil | MSc in HCI | Nielsen Norman Group UX Researcher | - | 6 | UX/Usability Designer |
| **E4** | Germany | MSc in HCI | - | - | 5 | UX/Usability Consultant |
| **E5** | America | Ph.D. in Cognitive Science | - | 4 | 8 | Senior UX Researcher |
| **Experiment II – www.ebay.com & Experiment III – www.booking.com** | | | | | | |
| **E1** | India | MSc in Computing | HFI-CUA | - | 6.5 | Senior UX Designer |
| **E2** | Sweden | MSc in Cognitive Science | HFI-CUA/CXA | 5 | 15 | UX/Usability Expert |
| **E6** | USA | MSc in Psychology | - | - | 7 | Usability Analyst |
| **E7** | USA | MSc in HCI | - | - | 5 | Interaction Designer |
| **E8** | Spain | MSc in HCI | - | - | 5 | UX/Usability Designer |

***Validation of the crowd's identified usability problems:*** Authors validated the crowd's identified usability problems by comparing them with expert issues identified. Only 23 usability problems did not match with experts' identified usability problems. For further validation, we sent these 23 usability problems to usability expert E1; see section 5.3.1 Analyzing usability inspection reports for more details.

Crowd workers often submit poor quality, incomplete, and fake work to claim payments. Gomide et al. (2014) suggested that the work of outliers (spammers) should be filtered out. However, we did not filter out any usability inspection reports to give a more objective view of the results.

Table 5.2 – Profiles of Novice Crowd Usability Inspectors

| | | | | Experiment I – WHO-EU Web Dashboard | | |
|---|---|---|---|---|---|---|
| **Resource ID** | **Country** | **English Language Skills** | **Qualification** | **Experience in Software Testing (Years)** | **Familiarity with HCI/Usability Experience** | **Familiarity with WHO-EU Web Dashboard (before Inspection)** |
| **C1** | Ukraine | Non-Native | Diploma in Computer Science | 2 | performed a Usability Assignment | Nil |
| **C2** | Armenia | Non-Native | Bachelors in IT | 1 | Familiar with Quality Assurance Testing | Nil |
| **C3** | Kenya | Non-Native | BSc. Computer Science | Nil | Studied HCI as Course/ No Experience | Nil |
| **C4** | India | Non-Native | BSc. Computer Applications | 2 | Studied HCI as Course/ No Experience | Nil |
| **C5** | Russia | Non-Native | BSc. Informatics | 2 | Familiar with Quality Assurance Testing | Nil |
| | | | Experiment II – www.ebay.com & Experiment III – www.booking.com | | | |
| **Resource ID** | **Country** | **English Language Skills** | **Qualification** | **Experience in Software Testing (Years)** | **Familiarity with HCI/Usability** | **Familiarity with Test Objects (before Inspection)** |
| **C6** | Turkey | Non-Native | Bachelor of Arts | 2 | Attended an HCI Course/ 2 years' Usability Testing Experience | Yes |
| **C7** | Pakistan | Non-Native | Bachelors in IT | 1.5 | Familiar with Quality Assurance Testing | Familiar with similar websites |
| **C8** | UK | Native | Bachelor of Business Administration | 2 | Nil | Not frequent user of www.eBay.com /No experience with www.booking.com |
| **C9** | Russia | Non-Native | BSc. Computer Science | 2 | Studied HCI as Course/ No Experience | Familiar with similar websites |
| **C10** | India | Non-Native | BSc. Computer Science | 1 | Studied HCI as Course/ No Experience | Familiar with similar websites |
| **C11** | Norway | Non-Native | BSc. Software Engineering | 1 | Studied HCI as Course/ No Experience | Familiar with similar websites |

## 5.3    ANALYSIS & DISCUSSION

A total of 32 usability inspection reports, including 17 crowd's and 15 experts' evaluations, were submitted to authors for further analysis. It is important to mention that a single usability inspector performed each usability inspection. It took three authors about 130 hours each to analyze a total of 508 comments. We employed Cohen's kappa as an inter-rater agreement measure to assess agreement reliability on coding usability problems (Carletta, 1996). The kappa (k) overall value was observed as 0.87, i.e., almost a perfect agreement. Subsection 6.3.1 discusses in detail the analysis of usability inspection reports. A total of 209 issues were uniquely reported, out of which 177 issues were reported by expert heuristic usability inspection and 121 issues by novice crowd usability inspection with an overlap of 98 issues (47%). The results contain 32 critical issues and 69 serious issues. An overview of the results is shown in table 10.3. The overlap ratio of the unique problems between novice crowd inspection and expert heuristic inspection for the first trial of the experiment, i.e., 54% (46 out of 85 unique issues), was higher compared to the second trial overlap ratio, i.e., 39% (25 out 64 unique issues) and third trial overlap ratio, i.e., 45% (27 out of 60 unique issues) of the experiment. One possible explanation for this difference is that the test object for the first trial of the experiment was a web dashboard containing a single-screen interface leading to a higher rate of finding similar usability issues. On the other hand, the test objects for the second and third trials of the experiment were two transactional websites, i.e., www.ebay.com, and www.booking.com, respectively, having multiple web pages with different functionalities, leading to lower chances of finding similar usability issues.

Table 5.3 – Overview of results

|  | WHO-EU Web Dashboard | www.ebay.com | www.booking.com |
|---|---|---|---|
| Original comments | 187 | 160 | 161 |
| Comments after splitting & combining | 193 | 163 | 169 |
| Total number of unique issues | 85 | 64 | 60 |
| Minor issues | 24 | 46 | 38 |
| Serious issues | 44 | 9 | 16 |
| Critical issues | 17 | 9 | 6 |
| Experts identified issues | 76 | 53 | 48 |
| Crowd identified issues | 55 | 33 | 33 |
| Issues overlapped (Crowd+Expert) | 46 | 25 | 27 |

## 5.3.1 Analyzing Usability Inspection Reports

We adopted a systematic approach to analyzing the individual usability inspection reports to code usability problems. Figure 5.2 shows a stepwise process for analyzing usability inspection reports.

**Step 1 – Identifying Problem Comments:**

The authors analyzed usability inspection reports for problem comments. In this regard, the authors encoded problem comments within the usability problem description. However, complex usability problem descriptions can have multiple unique problem comments and need further processing (Molich and Dumas, 2008).

**Step 2 – Splitting and Combining Problem Comments:**

The composite problem comments were split into multiple atomic problem comments (Molich and Dumas, 2008). They were split manually, extracting sentences topic-wise until they could not be further divided. Xu et al. (2015) also used the topic modeling technique to categorize crowd feedback. Moreover, related but non-identical problem comments were combined to represent a single problem comment. In this way, the usability inspection report becomes more understandable and manageable. However, each atomic comment was equal to the original comment reported in the usability inspection report. Moreover, each atomic problem comment has its solution to fix it without affecting any other atomic problem comments (Molich and Dumas, 2008). The following example demonstrates how a composite original problem comment was split into multiple atomic problem comments from the 1st trial of the experiment, i.e., WHO-EU Web Dashboard:

*The menu that appears after clicking "Select data" and the table with core health indicators list the same indicators. Selecting an option on the "Select data" menu does not affect the health*

*indicators table. It is not self-evident which way of selecting a health indicator a user should choose and how to start.*

1) *The menu that appears after clicking "Select data" and the table with core health indicators list the same indicators.*

2) *Selecting an option on the "Select data" menu does not affect the health indicators table.*

3) *It is not self-evident which way of selecting a health indicator a user should choose and how to start.*

A total of 34 original comments out of 508 were split into multiple atomic comments. In 46 cases, we combined multiple atomic comments in the same inspection report and found them identical. A total of 27 inspectors reported identical comments.

**Step 3 – Identifying & filtering false positives:**

A false alarm is an incorrectly identified usability problem (Gray and Salzman, 1998). In other words, a false alarm is a usability problem solving that would not improve the system's usability (Molich and Dumas, 2008). They further argued that if the myth that expert reviews find many false alarms were valid, there would have been an unevenly large number of usability problems reported by expert reviews than user-based methods (Molich and Dumas, 2008). Some authors believe that the usability issues reported by expert evaluations not confirmed by user-based methods are false alarms. This argument would have been reasonable if the results of user-based methods were reproducible (Molich and Dumas, 2008).

Molich and Dumas (2008) believe that the main reason for earlier false alarms in expert reviews was that too few user-based tests were conducted to verify the usability problems reported by expert reviews. Most of the time, there is only one usability test involving an evaluator of insufficient experience with few users and tasks.

However, three authors reviewed all the usability issues identified by usability inspectors for false positives, based on the definition provided in subsection 5.2.2.3 Instrumentation. A total of 58 false positives were identified, out of which eight were found in expert heuristic usability inspection while 50 were found in novice crowd usability inspection (see Table 5.4). False positives found in novice crowd usability inspection were substantially more than the expert heuristic usability inspection. The apparent reason for the substantial difference between the two UEMs regarding false positives is usability experts' profiles and novice crowd usability inspectors, i.e., skills, experience, and education.

**Step 4 – Combining similar usability problems at a higher abstraction level:**

Usability problems may be defined at different levels of abstraction by the same evaluator and multiple evaluators simultaneously in their inspection reports (Howarth et al., 2009).

Therefore, similar usability problems exist at different abstraction levels, and it is necessary to avoid over and under-reporting usability problems (Cockton and Lavery, 1999). For example, suppose one usability problem comment may report that a website is cluttered with overwhelming information. Another problem comment may describe that the property information page is cluttered with an over-dose of irrelevant information. Therefore, the master list should contain usability problems representing a higher abstraction level. In comparison, individual inspection reports containing lower abstraction levels may be mapped with a master list to describe the same usability problem.

**Step 5 – Filtering identical usability problems from Master-List:**

Finally, identical problem comments reported by multiple usability inspectors were reported uniquely in the master list. Two or more problem comments were considered identical if fixing the problem in one comment will fix the problem in other comments (Molich & Dumas, 2008). Few, sample identical problem comments from different usability inspectors, reported in www.booking.com, are listed below:

*www.booking.com:*

1. *Searching Accommodation: Within each property block, there are several rating systems. Which does a user follow? What does each mean? are they all related? It is quite confusing to understand the rating system.*

2. *Searching Accommodation: Different schemes of rating violate best practices of usability.*

3. *Searching Accommodation: Stars' rating option is confusing.*

4. *Searching Accommodation: There is a usability consideration over the meaning of star ratings. There is no easy way to get info on star ratings from the property page.*

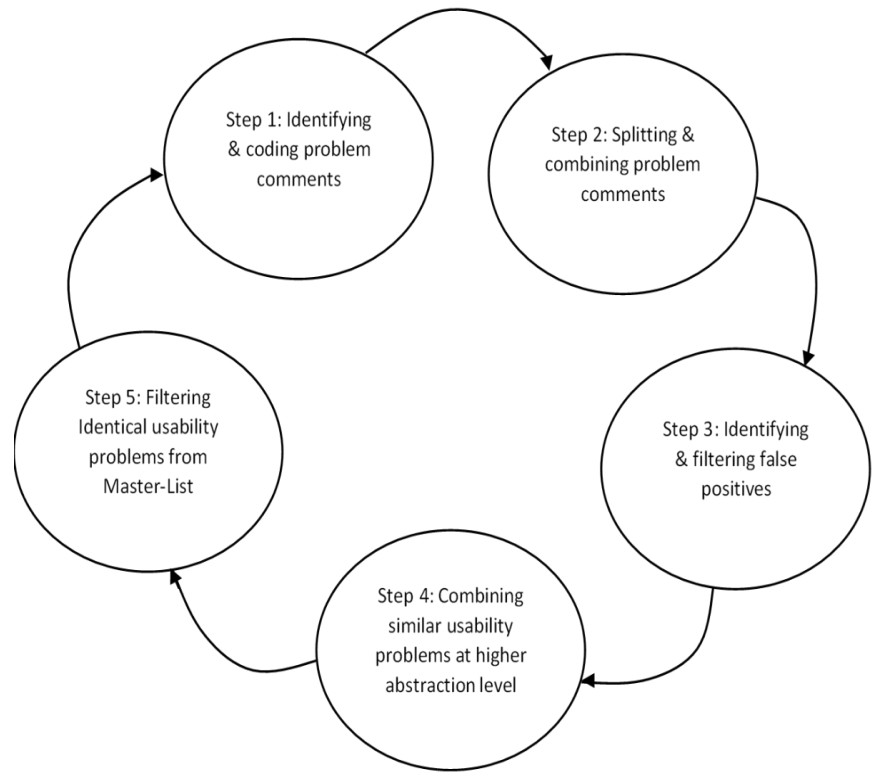

Figure 5.2 – Analyzing usability inspection reports.

### 5.3.2 Severity of Usability Problems

We used expert heuristic usability inspection as a benchmark for comparison. The severity of novice crowd inspector's reported issues was derived from experts' reported problems that matched with them, except for 23 crowd usability issues that did not match. Therefore, no contradiction was observed for the severity of the usability issues between novice crowd usability inspection and expert heuristic usability inspection.

In expert heuristic usability inspection, 51 issues were reported by multiple experts. For 34 out of 51 issues, the majority of the experts suggested the classification as serious, while deviations were only one degree away from the average classification, which we considered acceptable. For the rest of the usability issues, classification varied from minor to critical, and in this case, averaged severity was considered. No other contradictory classifications were observed.

### 5.3.3 Key Issues

Molich and Dumas (2008) suggested that if multiple experts categorize an issue as serious or critical, it indicates that such an issue may have considerable design consequences. Hence, we considered such issues as a quality measure based on the expert's judgment. Moreover, we considered a key usability issue as a repeatedly identified serious usability issue. In other words, a key usability issue is: 1) is at least reported by three usability experts or at least two usability experts and one or more crowd inspectors. 2) is identified as a serious or critical issue on average. We calculated average classification by associating values 1,2,3 to minor, serious, and critical issues, respectively. We then took an average of the values and rounded it off. A total of 30 key usability issues were reported uniquely. Details of the key usability issues reported by each inspector are shown in table 5.4. Some examples of the key usability issues are as below:

***www.booking.com:***

1. *The property detail page looks cluttered with too much information and a poorly organized content hierarchy.*

2. *Different rating symbols are displayed when you view any accommodation, i.e., stars, thumbs-up, and squares. These various rating schemes are not consistent and are quite confusing and difficult to comprehend.*

***www.ebay.com:***

1. *The user cannot select multiple items within one refiner category (e.g., select AT&T and T-Mobile under Network). It is a UX glitch that hinders the refiner experience. Users are immediately led to a page with the results before they can select any other preferences.*

2. *The page has focused on money/payment, not the product. I want to know if I want the product before I start to think about the price. They should place a description together with the product, not far below. There is a risk of missing the description and not bidding on the product of my interest.*

A list of all the key usability issues is provided in Appendix F. Most of the key usability issues were identified in the WHO-EU web dashboard 70% (21 out of 30). In comparison, 17% (5 out of 30) were found on booking.com, and the rest of the 13% (4 out of 30) were reported on eBay.com. The key usability issue identified by the maximum number of inspectors in the WHO-EU web dashboard, i.e., 7, including three experts and four novice crowd inspectors, was #12, violating heuristic, Minimizing occlusion. Likewise, key issue#2 received maximum hits (8 inspectors, including four experts and four novice crowd inspectors) in Booking.com, violating heuristic, Aesthetic and minimalist design. For ebay.com, the most identified key issue was #3 (identified by seven inspectors, including four experts and three novice crowd inspectors), violating heuristic, Consistency, and standards.

## 5.3.4 Resource Usage

How efficiently resources are used is an essential aspect of comparing UEMs. If we look at Figure 10.3, we can see that experts have outnumbered novice crowd inspectors to identify key issues per person. However, person-hours used to identify key usability issues are not free. The person-hours used to change the whole view altogether. The key usability issues identified per hour by each resource are shown in Figure 5.3. We can see that novice crowd inspectors are more efficient than experts in identifying key usability issues per hour. Figure 5.4 indicates that experts have identified more key usability issues at the cost of more time. Likewise, Figure 5.4 shows that experts' key usability issues were quite expensive compared to crowd inspectors. It is worth mentioning that novice crowd inspectors identified 27 out of the total 30 key usability issues.

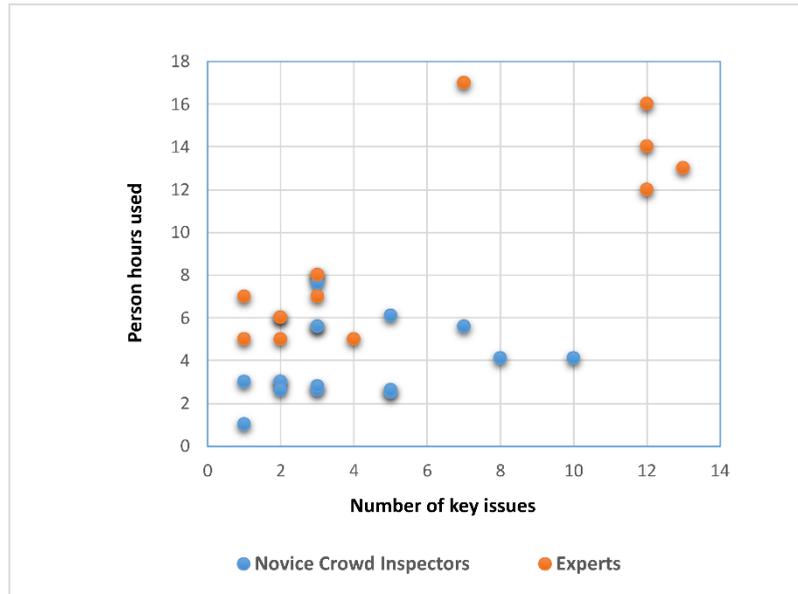

Figure 5.3 – Crowd vs. Experts – Person hours used.

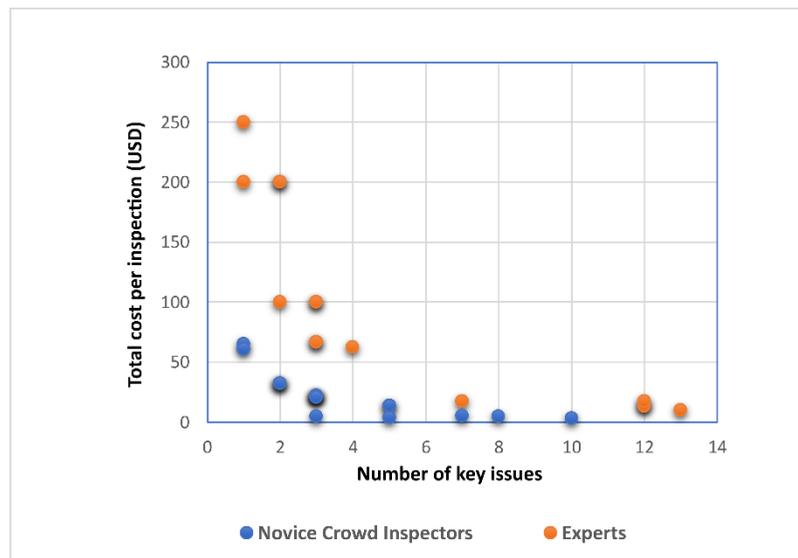

Figure 5.4 – Crowd vs. Experts – Cost per inspection

Table 5.4 – Detailed results

| Experiment I – WHO-EU Web Dashboard | | | | | | | | | | |
|---|---|---|---|---|---|---|---|---|---|---|
| **Resource** | E1 | E2 | E3 | E4 | E5 | C1 | C2 | C3 | C4 | C5 |
| **Total number of original comments** | 26 | 22 | 17 | 17 | 22 | 17 | 19 | 15 | 18 | 14 |
| **Total number of comments after splitting & combining** | 29 | 24 | 15 | 18 | 21 | 15 | 15 | 19 | 22 | 15 |
| **Total number of false positives** | 0 | 1 | 0 | 0 | 1 | 3 | 1 | 2 | 3 | 2 |
| **Total number of usability problems** | 29 | 23 | 15 | 18 | 20 | 12 | 14 | 17 | 19 | 13 |
| **Total number of minor problems** | 3 | 3 | 2 | 2 | 2 | 3 | 5 | 3 | 2 | 5 |
| **Total number of serious problems** | 19 | 15 | 10 | 13 | 10 | 4 | 3 | 11 | 13 | 4 |
| **Total number of critical problems** | 7 | 5 | 3 | 3 | 8 | 5 | 6 | 3 | 4 | 4 |
| **Total number of key usability issues** | 13 | 7 | 12 | 12 | 12 | 7 | 8 | 5 | 10 | 3 |
| **Person hours used** | 13 | 17 | 12 | 14 | 16 | 5.6 | 4.1 | 6.1 | 4.1 | 5.6 |
| **Payment made to each inspector (USD)** | 300 | 400 | 200 | 250 | 350 | 5 | 5 | 6 | 7 | 5 |
| **Total cost per inspection (USD)** | 300 | 400 | 200 | 250 | 350 | 65 | 65 | 66 | 67 | 65 |
| **Cost per issue (USD)** | 10.3 | 17.4 | 13.3 | 13.9 | 17.5 | 5.4 | 4.6 | 3.9 | 3.5 | 5.0 |
| **Cost per key issue (USD)** | 23.1 | 57.1 | 16.7 | 20.8 | 29.2 | 9.3 | 8.1 | 13.2 | 6.7 | 21.7 |
| **Issues per hour** | 2.2 | 1.4 | 1.3 | 1.3 | 1.3 | 2.1 | 3.4 | 2.8 | 4.6 | 2.3 |
| **Key issues per hour** | 1.0 | 0.4 | 1.0 | 0.9 | 0.8 | 1.3 | 2.0 | 0.8 | 2.4 | 0.5 |
| **Validity of usability problems** | 100% | 96% | 100% | 100% | 95% | 80% | 93% | 89% | 86% | 87% |
| **Thoroughness of usability problems** | 34% | 27% | 18% | 21% | 24% | 14% | 16% | 20% | 22% | 15% |

| Experiment II – www.ebay.com | | | | | | | | | | | |
|---|---|---|---|---|---|---|---|---|---|---|---|
| **Resource** | E1 | E2 | E6 | E7 | E8 | C6 | C7 | C8 | C9 | C10 | C11 |
| **Total number of original comments** | 20 | 17 | 15 | 13 | 8 | 17 | 15 | 18 | 14 | 16 | 07 |
| **Total number of comments after splitting & combining** | 21 | 16 | 14 | 11 | 8 | 19 | 20 | 22 | 14 | 18 | 0 |
| **Total number of false positives** | 0 | 1 | 1 | 0 | 2 | 5 | 3 | 4 | 3 | 3 | 4 |
| **Total number of usability problems** | 21 | 15 | 13 | 11 | 6 | 14 | 17 | 18 | 11 | 15 | 3 |
| **Total number of minor problems** | 7 | 11 | 12 | 6 | 3 | 5 | 4 | 11 | 3 | 6 | 0 |
| **Total number of serious problems** | 7 | 3 | 0 | 3 | 3 | 5 | 6 | 3 | 4 | 3 | 0 |
| **Total number of critical problems** | 7 | 1 | 1 | 2 | 0 | 4 | 7 | 4 | 4 | 6 | 0 |
| **Total number of key usability issues** | 3 | 2 | 1 | 1 | 3 | 1 | 3 | 3 | 2 | 3 | 1 |
| **Person hours used** | 8 | 5 | 5 | 7 | 8 | 3 | 2.6 | 7.6 | 3 | 2.8 | 1 |
| **Payment made to each inspector (USD)** | 300 | 400 | 250 | 200 | 200 | 5 | 6 | 7 | 5 | 6 | 1 |
| **Total cost for per inspection (USD)** | 300 | 400 | 250 | 200 | 200 | 65 | 66 | 67 | 65 | 66 | 61 |
| **Cost per issue (USD)** | 14.3 | 26.7 | 19.2 | 18.2 | 33.3 | 4.6 | 3.9 | 3.7 | 5.9 | 4.4 | 20.3 |
| **Cost per key issue (USD)** | 100 | 200 | 250 | 200 | 66.7 | 65 | 22 | 22.3 | 32.5 | 22 | 61 |
| **Issues per hour** | 2.6 | 3.0 | 2.6 | 1.6 | 0.8 | 4.7 | 6.5 | 2.4 | 3.7 | 5.4 | 3 |
| **Key issues per hour** | 0.4 | 0.4 | 0.2 | 0.1 | 0.4 | 0.3 | 1.2 | 0.4 | 0.7 | 1.1 | 1 |
| **Validity of usability problems** | 100% | 94% | 93% | 100% | 75% | 74% | 85% | 82% | 79% | 83% | 43% |
| **Thoroughness of usability problems** | 34% | 25% | 21% | 18% | 10% | 23% | 28% | 30% | 18% | 25% | 5% |

| Experiment III – www.booking.com | | | | | | | | | | | |
|---|---|---|---|---|---|---|---|---|---|---|---|
| **Resource** | E1 | E2 | E6 | E7 | E8 | C6 | C7 | C8 | C9 | C10 | C11 |
| **Total number of original comments** | 15 | 12 | 14 | 12 | 15 | 17 | 15 | 16 | 16 | 21 | 8 |
| **Total number of comments after splitting & combining** | 17 | 13 | 13 | 11 | 16 | 19 | 13 | 20 | 16 | 24 | 7 |
| **Total number of false positives** | 0 | 0 | 1 | 1 | 0 | 4 | 3 | 3 | 3 | 3 | 1 |
| **Total number of usability problems** | 17 | 13 | 12 | 10 | 16 | 15 | 10 | 17 | 13 | 21 | 6 |
| **Total number of minor problems** | 6 | 10 | 8 | 8 | 12 | 4 | 4 | 5 | 8 | 7 | 3 |
| **Total number of serious problems** | 8 | 1 | 4 | 2 | 4 | 9 | 5 | 9 | 4 | 11 | 2 |
| **Total number of critical problems** | 3 | 2 | 0 | 0 | 0 | 2 | 1 | 3 | 1 | 3 | 1 |
| **Total number of key usability issues** | 3 | 2 | 4 | 2 | 3 | 5 | 2 | 3 | 2 | 5 | 3 |
| **Person hours used** | 7 | 6 | 5 | 6 | 8 | 2.5 | 3 | 5.6 | 2.6 | 2.6 | 1 |
| **Payment made to each inspector (USD)** | 300 | 400 | 250 | 200 | 200 | 7 | 5 | 6 | 5 | 8 | 3 |
| **Total cost per inspection (USD)** | 300 | 400 | 250 | 200 | 200 | 67 | 65 | 66 | 65 | 68 | 63 |
| **Cost per issue (USD)** | 17.6 | 30.8 | 20.8 | 20.0 | 12.5 | 4.5 | 6.5 | 3.9 | 5.0 | 3.2 | 10.5 |
| **Cost per key issue (USD)** | 100 | 200 | 62.5 | 100 | 66.7 | 13.4 | 32.5 | 22 | 32.5 | 13.6 | 21 |
| **Issues per hour** | 2.4 | 2.2 | 2.4 | 1.7 | 2.0 | 6.0 | 3.3 | 3 | 5 | 8.1 | 6 |
| **Key issues per hour** | 0.4 | 0.3 | 0.8 | 0.3 | 0.4 | 2.0 | 0.7 | 0.5 | 0.8 | 1.9 | 3 |
| **Validity of usability problems** | 100% | 100% | 92% | 91% | 100% | 79% | 77% | 85% | 81% | 88% | 75% |
| **Thoroughness of usability problems** | 31% | 24% | 22% | 19% | 30% | 28% | 19% | 31% | 24% | 39% | 11% |

## 5.3.5        Statistical Analysis

To test the hypotheses (formulated above in subsection 5.2.2 Planning) statistically, we performed significance testing. The results for each hypothesis are discussed below:

**$H_a$: Usability Issues Found**

To find out whether novice crowd usability inspection finds the same usability issues (w.r.t. content & quantity) or different, we need to split this hypothesis into two parts, i.e., a) content b) quantity. As far as the content is concerned, we have already discussed that 47% (98 out of 209) of total unique issues were the same and identified by both novice crowd usability inspection and expert heuristic usability inspection. Moreover, in terms of key usability issues, 90% (27 out of 30) of total key issues overlapped between novice crowd usability inspection and expert heuristic usability inspection. To test the second part of the hypothesis, i.e., quantity, we compared the two samples for the number of usability issues identified, using a non-parametric test, the Mann-Whitney U test. The result (p=0.83) supports the alternate hypothesis $H_{a1}$, and there were no significant differences between the two groups. In other words, we can say that novice crowd usability inspection, on average, finds the same usability issues in terms of content & quantity as expert heuristic usability inspection.

**$H_b$: Cost-Effectiveness**

It is worth mentioning that the cost for novice crowd inspectors is aggregate; see table 6.4. And it includes the cost for expert heuristic usability inspection conducted by a single usability expert (E1). We divided this cost on all novice crowd inspectors, adding 60 USD for each crowd inspector in each experiment's trial. We calculated the cost for novice crowd usability inspection in this way since almost half of the use-cases were based on the problems identified by expert E1 in each trial.

To compare the UEMs for cost-effectiveness, we considered cost per issue identified for both samples, using the Mann-Whitney U test. The result, i.e., p=0.00001, supports the alternate hypothesis that novice crowd usability inspection is more economical than expert heuristic usability inspection is true.

**$H_c$: Time-Efficiency**

Our third hypothesis was based on the time spent on a usability inspection process. The time spent on the usability inspection process for each evaluation is shown in table 5.4. The time duration for novice crowd usability inspection is aggregate. It includes a) the time for novice crowd inspection, b) the time spent on expert heuristic usability inspection conducted by a single expert (E1), and c) the time for designing use-cases for crowd usability inspection (i.e., 3 hours).

The time duration for prerequisite expert review and use-cases' designing was divided on all novice crowd usability inspections for each experiment's trial.

The job completion time was considered by comparing the two samples. The result (p=0.00001) reveals that there is a significant difference between the two samples. Hence, the alternate hypothesis, i.e., $H_{c1}$ – Novice crowd usability inspection on average takes less time than expert heuristic usability inspection, is true. A summary of the hypothesis testing results is shown in table 5.5.

Table 5.5 – Hypothesis Testing Results

| S.No. | Hypothesis | P-value | Significance | Result |
|---|---|---|---|---|
| 1 | $H_a$: Usability issues found | p=0.83 | Not significant | $H_{a1}$ - Usability inspection using crowdsourcing on average finds the same usability issues (w.r.t. content & quantity) as a heuristic usability inspection |
| 2 | $H_b$: Cost-Effectiveness | p=0.00001 | Significant | $H_{b1}$ - Usability inspection using crowdsourcing incurs less cost than heuristic usability inspection. |
| 3 | $H_c$: Time-Efficiency | p=0.00001 | Significant | $H_{c1}$ - Usability inspection using crowdsourcing, on average, takes less time than heuristic usability inspection. |

## 5.3.6 Comparison of Results

Usability inspection employing novices has been investigated by researchers in the past (Koutsabasis et al., 2007; Følstad et al., 2010). The performance of the evaluators was often compared using two measures, i.e., validity and thoroughness. The validity of the results can be obtained by dividing the total number of correct problem predictions (real problems) by the total number of problem predictions, including false positives (All comments). The validity of the results shows the number of correct problem predictions, excluding false positives (Følstad et al., 2010). The thoroughness of the results can be measured by dividing the total number of correct problem predictions using all the real problems. The thoroughness of the results shows the degree to which real problems have been identified compared to all problems (Følstad et al., 2010).

The novice crowd usability inspection results are encouraging in terms of the validity of the results compared with expert heuristic usability inspection and other studies (Koutsabasis et al., 2007, Følstad et al., 2010) see table 5.4, 5.6. The validity of the novice crowd usability inspection results remained relatively consistent in all three experiments, ranging from 82% to 87%, with an average of 83% if we exclude the work of the crowd usability inspector, C11 (outlier). However, after including the work of C11, validity drops to a level ranging from 74% to 87%, with an average of 81%. In terms of thoroughness, novice usability inspection results are promising compared to expert heuristic usability inspection (see table 5.6). Compared with other studies, novice inspections are encouraging compared with Koutsabasis et al. (2007), while less thorough than the results of Følstad et al. (2010). One reason for the high level of thoroughness in the results of Følstad et al. (2010) could be that their team of work domain experts comprised three times more participants, i.e., 15, compared to our study, i.e., 5-6. The validity and thoroughness of crowd inspectors and experts' usability problems are also provided in table 5.4.

### 5.3.7 Challenges of Crowdsourcing and Solution Strategies

Over time, researchers have reported several challenges faced by crowdsourcing, including low quality of work (Zhao & Zhu, 2014, Wang et al., 2017), the cognitive load of examining the quality of work (Zhao & Zhu, 2014), the validity of self-reported demographic details (Liu et al., 2012), malicious (Wang et al., 2017) and anonymous submission of work multiple times to increase the reward, lack of appropriate reward (Guaiani, and Muccini, 2015, Garcia-Molina et al., 2016) and review mechanism of work.

Crowdsourcing has matured quite a bit in the field of computing (Ambreen and Ikram, 2016). Modern crowdsourcing platforms, i.e., Upwork, are quite established and offer different ways and means to cope with the challenges discussed earlier. We will discuss these features and support learned in this study while taking software usability evaluation into context.

Table 5.6 – Comparison of results

| Study Reference | Usability Evaluation Method | Teams Size | Validity | Thoroughness |
|---|---|---|---|---|
| (Koutsabasis et al., 2007) | Heuristic Evaluation | 3 | 94.4% | 24.3% |
| | Heuristic Evaluation | 3 | 60.7% | 24.3% |
| | Heuristic Evaluation | 3 | 100% | 20% |
| | Cognitive Walkthrough | 3 | 85.7% | 25.7% |
| | Cognitive Walkthrough | 3 | 70.8% | 24.3% |
| | Think-aloud Protocol | 3 | 90.5% | 27.1% |
| | Think-aloud Protocol | 3 | 94.4% | 24.3% |
| | Think-aloud Protocol | 3 | 82.4% | 20% |
| | Co-Discovery | 3 | 74.4% | 41.4% |
| (Følstad et al., 2010) | Work Domain Experts | 15 | 22.2% | 100% |
| | Work Domain Experts | 15 | 22% | 50% |
| | Work Domain Experts | 15 | 32% | 53% |
| | Work Domain Experts | 15 | 56% | 83% |
| This study | Crowd Inspection (Web Dashboard) | 5 | 87% | 18% |
| | Crowd Inspection (eBay.com) | 6 | 74% | 22% |
| | Crowd Inspection (Booking.com) | 6 | 81% | 25% |
| | Expert Inspection (Web Dashboard) | 5 | 98% | 25% |
| | Expert Inspection (eBay.com) | 5 | 92% | 22% |
| | Expert Inspection (Booking.com) | 5 | 97% | 25% |

**The validity of self-reported demographics.** Currently, crowdsourcing platforms maintain verified profiles of crowd-workers, e.g., Upwork issues identity badges for verified accounts (Upwork, 2022). Earning an identity badge requires two-stage verification is performed, a) video call verification, b) verification through government-issued ID. Moreover, maintaining crowd workers' profiles shows how they performed in previous projects and their feedback and ratings from employers (Upwork, 2022). Zhao and Zhu (2014) also suggested a public rating-based strategy to control the quality of the work. Besides, providing online tests to crowd-workers to prove their skills is another feature of crowdsourcing platforms (Upwork, uTest, 2022).

**Malicious and anonymous work.** Although some crowdsourcing platforms allow crowd-workers to submit their work anonymously (MTurk, 2022), the same crowd-worker may complete the same task multiple times to increase their reward. However, other crowdsourcing platforms (e.g., Upwork) are very strict about crowd workers' identity, and it is almost impossible to complete the same task multiple times to increase the reward.

**Poor quality of work.** Assessing crowd workers' quality is a major challenge in crowdsourcing (Zhao & Zhu, 2014). Especially, usability problem descriptions reported by novices need to be examined by usability experts for false positives (de Lima Salgado et al., 2018). Moreover, novice usability inspectors need to be supported with use-cases, etc. (de Lima Salgado et al., 2018). In our study, we have used a similar approach and supported novice crowd usability inspectors with use-cases. Besides, we compared the usability problem descriptions reported by novice crowd usability inspectors with experts to determine the severity of usability problems. Furthermore, the authors' panel examined the usability problem descriptions for false positives and duplicate comments (Molich et al., 2004).

**Lack of review and reward mechanism** Some crowdsourcing platforms (e.g., MTurk) do not allow to review of the work submitted by crowd workers, and crowd workers get paid for that work, regardless of the quality of work. However, this is not the case with all the crowdsourcing platforms. Some crowdsourcing platforms (e.g., Upwork) provide a proper mechanism to review the work submitted by crowd workers and ask them to revise the work submitted if required. And the payment is released only when the employer is satisfied with the quality of the work, and the employer has an option to discard the work altogether with no payments made to crowd workers. Moreover, employers can hire crowd workers on a fixed-price contract or per hour rate. Employers can divide the work into milestones and can release payments with milestones after reviewing the work. Hence, reward and review mechanisms are well controlled in modern crowdsourcing platforms (e.g., Upwork).

### 5.3.8 Wisdom of crowd and novice usability inspection

Current literature on usability/interactive design evaluation employing crowdsourcing has several limitations to the best of our knowledge. The existing literature does not address the challenges of crowdsourcing nor proposes any solution strategies in this regard, except comparing the performance of crowd-workers with the control group (see subsection 5.1.2.3 and 5.1.2.4 for more details). Finally, subsection 5.3.7 presents the experiences learned from this study to overcome crowdsourcing challenges while employing novice crowd usability inspectors for usability evaluation.

Moreover, the findings of our study imply that crowd-workers with HCI/interactive design skills and assistive tools can produce results comparable to experts. For example, we employed uses-case as assistive tools while Yuan et al. (2016) used a rubric of design principles. However, the study by Yuan et al. (2016) was based on a design critique of UI elements compared to our research based on comparative usability inspection. On the other hand, the studies that did not support crowd-workers with assistive tools and HCI/interactive design skills could not produce results comparable to experts (Xu et al., 2015, Bruun & Stage, 2015).

Besides, the studies conducted without employing crowdsourcing also support the argument that novices without experience and knowledge of usability would not produce comparable results to usability experts (Nielsen,1992; Borys and Laskowski, 2014). On the other hand, the studies that equipped experienced novices with assistive tools produced encouraging results (Howarth et al., 2009; Følstad et al., 2010). Some of the studies on novice usability evaluation did not practically validate their work. However, these studies also supported the argument that novices should be supported with assistive tools, i.e., use-cases, storyboards, design patterns, etc. (Botella et al., 2013; De et al., 2018).

Moreover, our study contributes to existing knowledge on crowd usability evaluation. We have not found any other research comparing novice usability inspectors (equipped with use-cases and HCI skills) with usability experts using crowdsourcing (see subsection 5.1.2.5 for more on the novelty of this study).

In the context of the above discussion and results of our study, it can be safely affirmed that the crowd's wisdom has the potential for novice usability inspection. Moreover, experienced novices/beginners with assistive tools can perform comparably to usability experts. Although novice crowd usability inspection gives comparable results to expert heuristic usability inspection, it is a hybrid method that involves a usability expert. We do not claim that crowd usability inspection can always be the best substitute for usability testing or expert usability inspection. It can be adopted as an economical alternate solution for budget-constrained software development organizations for usability evaluation. When it is expensive to hire 3-5 usability experts or conduct usability testing, novice crowd usability inspection may be employed as an alternative method for usability evaluation. Unlike expert usability inspection, novice crowd usability inspection involves the following overheads:

a) Designing use-cases to develop a questionnaire for novice crowd usability inspection.

b) Analyzing usability comments to extract usability problems, filtering false alarms, and determining the usability problems' severity.

In comparison to expert usability inspection, novice usability evaluators need to be supported with use-cases. However, designing the use-cases is time-consuming. Nevertheless, tasks are also identified in usability testing as well. Besides, analyzing the usability comments of novice crowd inspectors requires intellectual effort. However, the efforts for analyzing the usability comments can be mitigated using text-mining techniques using machine learning. A similar approach has been suggested by (Xu et al., 2015, Wang et al., 2017).

## 5.4    LIMITATIONS AND VALIDITY THREATS

A range of validity threats needs to be addressed for a comparative usability study (Gray and Salzman, 1998). We will discuss them one by one.

### 5.4.1  Statistical Conclusion Validity

There are three main issues in statistical conclusion validity: low statistical power and wildcard effect, random heterogeneity in sample, and too many comparisons to capitalize on chance factors. For expert evaluation, 3-5 experts are recommended (Jeffries and Deservers, 1992), and we employed five usability experts to perform a heuristic expert evaluation for each trial of the experiment. We did not notice a wildcard effect in expert heuristic evaluation results due to the random heterogeneity of experts' profiles. Therefore, we did not discard any results for expert heuristic evaluation. For novice crowd usability inspection, we analyzed the results for any outliers as adopted in other comparative usability studies (Chattratichart and Lindgaard, 2008; Gomide et al., 2014). The crowd usability inspector C11 identified significantly fewer usability problems (3 in 1$^{st}$ trial and 6 in 2$^{nd}$ trial of the experiment) than other crowd inspectors. The crowd inspectors, on average, identified 14 usability problems in each trial of the experiment. However, we did not exclude any usability inspection reports to give a more objective view of the results. Furthermore, we did not perform too many comparisons; therefore, there is no possibility of exploiting the chance factor. Besides, we did not rely on eyeball tests, i.e., averages, percentages; instead, we performed statistical tests to prove the validity of hypotheses.

### 5.4.2  Internal Validity

Internal validity threats concern whether the differences between the UEMs are causal rather than correlational. There are three main aspects of internal validity, i.e., instrumentation, selection, and setting. There are two main biases in instrumentation, i.e., how the severity of the usability problems is assessed and their categorization. To avoid biases in determining usability problems' severity, each expert independently identified the severity of usability problems. Moreover, the average severity of common usability problems belonging to the same key usability issue was considered regarding the key usability issues. The severity of the usability problems identified by novice crowd inspectors was derived from experts' identified usability problems, which matched with them, leaving only 23 usability problems that did not match. The severity of these 23 usability problems was assessed by the expert (E1), whose usability inspection report was used to design the use-cases for novice crowd usability inspection.

Novice crowd usability inspection consisted of use-cases that were based on a single expert evaluation. However, half of the total use-cases were general use-cases developed based on the test objects' functionality. Besides, novice crowd inspectors were not asked to categorize the usability problems in terms of heuristics or severity. Therefore, novice crowd inspectors identified usability problems irrespective of the expert evaluation, categorizing usability problems for severity and heuristics.

The novice crowd inspectors and experts were self-selected based on pre-defined criteria to overcome the selection bias. All the inspectors were hired using the same platform to avoid the problems with settings, i.e., Upwork, and the same platform was used to communicate with them.

### 5.4.3  Construct Validity

Construct validity deals with two main issues: a). Are we manipulating what we have claimed to manipulate (Causal construct validity)? b). Do we measure what we claimed to measure (effect construct validity)?

In causal construct validity, there are many threats. The most prominent threat is defining, operationalization, and understanding the UEM. Other threats include applying a flexible usability evaluation in only one way (mono-operation bias), applying a UEM to only one type of software (mono-method bias), and learning effect due to treatments' interaction.

To overcome the misinterpretation in the definition of UEMs, we clearly defined both expert heuristic usability inspection and novice crowd usability inspection. The expert evaluation was based on heuristics, while crowd evaluation was based on use-cases with clear instructions, how to perform it. Simple and clear instructions were given to evaluators to avoid mono-operation bias and variations in UEMs' operation. For example, all the experts were asked to perform evaluations independently without interacting with each other. To address the mono-method bias, we have generalized our findings at least to two websites and a web dashboard. The learning effect was neutralized by avoiding the interaction of the treatments. None of the participants was given both treatments in the experiment, i.e., expert heuristic usability inspection and novice crowd usability inspection.

In effect-construct validity, problems arise when results of empirical UEMs are compared with analytic UEMs. However, this is not the case in this study, and we compared intrinsic features and pay-off measures of expert heuristic usability inspection with novice crowd usability inspection separately. Moreover, we compared the UEMs based on the contents of usability problems identified rather than just the number of problems identified.

## 5.4.4 External Validity

This experiment was conducted using experts that had relevant knowledge and practical experience of heuristic usability inspection. Besides, novice crowd inspectors were also hired using a crowdsourcing platform, i.e., Upwork. Moreover, real-life systems, i.e., WHO-EU Web Dashboard and websites like www.booking.com and www.ebay.com, were employed for usability evaluations. Therefore, this research study's findings are valid for novice crowd usability inspection and expert heuristic usability inspection.

## 5.4.5 Conclusion validity

Our claims are consistent with the hypotheses that have been investigated in this research study. All the claims are compatible with the results of the study.

## 5.4.6 Limitations of the study

This study has the following limitations due to usability practitioner's effect:

1. The authors designed the questionnaires for novice crowd usability inspection based on a single expert evaluation. Hence, it may introduce limitations such as (a) The expert's usability inspection report may influence the results of crowd usability inspection, (b) The HCI knowledge, experience, skills of usability practitioners may affect the design of the questionnaire for novice crowd usability inspection, and hence results may vary.

2. The authors analyzed crowd usability inspection reports to extract usability problems. There might be a variation in results when a usability practitioner would extract usability problems from novice inspection reports.

3. Usability experts were not asked to have a consensus meeting. Hence, the lack of a consensus meeting might have influenced the results. Jakob Nielsen in 1992 found that double experts can perform better than regular experts. A consensus meeting among usability experts could improve the results, i.e., finding more key issues. However, further research work is required to investigate the effect of consensus meetings among experts.

4. Crowd inspectors were also not asked to have a consensus meeting. A consensus meeting may influence the results. Law and Hvannberg (2008) found that novice usability evaluation in collaborative settings results in inflation and deflation of usability problems.

5. The authors made sure that all the usability inspectors were fluent in the English language. However, we did not investigate the effect of nationality and cultural issues in usability evaluation in this study. Nationality, English language skills, and cultural issues may also influence the results. For example, usability inspectors from countries where English is not a native language may find it difficult to express usability problems. Besides, hiring usability inspectors to evaluate an online shopping website from a country where online shopping is not a cultural trend due to any reason (i.e., poor internet facility, lack of online payment facility, lack of trust in online shopping, etc.) would not be the best choice. Følstad et al. (2010) suggested that domain knowledge can help novice usability inspectors to perform better in usability evaluations.

# Chapter 6

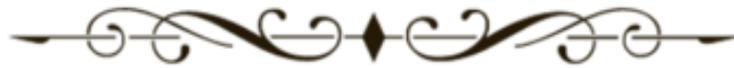

# 6. Study 3 - Framework for Crowd Usability Inspection: A Case Study

## 6.1 INTRODUCTION

Software usability is crucial for every user interface design. It makes the latest technology a pleasurable experience to use and determines which apps and websites we would use repeatedly. However, usability evaluation is expensive for small and medium-size budget-constrained software development organizations. Nevertheless, as Dr. Ralf Speth, CEO, Jaguar Land Rover, once said, "If you think good design is expensive, you should look at the cost of bad design" (Speth, R). Data collected from 863 design projects shows that if you devote 10% of the project's budget for usability, desired usability metrics will improve by 202% (Nielsen, 2003).

The usability Inspection method was introduced as cost-effective means of usability evaluation (Nielsen, J., 1989). However, usability inspection methods only work well with 3-5 usability experts (Jeffries and Deservers, 1992). Still, hiring 3-5 usability experts is not an economical solution for budget-constrained software engineering enterprises (Bak, Nguyen, Risgaard, & Stage, 2008; Ardito et al., 2011; Häkli, 2005). Researchers have tried new techniques that could reduce the cost of usability evaluation and make it a workable solution for small and medium-size software enterprises.

In this regard, Bruun and Stage's (2015) work is prominent, employing usability testing using crowdsourcing and the Barefoot technique. They suggested that local software practitioners may be given short training to perform usability testing and named this technique Barefoot. Moreover, they hired university students to perform remote usability testing while sharing minimalist training material online and named this technique usability testing using crowdsourcing. Later they compared the results of both techniques and found that Barefoot performs better than crowdsourcing. Moreover, they observed that usability testing using crowdsourcing does not work well without users' competencies in HCI, besides the need for end-user report analysis. Crowdsourcing over time has emerged as an economical and effective means for usability evaluation (Liu, Bias, Lease, & Kuipers, 2012; Zhao and Zhu, 2014; Bruun and Stage, 2015). In another study, we investigated crowdsourcing for usability inspection (Nasir, Ikram, & Jalil, 2022). We conducted multiple experiments to show that usability inspection using crowdsourcing is less expensive than expert usability evaluation with comparable results. We suggested that hiring the crowd workers with HCI competencies is economical to hire for usability inspection. Moreover, we proposed that crowd workers may be given use-cases to perform usability inspections based on the usability evaluation of a single expert. However, we did not apply our work in practical settings, i.e., the application of crowd usability inspection to develop software products inside an actual commercial entity.

### 6.1.1 Research Motivation

Wixon (2003) criticized the usability evaluation studies for comparing the usability evaluation methods merely focused on a single criterion, i.e., number of usability problems found, and that also in isolation from practical settings and context in which the method would be used. He argued that even if a usability evaluation method identifies 100% usability problems, it would still fail if it does not work within a software development process of an organization. He considered it short-sightedness if the proposed method's primary emphasis is identifying usability problems rather than fixing them. He suggested that the criterion for evaluating a usability method should be how it improved the usability of a product with available usability resources? He further suggested that the case study is the most effective approach for evaluating usability methods instead of experiments.

Moreover, Grey and Salzman (1998) also criticized the comparative usability studies for low statistical power, comparing UEMs based on several usability problems found without considering the nature & content of the problems. Grey and Salzman (1998) further recommended that the problem of low statistical power validity can be avoided by using a larger sample size or if using a small sample size, i.e., 3-5 evaluators, should replicate the same experiment with multiple sessions, using different participants, each time and possibly with different software systems as well.

Although, in our previous multi-experimental study, we took care of most of the critique on crowdsourcing except that we did not validate our work in practical settings using a case study (Nasir, Ikram, & Jalil, 2022). Moreover, we did not fix the usability problems found in test objects to see the impact of their work on the product's usability (Nasir, Ikram, & Jalil, 2022).

Although crowdsourcing has been investigated for usability testing and evaluation with varying degrees of success (Bruun and Stage, 2015; Liu, Bias, Lease, & Kuipers, 2012; Nasir, Ikram, & Jalil, 2022), still there is very little literature available for replicating the successful interventions for crowd usability testing and evaluation. In the absence of a guiding framework for crowd usability evaluation, crowd usability methods pose a higher risk of failure using a hit and trial strategy. A methodological framework for crowdsourcing in research has been proposed by Keating and Furberg (2013). Nevertheless, it is quite insightful. However, it is quite abstract to exploit it for crowd usability inspection specifically.

This study aims to fill this gap by proposing a framework for crowd usability inspection, based on the findings of previous research work on crowd usability inspection (Nasir, Ikram, &

Jalil, 2022). Later, we would validate this framework in an applied context using a case study with an iterative approach to fix the usability problems found.

### 6.1.2    Challenges of crowdsourcing

Crowdsourcing has encountered several challenges throughout the course of time, according to research. These challenges include low quality of work (Zhao & Zhu, 2014, Wang et al., 2017), the cognitive load of examining the quality of work (Zhao & Zhu, 2014), the validity of self-reported demographic details (Liu et al., 2012), malicious (Wang et al., 2017) and anonymous submission of work multiple times to increase the reward, lack of appropriate reward (Guaiani, and Muccini, 2015; Garcia-Molina et al., 2016) and review mechanism of work, see figure 6.1.

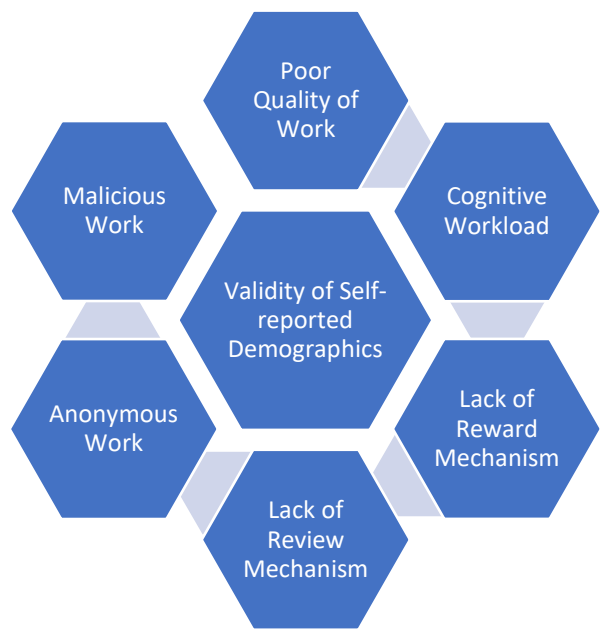

Figure 6.1 – Challenges of crowdsourcing

These challenges have been discussed in detail in our previous multi-experiment research study on crowd usability inspection (Nasir, Ikram, & Jalil, 2022). We discussed several solution strategies to overcome these challenges in (Nasir, Ikram, & Jalil, 2022). In this study, we would like to devise a framework for crowd usability inspection based on our findings and experiences learned in our earlier multi-experimental research study on crowd usability inspection.

### 6.1.3    Devising framework for crowd usability inspection

A usability practitioner would need to consider several aspects of the crowdsourcing to successfully exploit its benefits for usability evaluation. These aspects include, clearly defining the goal of usability evaluation (what would you measure and how?), identifying the tools for crowdsourcing (usability heuristics, classification scheme for usability problems, etc.), choosing the crowd (defining the profile of crowd usability inspectors, guidelines for crowd inspection, etc.), deciding about the platform for crowdsourcing, criteria for reviewing the work and payment method, and finally analyzing the usability evaluation reports.

The proposed framework for crowd usability inspection has six components in it, see figure 1. Each component of the proposed framework shows activities within it at an abstract level. All the components of the proposed framework are briefly discussed below.

**Goal**

The goal defines the basic ambition behind any methodological framework. Moreover, it determines how to measure success. For example, in crowd usability inspection, we can identify usability problems with the help of single expert evaluation and follow-up iterative crowd usability inspections and fix them to improve the quality of our product economically. The following aspects would be necessary for defining the goal:

- Define basic motivation for your framework

- Define success

- Define what you would measure?

- Define how you would measure?

**Instrumentation**

After defining the goal, we need to instrument the crowd usability inspection. For instance, we need to select the usability heuristics used in pre-requisite single expert usability evaluation. After performing the single expert evaluation, a usability practitioner would need to design use-cases for crowd usability inspection based on the findings of a single expert evaluation. The following points may be considered in the instrumentation phase:

- Be specific in choosing the set of usability heuristics as per your requirements of UI and software domain

- It would be worthwhile if the certifications/degrees etc. claimed by the expert are verifiable online

- The number of use cases may be a maximum of 20 as too many uses-cases would frustrate the crowd inspector leading to poor performance.

- Use-cases may be supported with graphics and web links, leading a crowd inspector to the exact course of action.

- Crowd inspectors may be asked to document their findings in a medium where they may have to save their work and continue later. Online forms that do not have the facility to save the work completed and continue later frustrate the crowd inspectors, coercing them to complete the job in one go. Eventually leading to poor performance.

**Crowd**

In this component of the framework, we would need to define the profile of the crowd usability inspector. Ideally, a crowd inspector may have an undergraduate degree in computing or a relevant field with preferably 1-2 years of experience in software testing or HCI. However, crowd inspectors' profiles may not exceed an expert's level, increasing the cost of crowd usability inspection. Besides, we need to have usability guidelines for crowd usability inspection. Usability guidelines could be usability heuristics, defined to help crowd inspectors identify usability problems. Following guidelines would be helpful in this phase:

- Pilot test the questionnaire containing use-cases to fix any possible flaws in it.

- Check out the profile of crowd inspectors for their past reviews and rating by previous employers and see whether the nature work they have done is relevant?

- Have a look at the profile of crowd inspectors to see whether their education is relevant?

**Platform**

Crowdsourcing platform plays a vital role in the successful execution of crowd usability inspection, and it affects it in several ways. For instance, payment method (credit card, western union, etc.) and tax collection mechanism that a crowdsourcing platform supports also restricts an employer and crowd inspectors. Besides, whether it allows reviewing the work submitted by crowd inspectors before making payment for that or not? What sort of contract its supports, i.e., per hour, fixed price? It also affects the crowdsourcing process. Ideally, a crowdsourcing platform should support reviewing work with different options, i.e., accept, revise, discard, before making final payment to maintain the quality of the work. Consider the following guidelines in this phase:

- A crowdsourcing platform should have the facility to maintain authentic profiles of the crowd-workers containing reviews and ratings by previous employers on the projects they have taken in the past

- Check out the payment methods that the crowdsourcing platform supports, i.e., Credit Card, Western Union, Direct fund transfers to bank accounts, etc., and make sure you can successfully send and receive payments at least through one method.

- A prospective crowdsourcing platform should allow reviewing the work submitted by crowd-workers to maintain the quality of the work.

- An ideal crowdsourcing platform should have a dispute resolution mechanism as well to deal with any possible conflicts between the crowd-workers and employers

- A prospective crowdsourcing platform should support different contract types, i.e., Fixed Price and Per Hour. A Fixed Price contract would be more suitable if you have a good assessment of the amount of work and its price. Make sure to divide work into milestones and release payments with milestones as; if you are not satisfied with any crowd worker, you can stop working with them early to avoid further waste of time and cost. Per Hour contract type is suitable when you are unsure about the amount of work to be done and you only have an idea about the per hour cost of hiring a resource.

**Agreement**

The contract to be agreed upon between crowd inspectors and employer is important in a way that it contains important information, i.e., instructions about what is to be done, and how, the criterion for reviewing the work, the deadline for submission of work, type of contract (Fix Price, Per Hour), etc. Once all these details are agreed upon, a crowd usability inspection project may be executed. Following suggestions can be helpful in this phase:

- Make sure that you provide clear instructions for what is to be done and in which file format, i.e., MS word doc, MS Excel sheet, etc. In this regard, you can provide a template with an example to report usability problems. Moreover, it would be wise to ask the crowd inspector to document the usability problems with related snapshots of the area of UI and web-link.

- It is important to mention for usability inspection that either you need to identify usability problems from crowd inspectors or suggestions to fix these problems. Both are two different aspects and need time and cost accordingly. In other words, just identifying usability problems is not that time-consuming and costly compared to suggesting a solution for them.

- It is wise to design payment criteria to thoroughly motivate the crowd inspectors to investigate the UI for usability problems thoroughly. For instance, you can offer a bonus for outstanding usability inspection reports.

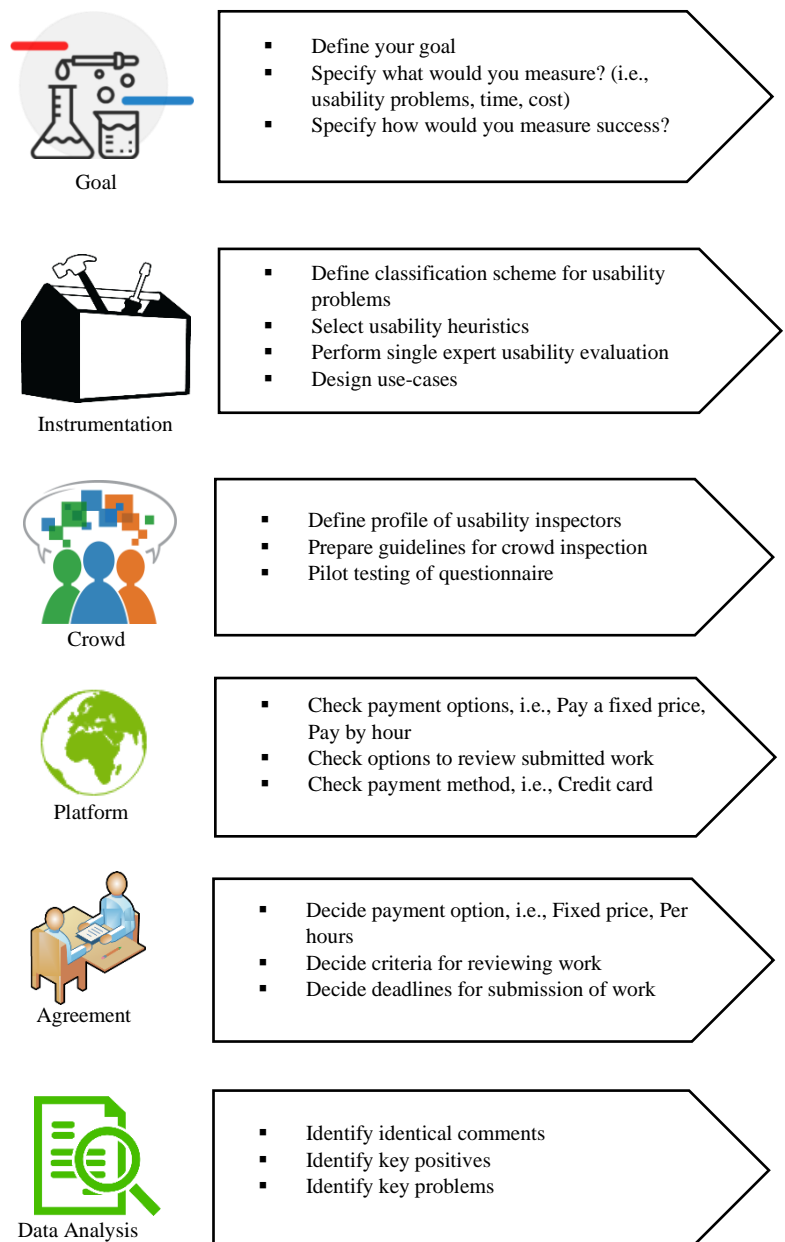

Figure 6.2 – Framework for crowd usability evaluation

**Data Analysis**

Once the crowd inspectors have submitted their usability inspection reports, further analysis is required to collect certain information elements, i.e., identical comments, false alarms, key problems, etc. After completing the analysis, usability findings may be used to fix the usability problems in the user interface. Subsequent crowd usability inspections may be performed after fixing the usability problems until there are no more problems. The following suggestions may be useful in this phase:

- Define the parameters for the analysis of crowd usability inspection reports, i.e., Identical comments, False Alarms, Bugs, key usability problems

- Define the severity levels for usability problems

▪ It would be wise to focus on key usability problems to reduce the time and cost of change

### 6.1.4 Process for improvement of the framework

We can improve this framework using a quality assurance model called PDCA (Plan-do-check-act Procedure) (Johnson, 2002).

**Plan:** Identify an opportunity for change and plan for it.

**Do:** Apply the change at a small scale

**Check:** Analyze the findings and figure out what you have learned.

**Act:** Act upon what you have learned. If the change did not result in success, repeat the process with a different strategy. If the change was successful, then integrate the change on a larger scale. Repeat the cycle for more improvements.

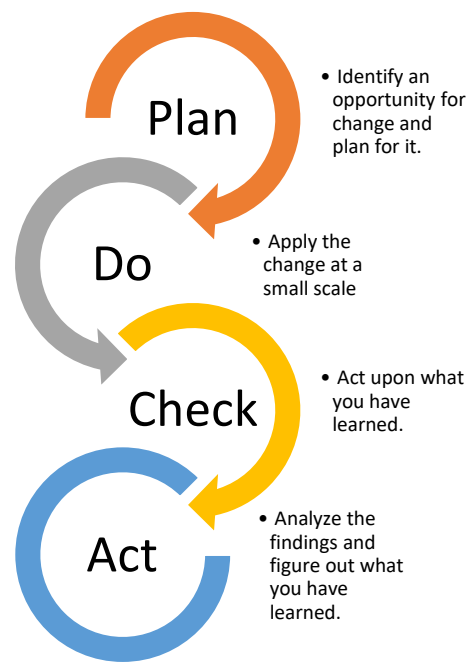

Figure 6.3 – Plan-do-check-act cycle

The rest of the chapter is organized as section 6.2 is a research method, section 6.3 is results and discussion, section 6.4 is validity threats and limitations, and section 6.5 is the conclusion and future work.

## 6.2    RESEARCH METHOD

In this research work, we would perform a case study to observe how our proposed framework for crowd usability inspection works in the real context, i.e., in a small/medium-sized budget software development organization, to improve the product's usability.

### 6.2.1    Research Question

The research question for this case study is as below:

*"Is crowd usability inspection effective enough for practitioners to translate its findings into a successful redesign in practical settings?"*

### 6.2.2    Preposition

The following proposition is used in this case study:

**RP1.** Novice crowd usability inspection can be used effectively by usability practitioners to redesign a software product in practical settings successfully.

### 6.2.3    Design

The design of this case study is single-case and embedded, as we would select a typical small/medium-sized budget software development organization with multiple units of analysis.

### 6.2.4    Unit of Analysis

The following units of analysis would be focused on in this case study:

1. Crowd usability inspection framework

2. Number of usability problems found after each iteration of crowd usability inspection

3. Number of usability problems fixed after each iteration of crowd usability inspection

4. Improvement in the usability of test object after each iteration of crowd usability inspection

5. Understanding and adaptation of crowd usability inspection framework by usability practitioners

### 6.2.5    Data Collection

The data collection source would be primary and direct using self-reported usability inspections. Data triangulation would be achieved by involving multiple crowd usability inspectors.

### 6.2.6    Selection Criteria for the Case Organizations and Test Objects

The proposed framework for novice crowd usability inspection was targeted at budget-constrained software development organizations. Besides, the use of the proposed framework has been investigated in a multi-experiment study for websites only (Nasir, Ikram, & Zakia, 2022). Therefore, we employed the following selection criteria to select the cases for this case study:

**a)** The case organization should be a small or medium-sized budget-constrained software development organization

**b)** The case object preferably be a website under development

**c)** The case organization should have a quality assurance team consisting of 1-2 usability practitioner(s)

The necessity to have a quality assurance team of practitioners in the case organization is to execute the crowdsourcing framework activities, i.e., analyzing novice crowd usability inspection reports, etc.

### 6.2.7    The Case – Usability Evaluation of a Transactional Website

This case study examines the working of the proposed framework for crowd usability inspection to enhance the usability of software products while promoting the usability evaluation practices in the budget-constrained software development organization. In this regard, we selected a small budget-sized software development organization with an annual budget of 64,000 USD, having a total 20 number of employees. Currently, they have three people working related to UI/QA; one of them has post-graduate with eight years of experience, while the other two are undergraduates with 1 to 2 years of experience. For the case study, we selected a transactional website (www.cobbpromo.com) as a test object, under development in this organization, with the final release due in a couple of weeks. This website provides the facility to order customized banners and flags online.

### 6.2.8    Executing the Framework Activities

We executed the case study by following the activities of the crowd usability inspection framework as below:

**Goal**

The goal of crowd usability inspection was to economically and effectively enhance the product's usability, i.e., www.cobbpromo.com. In this regard, we performed multiple sessions of crowd usability inspection and fixed those problems until there were no more critical usability problems found in the product.

**Instrumentation**

To achieve our goal, we needed to perform a single expert usability evaluation of the website. We hired a usability expert based on the following criterion:

1. Must have a degree in computing (BS, MS, Ph.D.)

2.  At least have five years of practical experience in usability evaluations

3.  Preferably have a certification in usability evaluation

An HFI CUA (Human Factors International Certified Usability Expert) was hired, having a masters' degree in cognitive sciences, with 15 years of experience in usability evaluation based on the above criterion. Following artifacts were used for expert usability evaluation of the website:

1.  10 Usability heuristics by Jakob Nielsen for usability inspection, see Annex A

2.  A classification scheme for usability problems, see Appendix G

Based on the findings of a single expert usability evaluation, the authors designed uses-cases for crowd usability inspection as the local UI/QA staff was too busy with other projects. However, local UI/QA personnel later reviewed the use-cases, and updates were made accordingly.

**Crowd**

We set the following selection criterion for hiring crowd usability inspectors:

1.  Must have an undergraduate degree in computing or a relevant field

2.  Preferably have 1-2 years of experience in software testing

The same set of usability heuristics was shared with crowd inspectors for usability inspection as with experts earlier, provided in Annex A. A set of use-cases designed earlier were drafted in a questionnaire to share with crowd usability inspectors (see Appendix H).

**Platform**

Considering the different factors discussed earlier in section 5.1.2, proposed framework, for crowdsourcing platform, i.e., payment method, work review mechanism, contract types, etc., we checked different crowdsourcing platforms, Amazon Mechanical Turk, Fiverr, uTest, and Upwork. We found Upwork more suitable as it allows reviewing the work submitted, and its payment policies and contract types are very flexible with user control and freedom.

**Agreement**

After selecting Upwork as a crowdsourcing platform, we posted our job consisting of all the contracts details. The payment method was a credit card, with contract type as a fixed price. We chose, a contract type fix price as crowd inspectors may not try to bill us unnecessarily. A set of usability heuristics (see Annex A) and a coding scheme to classify the usability problems (Appendix G) were provided as part of the contract document for expert heuristic usability inspection. Besides, a questionnaire consisting of use-cases was also part of the novice crowd usability inspection's contract document. The job was made available online through Upwork with a set deadline for submission of work.

**Data Analysis**

A total of six novice crowd inspectors were hired based on the criterion set earlier; their profiles are shown in table 6.1. We have assigned crowd inspectors identification letters starting from the letter "C1" to "C6". Three out of six novice crowd inspectors (C1-C3) participated in the first iteration of the novice crowd usability inspection, while the following three (C4-C6) participated in the second iteration. To categorize the usability problems, we used a classification scheme based on the work of Molich and Dumas (2008), provided in Appendix G.

Table 6.1 – Profiles of Crowd Usability Inspectors

| Phase – I | | |
|---|---|---|
| **Resource ID** | Education | Experience in Software Testing (Years) |
| C1 | Under-Graduate (Computing) | 1 |
| C2 | Under-Graduate (Computing) | 2 |
| C3 | Under-Graduate (Electronics) | 2 |
| **Phase – II** | | |
| **Resource ID** | Education | Experience in Software Testing (Years) |
| C4 | Under-Graduate (Computing) | 1 |
| C5 | Under-Graduate (Computing) | 0 |
| C6 | Under-Graduate (Computing) | 1 |

## 6.3    RESULTS & DISCUSSION

The crowd usability inspection employing novice inspectors was conducted in multiple iterations. Therefore, we would discuss the results for each iteration-wise subsequently.

### 6.3.1    First Round of Novice Crowd Usability Inspection

A total of four usability evaluation reports, including three novice crowd usability inspections, and a single expert evaluation, were received in the first phase of usability inspection for further analysis. The authors discussed the usability inspection reports with the UI/QA team of www.cobbpromo.com. It took 30 hours to analyze 82 comments by a panel of three authors and a QA/UI team consisting of 3 members. A consensus was easily developed on 85% of the comments, while the rest were sorted out with a careful discussion. A total of 51 unique comments were reported, out of which 22 were false positives. The rest of the 29 comments were unique issues, consisting of 18 minor problems, five serious problems, one critical problem, and five bugs; see table 2 for an overview of results. An overlap ratio of 66% (16 out 24 problems issues) was observed between single expert usability inspection and crowd usability inspection. Besides, all the severe and critical usability issues identified in a single expert evaluation were also reported by crowd usability inspection.

Table 6.2 – Overview of results

| | Phase – I | Phase – II |
|---|---|---|
| Original comments | 82 | 3 |
| Comments after splitting & combining | 51 | 3 |
| Total number of unique issues | 24 | 3 |
| Minor issues | 18 | 3 |
| Serious issues | 5 | 0 |
| Critical issues | 1 | 0 |
| Bugs | 5 | 0 |
| False Positives | 22 | 0 |

### 6.3.1.1 Identical Comments

A total of 36 comments out of 82 were identified as identical comments. Identical comments were merged into atomic comments. Moreover, compound comments were split into simple atomic comments. Following is an example of a compound comment that has been split into multiple atomic comments:

*If I want to read a review about a product, I cannot see if there is one on the tab; I have to click the tab and see any review. It looks like I only can write a review on the review page, or is it because there is no review. You do not know about this design. I can make a review of the product before I buy it. How trustworthy are these reviews?*

*1) If I want to read a review about a product, I can't see if there is one on the tab, I have to click the tab and see if there is any review*

*2) I can make a review of the product before I bought it? How trustworthy are these reviews?*

In 16 cases, compound comments were split into simple atomic comments.

### 6.3.1.2 Key Issues

Molich and Dumas (2008) proposed that if multiple usability inspectors have identified an issue as serious or critical, then such an issue may considerably impact user interface design. They considered such issues as a quality measure and termed them as key issues. Likewise, in this case, study, a single expert evaluation combined with crowd usability inspection identified six key usability issues. Five of these six key issues were identified as serious, while one was of critical severity. Following is an example of a key issue:

*User can't change billing or shipping address without changing both to the same address. There is only one address field when I edit, and that edit, change both addresses.*

### 6.3.1.3 Classifications

Expert usability evaluation (single expert based) used the classification scheme provided in Appendix A. On the other hand, crowd usability inspectors' identified usability issues were matched with expert usability inspection, and their severity was determined, except for three usability issues that did not match with expert evaluation. A usability expert later determined the severity of these three usability issues as minor. Besides, no other contradictions were noticed.

**6.3.1.4 False Positives**

The false positive is a mistakenly identified usability issue (Gray and Salzman, 1998). Moreover, if fixing a usability issue does not improve the usability of the product, then this usability issue would be considered a false positive (Molich and Dumas, 2008). Besides, (Molich and Dumas, 2008) also contested the hypothesis that usability problems identified by experts that are not confirmed by user-based methods are false positives. They believed that such a hypothesis would have been more plausible if the results of the user-based methods had been reproducible. However, the panel dropped out such comments that are not reproducible or fixing those would not improve the website's usability. A total of 7 comments were classified as false positives, see table 3. Following are the sample false positives identified.

1. *"Compare Products" options should not be on global navigation (It is not clear that if we remove the "Compare Products" option from global navigation, how it would increase the usability of the website)*

2. *Breadcrumb flow is not correct (However, we did not find any inconsistency in this regard)*

**6.3.2   Second Round of Novice Crowd Usability Inspection**

In the first round of novice crowd usability inspection, five out of the total six key issues were identified as serious, while one was of critical severity. The development team was provided a set of changes in the website based on the key issues identified in the first round of the novice crowd usability inspection. The development team made changes to fix the six key usability issues in the website. Later, the UI/QA team designed a questionnaire based on the six key usability issues identified in the first round of the novice crowd usability inspection to validate earlier changes. In this regard, three more crowd inspectors were hired. As shown in table 6.4, the results showed that novice crowd usability inspection did not come across any new severe or critical usability problems, nor any key issues identified in the first round. Therefore, there was no need to conduct a third round of crowd usability inspection.

Table 6.3 – Detailed results

| Resource | E1 | C1 | C2 | C3 | C4 | C5 | C6 |
|---|---|---|---|---|---|---|---|
| The Case – www.cobbpromo.com | | | | | | | |
| Total number of original comments | 39 | 24 | 24 | 28 | 2 | 1 | 0 |
| Total number of comments after splitting & combining | 31 | 17 | 19 | 25 | 2 | 1 | 0 |
| Total number of minor problems | 14 | 7 | 7 | 10 | 2 | 1 | 0 |
| Total number of serious problems | 5 | 3 | 3 | 4 | 0 | 0 | 0 |
| Total number of critical problems | 1 | 1 | 1 | 1 | 0 | 0 | 0 |
| Total number of key problem issues | 6 | 4 | 4 | 5 | 0 | 0 | 0 |
| Total number of bugs | 3 | 1 | 2 | 3 | 0 | 0 | 0 |
| False Positives | 2 | 1 | 2 | 2 | 0 | 0 | 0 |

## 6.4    FEEDBACK ON THE FRAMEWORK

To get feedback on the framework for crowd usability inspection, an interview-based survey was conducted in the case organization. All three quality assurance team members participated in the survey. A predefined interview template was used to conduct the interview, see Appendix I. In terms of the understandability, all the participants on average rated the framework for crowd usability inspection as easily understandable, i.e., 9/10. To design the instrument for crowd usability inspection, one participant mentioned that searching for a suitable usability expert is a tedious process. One must send job request invitations to several prospective candidates. He also mentioned that some crowdsourcing platforms, i.e., Upwork restricts the number of invitations that an outsourcer can send to potential crowd workers using a basic free account. Besides, highly rated usability inspectors are already occupied with several other projects and their demands for payment are also very high. Another UI/QA team member pointed out that designing use-cases based on the expert usability inspection report are a time-consuming process. No problem was reported regarding hiring novice crowd usability inspectors, drafting the agreement/contract, and using a crowdsourcing platform, i.e., Upwork. The project manager when asked, rated the framework for crowd usability evaluation in terms of cost as quite economical, i.e., 8/10. Besides, the project manager also mentioned that it was easy to convince the customer for crowd usability evaluation. A couple of developers when asked, pointed out that it was easy to fix the usability issues identified by crowd usability evaluation as they were most relevant to user interface design. Moreover, one developer mentioned that since it was not the responsibility of the development team to identify usability issues except to fix them, therefore, it makes their job easy in this way. On average, development team participation in fixing the usability issues was rated as high, i.e., 9/10. To integrate the framework for crowd usability inspection in the development process, when asked, participants found it quite easy and useful and reported no issues. The framework for crowd usability inspection was applied just before the final software release to find any usability issues. However, the project manager pointed out that the application of this framework is equally useful for initial prototypes in early software development phases, i.e., system analysis and design.

## 6.5    VALIDITY THREATS & LIMITATIONS

### 6.5.1    Validity Threats

We considered four aspects of the validity threats identified by Runeson & Host (2009) for case studies in software engineering. We will discuss each below:

### 6.5.1.1 Construct Validity

Construct validity threat refers to the extent to which operational measures employed reflect what the researcher intended to investigate compared to what was analyzed according to research questions. To ensure that usability heuristics are interpreted and understood by usability experts and crowd usability inspectors in the same way, each usability heuristic was explained with an example, see Annex A. Moreover, to ensure that all usability inspectors classify the usability problems the same way, a common classification scheme was shared with all usability inspectors (see Appendix G). Besides, the tasks in the questionnaire were accompanied by snapshots to help crowd usability inspectors to interpret the scenarios in the same way, see Appendix H.

### 6.5.1.2 Internal Validity

Internal validity threat exists when causal relationships are investigated, and a researcher is not aware of other factors or their influence on the studied factors. Our central unit of analysis was the crowd usability inspection framework, whose effect was examined in terms of several usability problems found and fixed after each iteration of crowd usability inspection. Nevertheless, there could have been other factors that affected the usability evaluation process. In this regard, we neutralized the effect of the profile of crowd usability inspectors' and experts by fixing them to set criteria, discussed in subsection Instrumentation. Moreover, to overcome any biases in the analysis of usability inspection reports, we discussed each usability inspection report submitted with each inspector to avoid any misinterpretation.

### 6.5.1.3 External Validity

External validity refers to how far findings are generalizable to other cases and valuable for the people outside the investigated case. This case study established that small and medium-size, budget-constrained software development organizations can successfully employ a crowd usability evaluation framework to identify and fix usability problems. Therefore, the findings of this case study are useful and generalizable to small and medium-size budget-constrained software development organizations.

### 6.5.1.4 Reliability

This aspect of validity threat is about how much the data and analysis are dependent on the researchers and whether the same results are reproducible by other researchers, following the same research methodology. In this regard, a research method was clearly defined and followed, consisting of the research question, prepositions, units of analysis, and crowd usability inspection framework's activities. Besides, three researchers performed this study, involving a team of UI/QA team members in data analysis to enhance the reliability of the study. Similarly, the usefulness of crowd usability inspection has already been established by our previous multi-experimental study (Nasir, Ikram, & Jalil, 2022). Moreover, this study has been designed so that its results are self-validated, i.e., The usability problems found and fixed in the current usability inspection cycle do not appear in subsequent usability inspection cycles.

### 6.5.2    Overheads of the Framework

The framework for novice crowd usability inspection involves the following overheads:

a)  A significant amount of effort would be required to define goals instruments, hire crowd inspectors, forge agreements, and make payments.

b)  Designing use-cases to develop a questionnaire for novice crowd usability inspection.

c)  Analyzing usability comments to extract usability problems, filtering false alarms, and determining the usability problems' severity.

Compared to expert usability inspection, novice usability evaluators need to be supported with use-cases. However, designing the use-cases is time-consuming. Nevertheless, tasks are also identified in usability testing as well. Besides, analyzing the usability comments of novice crowd inspectors requires intellectual effort. However, the efforts to analyze the usability comments can be mitigated using machine learning text-mining techniques. A similar approach has been suggested by (Xu et al., 2015, Wang et al., 2017).

### 6.5.3    Limitations of the Framework

The framework for novice crowd usability inspection has a few limitations:

a)  Software projects with confidentiality & privacy concerns may not be suitable for crowd usability evaluation. Primarily the software products related to the military & defense-related industry require secrecy.

b)  Software projects that are not yet publicly released may not be suitable candidates for crowd usability evaluation. The crowd usability of such projects can expose their design and architecture to competitors, leading to the risk of plagiarism and copyrights concerns.

c)  The crowd usability evaluation is significantly dependent on the skills and experience of the available crowd. The unavailability of crowd usability inspectors with appropriate skills and experience can also affect the outcome of the framework for crowd usability inspection. Sometimes, the available crowd inspectors may not be the best candidate for crowd usability inspection.

d)  Software projects with strict time constraints may not be suitable for crowd usability inspection. Sometimes, crowd inspectors do not complete their work on time. Eventually, one may have to look for and hire new crowd inspectors that can substantially increase the efforts and time to complete the job.

# Chapter 7

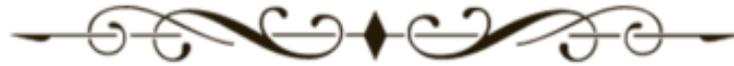

# 7.    Results & Discussion

This Thesis aimed to make usability practices more affordable yet equally effective for budget-constrained software development organizations to help them in improving the usability of their products. We achieved this objective in three phases, a) Identified and examined the current research evidence in usability inspection methods, b) Examined the wisdom of the crowd for usability evaluation by employing novice crowd inspectors in comparison with experts, c) Devised a framework for crowd usability inspection and validated it in practical settings by a case study. We would further discuss these three phases in the perspective of research questions proposed in the thesis synopsis.

## 7.1    RQ1: What evidence has been reported in the literature regarding usability inspection methods?

A systematic mapping study (Chapter 4) was conducted to aggregate usability inspection methods to answer this question. This mapping study allowed us to identify different trends in the use of usability inspection methods, a) Heuristic evaluation is the most widely employed usability inspection method, b) Web applications are the most frequently evaluated application types for usability, c) The most frequently violated usability heuristics are consistency & standards, and visibility of the system status, d) An increase in the tendency to employ tool-support is observed in user testing compared to usability inspection methods, e) An increasing trend for usability evaluation studies was observed for mobile applications; therefore, usability evaluation methods/heuristics for mobile applications may also be a potential research direction.

The number of publications for web applications in our data set is slightly less than Fernandez et al.'s systematic review results from 2004 to 2007 (see figure 4.13) (Fernandez et al., 2011). This slight decrease is that we only included publications for Usability Inspection Methods while Fernandez et al. included publications for User Testing. Besides, Fernandez et al. added papers through manual search in various relevant conferences and journals (Fernandez et al., 2011). Therefore, increasing the number of publications in their study for web applications is justified. Fernandez et al. did not cover the whole year of 2008 as they performed their search in March 2008 (Fernandez et al., 2011). Therefore, the number of publications reported by Fernandez et al. in 2008 is significantly less than our results (see figure 4.15).

Besides that, Fernandez et al. reported that 69% of studies employed manual usability evaluation methods while only 31% reported using tool-supported usability evaluation methods. However, our study shows that 91% of studies employed manual usability evaluation methods while only 9% employed tool-supported methods. One reason for the increase in the number of tools supported by Fernandez et al.'s systematic review results is that they included studies using User Testing, Metric based Usability Evaluation methods, and Formal Usability Evaluation Methods. It is relatively easy to employ tool support in User Testing than in Usability Inspection methods that evaluate the usability of a software system based on usability experts' knowledge, skills, and experience. Usability Testing methods often employ tool support through performance measurement, user activity analysis, etc. (Ivory & Hearst, 2001). To validate it further, we went through all the 26, usability inspection-based studies in Fernandez et al.'s data set identified by Rivero & Conte, (2012). We observed that only 19% of studies were tool-supported, while 79% employed manual usability inspection. Therefore, we can infer that usability inspection-based studies used less tool support than other usability evaluation-based studies.

Likewise, Fernandez et al.'s study reported that 53% of studies empirically validated usability evaluation methods while 47% did not employ empirical validation. Our results show that 94% of studies empirically validated the usability evaluation method, while only 6% were not validated empirically. The trend of empirical validation is gradually increasing in usability evaluations, see figure 4.16. The sudden decline in the graph in 2010, 2012, and 2013 is due to situational factors discussed above. Fernandez et al.'s Systematic Mapping Study included publications from 2009 (Fernandez et al., 2011). Therefore, we could not compare our results for recent years with Fernandez et al.'s work (Fernandez et al., 2011).

Additionally, the finding that the Heuristic and Cognitive methods are the most frequently investigated usability inspection methods is common in our results and systematic mapping study conducted by Rivero & Conte (2012). However, Rivero et al.'s Mapping Study on Usability Inspection methods was based on 26 publications extracted from Fernandez et al.'s Systematic Review data set. Moreover, we did not find any other Systematic Review or Mapping Study to compare our results.

## 7.2    RQ2: Is the wisdom of crowd effective for usability evaluation?

To address the question, we examined the wisdom of the crowd for usability evaluation. In this regard, novice usability inspectors were hired from crowdsourcing platforms to perform usability inspections compared to usability experts. The novice crowd inspectors were equipped with use-cases designed based on a single expert's heuristic usability inspection. Study 2 (Chapter 5) of Thesis concluded that novice crowd usability inspection on average could give comparable results to an expert heuristic evaluation. The results showed that novice crowd inspectors, on average, find the same usability problems with respect to content and quantity.

Moreover, novice crowd usability inspection is more economical and less time-consuming than expert heuristic usability inspection. The objects of the study included a web dashboard and two other transactional websites to generalize the findings.

How efficiently resources are used is an essential aspect of comparing UEMs. If we look at Figure 10.3, we can see that experts have outnumbered novice crowd inspectors to identify key issues per person. However, person-hours used to identify key usability issues are not free. The person-hours used to change the whole view altogether. The key usability issues identified per hour by each resource are shown in Figure 5.3. We can see that novice crowd inspectors are more efficient than experts in identifying key usability issues per hour. Figure 5.4 indicates that experts have identified more key usability issues at the cost of more time. Likewise, Figure 5.4 shows that experts' key usability issues were quite expensive compared to crowd inspectors. It is worth mentioning that novice crowd inspectors identified 27 out of the total 30 key usability issues.

This Thesis also contributed a thorough discussion on the challenges faced by crowdsourcing in usability evaluation and proposed solutions for it. In other words, this Thesis demonstrated the effective use of the crowd's wisdom for usability inspection.

## 7.3    RQ3: Is the crowd usability inspection framework effective in practical settings?

In the light of the findings of the experimental study in study 2 (Chapter 5) of this Thesis, a framework for crowd usability inspection was proposed. The framework's application was then demonstrated in study 3 (Chapter 6), using a case study. The case selected was a medium-sized (in terms of a budget, team) software development organization. An ongoing software project, i.e., a transactional website, was subjected to crowd usability inspection. The study showed that 3-5 usability inspectors could successfully eliminate key usability issues in a software product within 2-3 rounds of crowd usability inspections. This Thesis successfully demonstrated the use of the crowd usability inspection framework in practical settings.

The crowd usability inspection employing novice inspectors was conducted in multiple iterations. A total of four usability evaluation reports, including three novice crowd usability inspections, and a single expert evaluation, were received in the first phase of usability inspection for further analysis. The authors discussed the usability inspection reports with the UI/QA team of www.cobbpromo.com. It took 30 hours to analyze 82 comments by a panel of three authors and a QA/UI team consisting of 3 members. A consensus was easily developed on 85% of the comments, while the rest were sorted out with a careful discussion. A total of 51 unique comments were reported, out of which 22 were false positives. The rest of the 29 comments were unique issues, consisting of 18 minor problems, five serious problems, one critical problem, and five bugs; see table 2 for an overview of results. An overlap ratio of 66% (16 out 24 problems issues) was observed between single expert usability inspection and crowd usability inspection. Besides, all the severe and critical usability issues identified in a single expert evaluation were also reported by crowd usability inspection.

In the first round of novice crowd usability inspection, five out of the total six key issues were identified as serious, while one was of critical severity. The development team was provided a set of changes in the website based on the key issues identified in the first round of the novice crowd usability inspection. The development team made changes to fix the six key usability issues in the website. Later, the UI/QA team designed a questionnaire based on the six key usability issues identified in the first round of the novice crowd usability inspection to validate earlier changes. In this regard, three more crowd inspectors were hired. As shown in table 1.4, the results showed that novice crowd usability inspection did not come across any new severe or critical usability problems, nor any key issues identified in the first round. Therefore, there was no need to conduct a third round of crowd usability inspection.

# Chapter 8

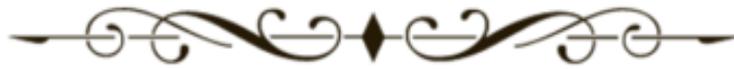

# 8.     Validity Threats and Limitations

We employed different research methods in each study conducted as part of this thesis work. Each study was exposed to various validity threats. The validity threats and the preventive measures to address them are discussed in detail in each study. Study 2 (Chapter 5) reports a comparative usability study based on a series of experiments. Gray and Salzman, in 1998, reported several validity threats for comparing usability evaluation methods. Study 2 (Chapter 5) has addressed the validity threats identified by (Gray and Salzman, 1998). Study 3 has reported a case study and has managed the validity threats Runeson & Host (2009) identified for conducting case studies in software engineering. However, here we would provide a collective discussion on validity threats to the Thesis based on the classification of Wohlin et al. (2012) and the measures are taken to address them.

## 8.1     Construct validity

This type of validity represents that to what degree operational study measures represent what the researchers intended to investigate, compared to what they investigated as per the research questions. The experiment presented in study 2 (Chapter 5) addressed the causal construct validity in four different aspects, i.e., a) Defining, operationalizing, and understanding the UEM, b) Applying the UEM in only one way (mono-operation bias) c) Examining the UEM on only one type of software (mono-method bias) d) Learning effect as a result of the interaction of the treatments.

In study 2 (Chapter 5) and study 3, to overcome any misunderstanding about the UEMs and mono-operation bias, we provided clear definitions for both UEMs with different artifacts, usability heuristics with guidelines, use-cases with guidelines. The expert heuristic evaluation employed usability heuristics, while novice crowd usability evaluation employed use-cases with self-explanatory instructions on how to execute them. Study 2 (Chapter 5) applied the UEMs to two different websites and a web dashboard to mitigate the mono-method bias. None of the participants were given both treatments to eliminate the learning effect. In study 3, a standard classification scheme was shared with them to ensure that all usability evaluators consistently classified the usability problems. Besides, all the participants completed their usability evaluations individually without interacting with each other.

## 8.2     Internal validity

This validity category represents that a researcher is not aware of other factors that might influence the results. Internal validity is relevant when causal relationships are examined. In study 2 (Chapter 5), to ensure that the differences between the UEMs are causal instead of correlational, three key aspects of the internal validity were focused on, i.e., instrumentation, selection, and setting.

In study 2 (Chapter 5) and study 3 (Chapter 6), the two main biases in instrumentation, i.e., the severity of usability problems and classification, were addressed by ensuring that each expert performs usability evaluation independently. The severity and classification of crowd-identified usability problems were determined by comparing crowd inspections with an expert's. The novice crowd inspectors identified usability issues independently by executing use-cases and were not asked to categorize the usability issues based on severity or heuristics. The experts and the crowd inspectors were self-selected based on set criteria to mitigate the selection bias. In the case of a systematic mapping study (Chapter 4), we mitigated the selection bias, and we did not miss the relevant literature by searching all the major databases, including Compendex EI, IEEE explorer, ACM Digital Library, and Science Direct.

## 8.3    External validity

This aspect of the validity concerns how much we can generalize the study's findings and to what extent these findings are important to people outside the examined case. In study 2, the study's findings are limited to usability experts and novice crowd inspectors selected as per predefined profiles. Moreover, the results of the study are generalizable to websites only. In study 3, the study's findings are only applicable to small and medium-sized-budget organizations that find it challenging to hire expensive usability experts to improve their products' usability.

## 8.4    Reliability

This aspect of validity refers to the extent to which data and the analysis rely on the specific researchers. In other words, if another researcher replicates the same research, then the results should be the same. The dependence of the data and analysis on specific researchers makes the reliability of the study questionable. The experiment reported in study 2 (Chapter 5) was repeated three times with three different subjects to make results more reliable. However, the authors analyzed the novice crowd usability inspection reports. The results may vary when actual practitioners would analyze novice crowd inspectors' usability inspection reports. In study 3, the case study was designed so that results are self-validated, i.e., usability problems reported in the current usability inspection do not appear in the subsequent iteration.

To ensure the reliability of the study in Chapter 8, we followed guidelines for conducting a systematic mapping study (Kitchenham & Charters, 2007). We set and used a proper search string and inclusion/exclusion criteria to extract all the relevant studies.

# Chapter 9

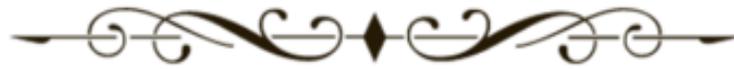

# 9. Conclusion and Future Work

In this Thesis, I have investigated crowdsourcing in the context of usability evaluation. In this regard, current research evidence in literature was identified, and different trends in usability evaluation inspection were reported. I also identified that novice crowd usability inspectors, with the help of use-cases (guided by a single expert's heuristic usability evaluation), give comparable results to usability experts.

## 9.1 Findings & Contributions & Conclusion

The prominent findings of this Thesis are as below:

- The heuristic usability inspection is the most widely employed usability evaluation method, with websites as the most frequently evaluated application type for usability.
- Novice crowd usability inspection guided by a single expert's heuristic inspection gives promising results in comparison with expert heuristic usability evaluation, in terms of
    - o Usability problems are found with respect to content and quantity.
    - o The severity of the usability problems
    - o Time efficiency
- The proposed framework for crowd usability inspection employing novice crowd usability inspection is effective and economical for budget-constraint software development organizations and works well in practical settings.

Besides, other contributions of this Thesis include:

- A comprehensive discussion was provided, analyzing the challenges faced by usability practitioners, employing crowdsourcing for usability inspection.
- A method to design use-cases to help novice crowd usability inspectors was proposed.
- A method to analyze novice usability inspectors' usability evaluations was proposed.

In the light of the findings of this Thesis, we can safely conclude that wisdom of the crowd has potential for usability evaluation. Moreover, novice crowd usability inspection guided by a single expert's usability inspection is a potential economical and effective alternative for expert heuristic usability evaluation for resource-constrained software engineering organizations.

## 9.2    Future Work

In the future, this research work can be extended in the following directions.

- An updated systematic literature review investigating the usability evaluations methods may be conducted to shed light on current research trends in usability evaluation methods.

- The effect of consensus meetings among novice usability inspectors may be analyzed in terms of the performance.

- The performance of usability testing employing crowdsourcing may be compared with traditional usability testing considering the limitations of the existing literature.

- A variation of the proposed framework may be investigated to hire usability experts and compared with traditional expert usability inspection.

- The use of the proposed framework for novice crowd usability inspection may be validated with more industrial cases.

- The validation of the proposed framework for novice crowd usability inspection can be extended using a focus group of usability experts.

# APPENDIX A: LIST OF STUDIES INCLUDED

[Web link]

https://drive.google.com/file/d/0B4iBF34GTmF7ZVhLbFk0SGRzUjA/view?usp=sharing

# APPENDIX B: PUBLICATION CHANNELS

Table B.1 – Publications Channels

| Type | Name of Channel | No. of Publications | Studies |
|------|-----------------|---------------------|---------|
| Conference | Conference on Human Factors in Computing Systems (CHI) | 8 | S1, S18, S58, S60, S68, S69, S80, S87 |
| | Information Technology: New Generations (ITNG) | 7 | S105, S108, S111, S112, S114, S115 |
| | Human-Computer Interaction–INTERACT | 7 | S3, S17, S45, S49, S72, S74, S81 |
| | ACM International Conference Proceeding (AICPS) | 6 | S21, S52, S66, S76, S79, S85 |
| | Nordic Conference on Human-Computer Interaction (Nordic CHI) | 5 | S66, S76, S79, S85, S94 |
| | Empirical Software Engineering and Measurement (ESEM) | 3 | S6, S7, S96 |
| | Symposium on Software Engineering (SBES) | 3 | S5, S29, S53 |
| | In Web Information Systems Engineering–WISE | 3 | S56, S67, S75 |
| | In Human-Centered Software Engineering | 3 | S93, S8, S12 |
| | In Human Centered Design | 2 | S39, S40 |
| Journals | Interacting with Computers | 4 | S83, S84, S126, S137 |
| | International Journal of Medical Informatics | 4 | S120, S127, S129, S136 |
| | International Journal of Human Computer Studies | 2 | S123, S132 |
| | International Journal of Information Management | 2 | S124, S136 |
| | Journal of Biomedical Informatics | 2 | S133, S141 |
| | International Journal of Human-Computer Interaction | 1 | S138 |
| | Journal of Systems and Software | 1 | S119 |

# APPENDIX C: WEB-LINKS FOR TEST OBJECTS

1. WHO-EU Web Dashboard [Weblink]: https://gateway.euro.who.int/en/
2. [Weblink]: www.ebay.com
3. [Weblink]: www.booking.com

## APPENDIX D: USABILITY HEURISTICS FOR WEB DASHBOARD

Heuristics for Web Dashboard & Information Visualization [Web-Link]:

https://drive.google.com/file/d/1vYkaKHegMLqUHlk4KklTImlwnvUzYyI6/view?usp=sharing

## APPENDIX E: QUESTIONNAIRE FOR CROWD USABILITY EVALUATION

1. Questionnaire for WHO-EU Web Dashboard [Web-Link]:

   https://docs.google.com/a/iiu.edu.pk/forms/d/1I9-KCINkMWrOPhtwxnzGBr6TOErb-IQeH0u390IX_60/viewform

2. Questionnaire for www.ebay.com [Web-Link]:

   https://drive.google.com/file/d/1z5lLGCn_FPeVlgMgXw3i8s5ptrK00aSn/view?usp=sharing

3. Questionnaire for www.booking.com [Web-Link]:https://drive.google.com/file/d/1Xp-FqP5E_yCUajkhlg1BhYQxYwOTPCpg/view?usp=sharing

## APPENDIX F: KEY USABILITY ISSUES

https://drive.google.com/file/d/13nwNSdVi7sdxNHDsY3bCpwb4jMzcyGt7/view?usp=sharing

## APPENDIX G: A CODING SCHEME FOR USABILITY PROBLEM DETECTION & CLASSIFICATION

Table G.1 – Coding Scheme

| Code | Short Description | Definition |
|------|------------------|------------|
| MP | Minor problem | Causes a few seconds delay to extract desired information or understanding it |
| SP | Serious problem | Causes 1-5 minutes delay and prevents a user from understanding or extracting desired information, but the user somehow manages the information flow |
| CP | Critical problem | Prevents the user from obtaining desired information or understanding it, leading the user to complete failure |
| FP | False-positive | A usability issue is considered false positive, either if it is not reproducible or fixing that would not improve the usability of the system |
| CM | Comment | A comment can be a problem or a bug reported by a usability evaluator |
| P-CM | Problem comment | A comment that is classified as a minor, serious, or critical problem |
| A-CM | Atomic comment | A comment that cannot be further divided into more comments. Atomic comments can be resolved without affecting other usability issues |
| I-CM | Identical comment | Two or more comments are identical if fixing one comment will fix other comments as well |
| IS | Issue | A set of one or more identical atomic comments |
| K-IS | Key issue | An issue that is reported by three experts or two experts and at least one crowd usability inspector, and it is, on average, marked as important, i.e., serious or critical problem. |
| C-CL | Contradictory classifications | This occurs when the same underlying usability finding is classified differently by more than one usability evaluator. For example, one expert may classify the finding as a serious usability problem while others may mark it as a minor usability problem. |

# APPENDIX H: QUESTIONNAIRE FOR CROWD USABILITY INSPECTION

1.    First round:

https://drive.google.com/file/d/1cd07EPzVm5Q0MsqjVJC8-JMOT0kY6WYy/view?usp=sharing

2.    Second round:

https://docs.google.com/document/d/11GOvkfS1z8DvhtexvcirIOifei83uIHc/edit?usp=sharing&ouid=112764386028535750980&rtpof=true&sd=true

# APPENDIX I: INTERVIEW TEMPLATE FOR FEEDBACK ON PROPOSED FRAMEWORK

| Interviewee Details | | |
|---|---|---|
| **Company Name:** | **Country:** | **Date:** |
| **Interviewee Name:** | | |
| **Interviewee Designation:** | **Software Development Experience:** | |
| **Education:** | **Practical Usability/Quality Assurance Experience:** | |
| **HCI Skills/Certification:** | **Total Experience:** | |
| **English Language Skills:** | **Software Project Type: (Web/Mobile etc.)** | |

| Questions |
|---|

**Question# 1:**   How would you rate the framework for crowd usability evaluation in terms of **understandability**? Please mention if any problems you came across while understanding it?

Not understandable <-                                    -> Highly understandable
                    1     2     3     4     5     6     7     8     9     10

Answer: ______________________________________________

**Question# 2:**   Did you come across any problems while designing the **instruments** for crowd usability evaluation? Please mention if you came across any problems?

Answer: ______________________________________________

**Question# 3:**   Did you come across any problems while **hiring crowd usability inspectors**? Please mention if you came across any problems?

Answer: ________________________________

**Question# 4:** Did you come across any problems while **drafting a contract/agreement** with crowd usability inspectors? Please mention if you came across any problems?

Answer: ________________________________

**Question# 5:** Did you come across any problems while **using crowdsourcing platform**? Please mention if you came across any problems?

Answer: ________________________________

**Question# 6:** Did you come across any problems while **reviewing/analyzing crowd usability inspection reports**? Please mention if you came across any problems?

Answer: ________________________________

**Question# 7:** How would you rate the framework for crowd usability evaluation in terms of **cost incurring in usability evaluation**?

Answer:
Highly costly <-                                    -> Highly economical
1     2     3     4     5     6     7     8     9     10

**Question# 8:** How would you rate the framework for crowd usability evaluation in terms of **Customer participation**?

Answer:
Low participation <-                                    -> High participation
1     2     3     4     5     6     7     8     9     10

**Question# 9:** How would you rate the framework for crowd usability evaluation in terms of **developer mindset/participation to fix usability issues**?

Answer:
Low participation <-                                    -> High participation
1     2     3     4     5     6     7     8     9     10

**Question# 10:** Did you come across any problems while **integrating the crowd usability inspection in organizational development process**? Please mention if you came across any problems?

Answer: ________________________________

**Question# 11:** Did you come across any other problems while using framework for crowd usability inspection? Please mention if you came across any problems?

Answer: _______________________________________________

**Additional Notes**

Enter Additional Notes.

## ANNEX A: USABILITY HEURISTICS BY JAKOB NIELSEN

10 Usability Heuristics by Jakob Nielsen for User Interface Design [Web-Link]:

https://drive.google.com/file/d/1A6h0hTGUeGatH1XHynHMG29UI5ezmrxq/view?usp=sharing